\documentclass{svjour3}                     
\smartqed  

\usepackage{amsmath}
\usepackage{pifont}
\usepackage{graphicx}
\usepackage{color}
\usepackage[colorlinks=true,linkcolor=blue,citecolor=blue,urlcolor=blue]{hyperref}
\usepackage{upgreek}
\renewcommand{\mu}{\upmu}

\usepackage{natbib}
  \setcitestyle{aysep={}} 
\usepackage{txfonts}
\usepackage{lscape}

\newcommand{\mic}{$\mu$m}

\newenvironment{system}%
{\left\lbrace\begin{array}{@{}l@{}}}%
{\end{array}\right.}

\journalname{The Astronomy and Astrophysics Review}

\begin{document}

\title{The formation and cosmic evolution of dust in the early Universe. I. Dust sources}

\titlerunning{Dust formation and evolution. I. Dust sources}        

\author{Raffaella Schneider \and
        Roberto Maiolino 
}


\institute{R. Schneider \at
              Dipartimento di Fisica, Sapienza Universit{\'a} di Roma, Piazzale Aldo Moro 5, 00185 Rome, Italy \\
              \email{raffaella.schneider@uniroma1.it}           
           \and
           R. Maiolino \at
           Cavendish Laboratory, University of Cambridge, 19 J. J. Thomson Ave., Cambridge CB3 0HE, UK \\
           Kavli Institute for Cosmology, University of Cambridge, Madingley Road, Cambridge CB3 0HA, UK\\
           \email{r.maiolino@mrao.cam.ac.uk}
}

\date{Received: date / Accepted: date}

\maketitle

\begin{abstract}

Dust-obscured star formation has dominated the cosmic history of star formation since $z \simeq 4$. However, the recent finding of significant amount of dust in galaxies out to $z \simeq 8$ has opened the new frontier of investigating the origin of dust also in the earliest phases of galaxy formation, within the first 1.5 billion years from the Big Bang. This is a key and rapid transition phase for the evolution of dust, as galaxy evolutionary timescales become comparable with the formation timescales of dust. It is also an area of research that is experiencing an impressive growth, especially thanks to the recent results from cutting edge observing facilities, ground-based and in space. Our aim is to provide an overview of the several findings on dust formation and evolution at $z > 4$, and of the theoretical efforts to explain the observational results. We have organized the review in two parts. In the first part, presented here, we focus on dust sources, primarily supernovae and asymptotic giant branch stars, and the subsequent reprocessing of dust in the interstellar medium, through grain destruction and growth. We also discuss other dust production mechanisms, such as Red Super Giants, Wolf--Rayet stars, Classical Novae, Type Ia Supernovae, and dust formation in quasar winds. The focus of this first part is on theoretical models of dust production sources, although we also discuss the comparison with observations in the nearby Universe, which are key to put constraints on individual sources and processes. While the description has a general applicability at any redshift, we emphasize the relative role of different sources in the dust build-up in the early Universe. In the second part, which will be published later on, we will focus on the recent observational results at $z > 4$, discussing the theoretical models that have
been proposed to interpret those results, as well as the profound implications for
galaxy formation.


\end{abstract}

\setcounter{tocdepth}{3} 
\tableofcontents

\newpage

\section{Introduction}
\label{intro}

Dust is a fundamental component of the interstellar medium. Dust extinction and
reddening at optical and UV wavelengths, as well as its thermal emission at
infrared and sub-millimetre wavelengths, have important implications on the
observational properties and detectability of galaxies, especially at high
redshift. Dust has a fundamental role in the cooling of the
interestellar medium and, therefore, facilitating the gravitational collapse, hence the formation of stars across a broad range of masses.

Dusty galaxies have been extensively studied in the local Universe and across the
cosmic epochs. Obscured systems are found to dominate the cosmic star formation
budget out to $z \simeq 4$ \citep{zavala2021}. The new frontier that has been opened in recent
years had been the finding of significant amount of dust in galaxies beyond
$z \simeq 4$ and out to $z \simeq 8$, i.e. within the first 1.5 billion years after Big Bang
\citep[e.g.][]{Laporte2017,Inami2022,Wang2021,witstok2023,Banados2018,Tamura2017,Popping2017}. This is an interesting timescale from a theoretical perspective, indeed it is comparable with the timescales of some of the most
prominent star formation processes, hence opening different scenarios on
the relative contribution of various dust formation channels at such early
epochs. Therefore, the first 1.5 Gyr after Big Bang represent a key and rapid transition phase in the
production and processing of dust grains.

This area of research has recently experienced an impressive growth, expecially thanks to the several
recent observational results from the \textit{Atacama Large Millimeter/submillimeter Array} (ALMA) and from the \textit{James Webb Space
Telescope} (JWST), which have triggered the development of several models to explain the content and
properties of dust in the early Universe, as well as their implications for galaxy formation and
observability.

Although this field is evolving very rapidly, we believe it is now a proper time to provide an overview of
the several findings on early dust formation and evolution, and of the theoretical efforts to explain
the observational results. As this is a massive area of research, we have organized the review in two parts.

\begin{figure*}
\centering
\includegraphics[width=0.9\textwidth]{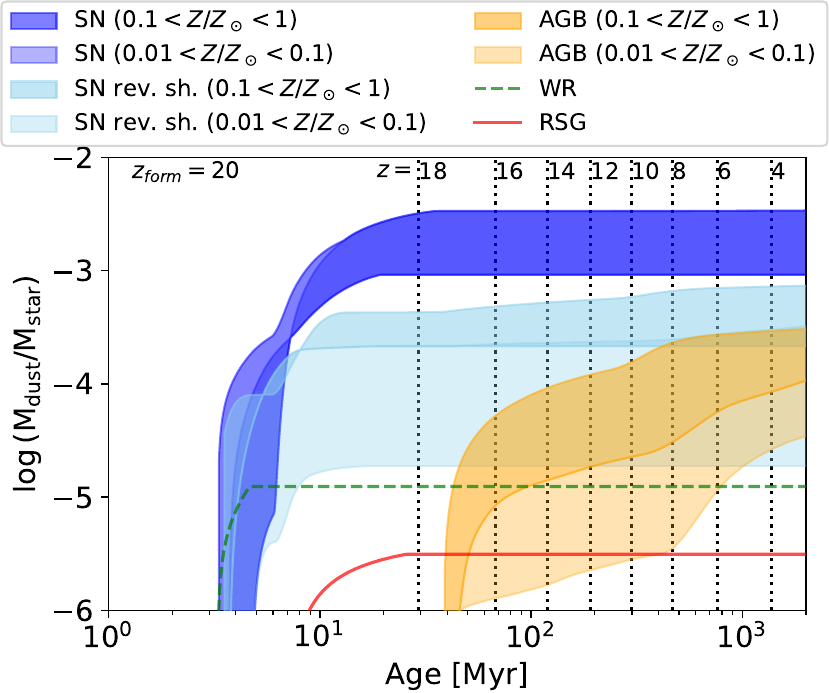}
\caption{Summary of the typical dust enrichment timescales for some of the dust production mechanisms discussed in this review. We have assumed that all stars are formed in a burst at
$z_{\rm form} = 20$ ($t_{\rm age} = 0$) with a Salpeter initial mass function in the mass range $[0.1 - 100] ~ M_\odot$, 
and we show the time-dependent cosmic dust yield, i.e. the dust mass released normalized to the total stellar mass formed. For supernovae, we adopt the yields for non rotating stellar progenitors with initial mass in the range 
$[13 - 120] ~ M_\odot$ from \citet{marassi2019} 
with no (cyan) and with (blue) the effect of the partial dust destruction due to the reverse shock (adopting a circumstellar medium density of 
$n_{\rm ISM} = 0.6 ~ {\rm cm}^{-3}$,
see Sect.~\ref{sec:snrevshock} and Table \ref{table:RSmassfrac}); for AGBs, we show the yields from the ATON model for stars with initial mass in the range 
$[1 - 8]~M_\odot$
from \citet{ventura2012a, ventura2018, dellagli2019}. For SNe and AGBs, the dark shaded regions show metallicities in the range $\rm 0.1<Z/Z_\odot <1$, while lighter shades show metallicities in the range $\rm 0.01<Z/Z_\odot <0.1$. 
We also show the contribution of Wolf--Rayet stars (WR, green dashed) and Red Super Giants (RSG, red solid), assuming that they come from stars with masses 
$\geq 40 ~ M_\odot$, and in the range $[10 - 25]~ M_\odot$, 
respectively. For more details on the adopted dust yields we refer to Sect.~\ref{sec:snmodels} for SNe, Sect.~\ref{sec:agbmodels} for AGBs, and Sects.~\ref{sec:rsg}-\ref{sec:wr} for RSGs and WR stars. In Sect.~\ref{sec:relativerole} we illustrate the relative importance of AGBs and SNe adopting different sets of yields, star formation histories, stellar initial metallicity, and stellar initial mass function. Finally, in Sect.~\ref{sec:qsodust}, we quantify the dust mass formed in quasar winds.
}
\label{fig:dustevo_summary}
\end{figure*}

The first part is presented here and aims at reviewing the landscape of the theoretical models of the 
possible sources of dust in the early Universe. This is meant to provide the essential backbone required 
for understanding the observations at high redshift, as well as the key ingredient for the models specifically aimed at interpreting in detail the high redshift observations.
In this part we focus on the theoretical scenarios describing nucleation and growth of dust grains in different sources of
dust, primarily various models for dust formation in atmospheres of Asymptotic Giant Branch (AGB) and Supernova (SN) ejecta.
However, we will also discuss models of additional
sources of dust that might be relevant in the early phases of galaxy formation, such as Red Super Giants, Wolf--Rayet stars, and
also dust formed in the quasar-driven winds, and we will also discuss the dust reprocessing in the interstellar medium. Our presentation can be useful to describe dust formation and evolution at any redshift. However, we will emphasize their specific role in the context of the timescales available in the early Universe.
Figure \ref{fig:dustevo_summary} gives a quick glimpse of the timescales involved in the dust
formation associated with some of these sources (see Sect.~\ref{sec:relativerole} for more details); while the figure is highly incomplete, it serves to illustrate the
timescales at play and why these sources of dust are relevant in different stages of galaxy formation in the early Universe, hence the motivation for this part of the review.

In this part of the review we do not cover in detail the observational aspects associated with the sources of dust, through the extensive observational studies of dust formation and destruction in various classes, which would require a separate review, and out of the scope of our primary goals of providing the information for the early universe. However, we do briefly compare the
expectations of different theoretical models with observations, which are mostly confined in the nearby Universe, for each category of dust sources.

In the second part of the review, which will be published later on, we will focus on the recent observational results investigating the dust content and properties in different classes of galaxies at $z > 4$, out to the earliest epochs for which such constraints have been obtained. We will also discuss the theoretical models and cosmological simulations that have
been proposed to interpret those results, as well as the profound implications for
galaxy formation. 

We clarify that this is not, by any means, the first review on the dust sources and dust reprocessing. Many other extensive review have been presented in the past, starting from \citet{draine2003}, which discussed the observed properties of interstellar dust grains (wavelength dependent extinction, scattering, emission, and polarization), and their implications for dust models \citep{draine2009}; dust production by supernovae \citep{gall2011, sarangi2018}, and its subsequent processing and survival in supernova remnants \citep{micelotta2018}; dust formation and mass loss of stars on the asymptotic giant branch \citep{hofner2018};
and the properties of dust in the interstellar medium of nearby galaxies \citep{galliano2018}, which provide an invaluable laboratory to explore fundamental dust processes across a diversity of environmental conditions (metallicity, star formation activity, etc.), hence constituting a necessary intermediate step towards understanding distant galaxies.

Here we leverage on those reviews, by expanding and updating them in some areas, with the specific focus of exploring the nature and origin of dust in the early universe. 

\section{Supernovae}
\label{sec:1}

\subsection{Models of dust formation in supernova ejecta} 
\label{sec:snmodels}

Since the explosion of SN1987A, direct observations of dust formation in core-collapse SNe have motivated theoretical investigations of dust condensation in SNe. Three main approaches have been followed, with increasing degree of complexity. 

\subsubsection{Classical nucleation theory}
The simplest approach adopts the so-called classical nucleation theory (CNT), which was first applied by \citet{kozasa1989, kozasa1991} to model dust formation in SN 1987A. In CNT, when a gas becomes supersaturated, particles (monomers) aggregate in a seed cluster that subsequently grow by accretion of other monomers.

For grain materials whose molecules are not present in the gas phase, the rates of nucleation and grain growth are controlled by one chemical species, which is referred to as the key species. This is the species of the reactants that has the least collisional frequency onto a target nuclei \citep{kozasa1987}.  Under these conditions, the steady-state nucleation rate (that is the number of critical clusters formed per unit volume and unit time) is given by:
\begin{equation}
   J = \alpha \, \Omega \left( \frac{2 \, \sigma}{\pi \, m_{\rm k}} \right)^{1/2} \, c^2_{\rm k} \, {\rm exp}\left[ - \frac{4 \, \mu^3}{27 ({\rm ln} S)^2}\right], 
   \label{eq:nucl_rate} 
\end{equation}
\noindent
and the grain growth rate is:
\begin{equation}
    \frac{dr}{dt} = \alpha \, \Omega \, v_{\rm k} \, c_{\rm k}.
    \label{eq:growth_rate}
\end{equation}
\noindent
In these expressions, $\alpha$ is the sticking coefficient (the probability that when a collision occurs, the collider sticks to the target), $\Omega = 4/3 \pi a_0^3$ is the volume of the monomer of the key species in the condensed phase, $\sigma$ is the surface tension of the condensed material, $m_{\rm k}, c_{\rm k},$ and $v_{\rm k}$ are the mass, concentration and velocity of the key species monomers, $\mu = 4 \pi a_0^2 \sigma/(k_B T)$ is a parameter, $T$ is the ejecta temperature and $S$ is the super-saturation ratio, expressed as:
\begin{equation}
    {\rm ln} S = - \frac{\Delta G_{\rm r}}{k_B T} + \Sigma_i \nu_i P_i,
\end{equation}
\noindent
where $\Delta G_{\rm r}$ is the Gibbs free energy for the condensation reaction $\Sigma_i \nu_i \, A_i = $ solid ($A_i$ are the chemical species of the reactants and products in the gas-phase and $\nu_i$ are the stoichiometric coefficients, which are positive for the reactants and negative for the products), and $P_i$ are the partial pressures of the $i$-th species. 

\begin{figure*}
\centerline{\includegraphics[width=\textwidth]{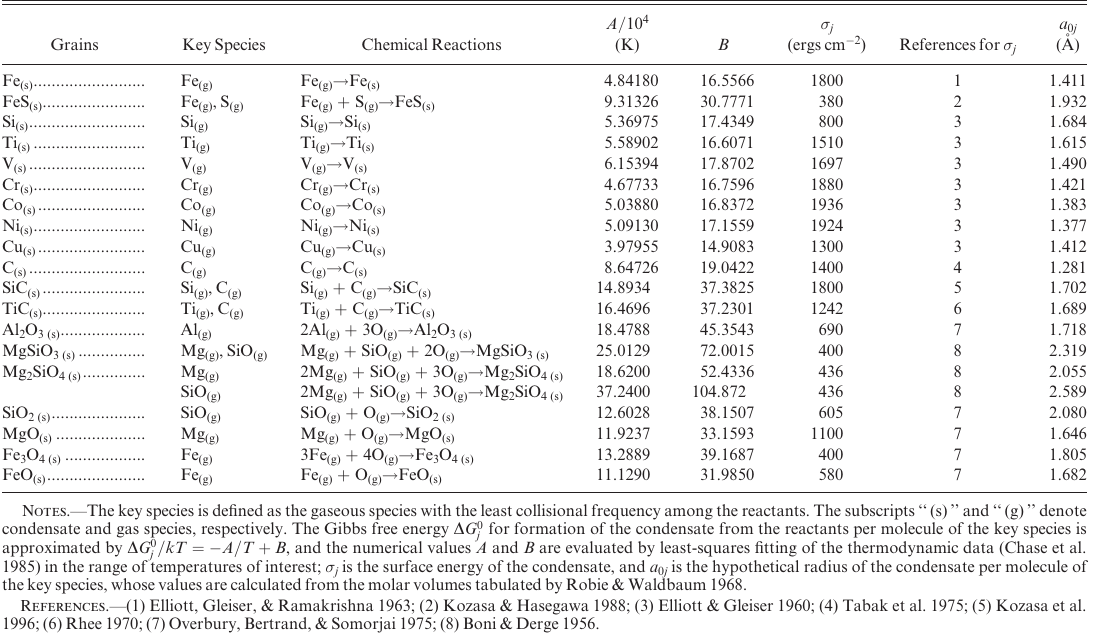}}
\caption{Grain species and properties that are generally included in SN dust formation models. The table reports the species name, the key species, the chemical reactions, the numerical values of the coefficients required to compute the Gibbs free energy, approximated as $\Delta G_{\rm r}/k_{\rm B}T = -A/T+B$,
the surface energies of the condensates, and the monomer radii (see also the Notes at the bottom of the figure for additional references). Table reproduced with permission from \citet{nozawa2003}, copyright by AAS.}
\label{table:SNdust}
\end{figure*}

The grain properties that are generally included in SN dust formation models are reported in Fig.~\ref{table:SNdust}. 

The presence of CO and SiO molecules in SN ejecta is very important for dust formation, because carbon atoms bound in CO molecules are not available to form Amorphous Carbon (AC) grains and SiO molecules take part in the reactions that lead to the formation of oxide grains, such as MgSiO$_3$, Mg$_2$SiO$_4$ and SiO$_2$. In some of the models, the CO and SiO abundance is computed under the assumption of chemical equilibrium, balancing radiative association rates with destruction rates through collisions with energetic electrons and, for SiO, charge transfer with positive Ne ions \citep{todini2001, schneider2004}. Other models have computed the CO and SiO abundance under non-steady state conditions \citep{bianchi2007}, including in the reaction network additional species, such as C$_2$ and O$_2$, and bimolecular neutral-neutral reactions \citep{marassi2014, marassi2015, marassi2019}. When applied to the same SN progenitor, the CO abundance predicted by this upgraded molecular network at the onset of dust nucleation is in good agreement with the result of chemical kinetic model (\citealt{sarangi2013}, see Sect.~\ref{sec:chemicalkineticmodel}). Finally, some of the models perform dust formation calculations assuming that the formation of CO and SiO molecules is complete, so that
no carbon-bearing grains condense if C/O $< 1$, and no Si-bearing grains -- except for oxide grains -- condense if Si/O $< 1$ \citep{nozawa2003, nozawa2010}.

In general, at the beginning of the nucleation process the gas is moderately supersaturated, the nucleation rate is small and large seed clusters, made of $N$ monomers, tend to form. 
Due to the expansion, the volume of the ejecta increases, the supersaturation rate grows and 
smaller clusters form with a larger formation rate. This occurs until the gas becomes sufficiently
rarefied (because of expansion and/or exhaustion of monomers in the gas phase), and the formation rate drops. This sequence of events, together with accretion, results in a typical log-normal-like grain size distribution \citep{bianchi2007}. 

Thanks to its relative simplicity, CNT has been applied to perform systematic explorations of dust condensation in 1D spherically symmetric SN explosion models with varying progenitor mass, metallicity, rotation rate, explosion energy, and supernova type. \citet{todini2001} used it to model dust formation in core collapse supernovae starting from the grid of explosion models by \citet{woosley1995}, hence assuming progenitors masses in the range $[12 - 40] \, M_\odot$, and initial metallicities 
$Z = 0, 10^{-4} Z_\odot, 10^{-2} Z_\odot, 1 Z_\odot$. Using the same grid of SN explosion models, \citet{bianchi2007} explored the additional effect of the partial destruction by the SN reverse shock (see Sect.~\ref{sec:snrevshock} for more details). CNT has been also applied to explore dust formation in pair-instability \citep{schneider2004} and faint SN explosions \citep{marassi2014}
with massive metal-free (Population III) stellar progenitors, to provide a formation pathway of iron-poor stars in the Galactic halo (see e.g. \citealt{debennassuti2017}), and to explain their observed surface elemental abundances. Finally, \citet{marassi2019} has applied CNT on a new extensive grid of core-collapse SN models \citep{limongi2018}\footnote{The grid of SN models comprises progenitor masses in the range [13 - 120] $M_\odot$ with initial equatorial rotational velocities of $v = 0$ and $v = 300$ km/s, and four different initial progenitor metallicities, $Z = 10^{-3} Z_\odot, 10^{-2} Z_\odot, 10^{-1} Z_\odot, 1 Z_\odot$ (see \citealt{marassi2019} for more details).} to investigate how metallicity, rotation, and fallback impact the nucleosynthetic output of the explosion, and the total mass, size, and composition of dust formed in the ejecta.

The applicability of CNT in astrophysical environments has been questioned
due to the lack of chemical equilibrium resulting from the low number density
(and collision rate) of monomers \citep{donn1985}. However, recent calculations
show that even when the number of critical clusters was artificially depressed
far below than the value predicted by CNT, the resulting grain size distribution
and mass are little affected, with changes in the mean grain radius
smaller than 15\% \citep{paquette2011}. In addition, \citet{nozawa2013} have 
demonstrated that a steady-state nucleation rate is a good approximation in SN ejecta, 
at least until the collisional timescales of the key species $\tau_{\rm coll}$ is 
much smaller than the timescale with which the supersaturation ratio increases, $\tau_{\rm sat}$,
otherwise the effects of non-steady state lead to lower condensation efficiencies and larger average
radii of newly formed grains. Since the dust destruction efficiency by the SN reverse shock
heavily depends on the grain size distribution (see Sect.~\ref{sec:snrevshock}), the knowledge of
the size distribution of newly formed dust is critical to predict the mass and sizes of grains that
survive and enrich the interstellar medium (ISM). The analysis performed by \citet{nozawa2013} shows that the 
steady state nucleation rate is applicable only if $\Lambda (t_{\rm on}) \equiv \tau_{\rm sat}(t_{\rm on})/\tau_{\rm coll}(t_{\rm on}) \gtrsim 30$,  and 
$\Lambda (t_{\rm on})$ can be expressed as a function of the gas density and temperature at the time 
$t_{\rm on}$ when dust formation starts\footnote{\citet{nozawa2013} also provide fitting formulae 
to their non-steady state models, that express the
final average grain size and condensation efficiency as a function of $\Lambda (t_{\rm on})$ and that 
can be used to estimate the typical size and mass of newly formed grains formed in different astrophysical
environments.}. When applied to the physical conditions predicted by Type-IIp or Type-IIb SN models,
$\Lambda (t_{\rm on})$ is generally found to be $\gtrsim 100$, and the steady state approximation is found to be appropriate.

\subsubsection{Kinetic nucleation theory}
A second method to model dust formation in SN ejecta is the so-called kinetic nucleation theory (KNT).
Compared to CNT, KNT is more realistic as it does not assume a steady state between condensation and evaporation: the condensation rate of clusters of $N \ge 2$ atoms is computed from kinetic theory and the evaporation rate by applying the principle of detailed balance. The method is fully described in \citet{nozawa2003} where it has been applied to model dust formation in Population III core-collapse SN explosions with progenitor masses $[13 - 20] \, M_\odot$ and in pair instability SN explosions with progenitor masses of $170$ and $200 \, M_\odot$. 

Dust formation is expected to depend on the type of core-collapse SN explosion, and in particular on the mass of the outer H-rich envelope \citep{kozasa2009}. A less massive outer envelope leads to larger expansion velocities of the He-core, causing a rapid decrease in the temperature and density of the expanding ejecta. Investigation of dust formation applying KNT to a SN-Ib explosion model similar to the observed SN 2006jc \citep{Nozawa2008}, and to a SN-IIb explosion model similar to Cas A \citep{nozawa2010}, show that dust formation can occur earlier than in Type-IIp SN explosions, the total dust mass formed is comparable but the grain sizes are strongly reduced, with important implications for their destruction by the reverse shock (see Sect.~\ref{sec:snrevshock}).  An exploration of the dependence of dust formation on the properties of the SN explosions (progenitor mass, explosion energy) has been recently carried out by \citet{brooker2022}, who applied KNT to a large database of SN explosion models based on the work by \citet{fryer2018}, with 15, 20, and 25 $M_\odot$ progenitor masses and covering a wide range
of explosion energies. They generally find that the bulk of dust production, irrespective of individual grain species, 
occurs earlier for more energetic explosions, as these explosions evolve more rapidly owing to higher initial kinetic velocity. As a result, for a given progenitor mass, there is also a clear dependence of grain size of individual species on the explosion energy, where less energetic models ultimately produce larger dust grains, as their ejecta experience the physical conditions amenable to dust production over a longer period of time. Because the energy of the SN explosion sensitively impacts the resulting nucleosynthesis, both the dust composition and dust mass are found to depend on the explosion energy, consistent with previous findings \citep{marassi2019}.

Despite the encouraging results discussed above, CNT and KNT do not consider the actual chemical pathway that leads to the formation of the molecular precursors and seed nuclei. To overcome this limitation, \citet{lazzati2016} developed a formalism that is able to join the chemical phase with the grain growth phase using KNT. As a proof of concept, they applied this hybrid approach to the formation of carbonaceous grains in the ejecta of a 15 $M_\odot$ SN explosion with initial solar metallicity. Compared to CNT, they find a more gradual dust formation, extending from a few months up to a few years after the explosion, in closer agreement with observations of local SN remnants (see Sect.~\ref{sec:snobservations}).

\subsubsection{Molecular nucleation theory}
\label{sec:chemicalkineticmodel}
The third method to compute dust formation is the chemical kinetics model or molecular nucleation theory (MNT), where the chemical pathway proceeds through simultaneous phases of nucleation and condensation. The nucleation phase, which leads to the formation of molecular and cluster precursors, is described by an extended non equilibrium chemical network. In the condensation phase, the small clusters formed in the gas phase condense through coagulation and coalescence to form large grains, provided suitable conditions are met. The method was introduced by \citet{cherchneff2008}, which applied it to investigate the nucleation phase in the ejecta of a Population III pair-instability supernova explosion with progenitor mass of 170 $M_\odot$. \citet{cherchneff2009} and \citet{cherchneff2010} applied the same approach to investigate the formation of molecules and early dust precursors in the ejecta of Population III supernova explosions with progenitor masses of 20 and 170 $M_\odot$, studying the effect of different levels of heavy element mixing in the ejecta. The model was then applied to the stratified ejecta of Type-IIp SN explosions with progenitor masses of 12, 15, 19, and 25 $M_\odot$ and initial solar metallicity \citep{sarangi2013}, and then extended from the nucleation phase to the condensation phase by \citet{sarangi2015}. In this approach, the condensation phase occurs through coagulation between small clusters rather than grow through adsorption of gas monomers or molecules, as in CNT, KNT and in the hybrid model by \citet{lazzati2016}. In the latter model, it is found that monomers are more abundant than
clusters, and, being lighter, have a larger thermal velocity that makes collisions more frequent. The formation of large grains likely requires coagulation and growth to be taken into account simultaneously in the condensation phase (see \citealt{sluder2016} and the discussion below).

In all the models described above, the physical properties of the expanding SN ejecta were based on fully mixed one zone models or on 1D spherically symmetric models where the elemental abundances are distributed in concentric shells, with different degrees of mixing and a uniform or clumpy gas distribution. Attempt to incorporate dust formation into more sophisticated description of the ejecta have been made by \citet{sluder2016}, who developed a model to account for anisotropic $^{56}$Ni dredge-up, the so-called “nickel bubbles”, that arise as a consequence of the strongly a-spherical explosion geometry (see \citealt{sluder2016} and references therein). Using MNT in a framework where the nucleation phase is joined to the condensation phase through both coagulation and grain growth, they modelled dust formation in SN1987A adopting a $25 M_\odot$ core-collapse SN model with LMC initial metallicity ($Z = 0.007$). 

\subsubsection{Models comparison}
\begin{figure}
\centerline{\includegraphics[width=12cm]{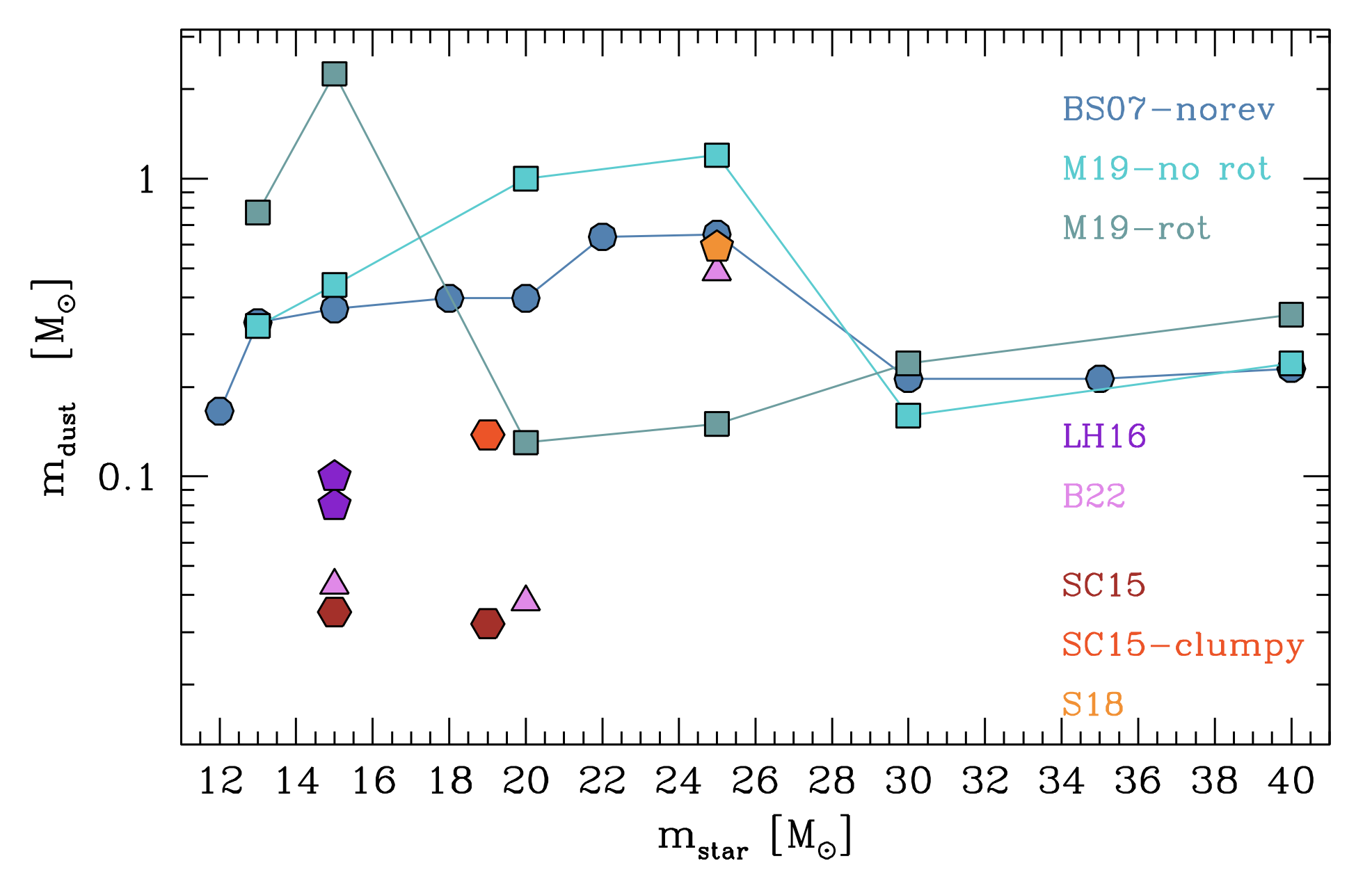}}
\caption{Total mass of dust formed in SN explosions for different stellar progenitor masses predicted by various theoretical models, as indicated in the legend. Here we report the mass of dust at the end of the condensation phase, before the destruction due to the reverse shock. The blue colour palette indicates models based on CNT (see text). In particular, the light blue dots refer to the model by \citet[][BS07-norev]{bianchi2007}, cyan (gray) squares to the non rotating (rotating) models by \citet[][M19]{marassi2019}. All these models assume a fixed explosion energy of $1.2 \times 10^{51}$ erg. The red colour palette refers to models based on MNT. In particular, dark red and red points represent the predictions of \citet[][SC15]{sarangi2015} for a $15 \, M_\odot$ progenitor (standard case, with a $^{56}$Ni mass in the ejecta of $0.075 \, M_\odot$,
taken from their Table 3), and for the $19 \, M_\odot$ case assuming a homogeneous and a clumpy ejecta (taken from their Table 5). In their analysis, they adopt SN explosion models from \citet{rauscher2002} for a fixed explosion energy of $10^{51}$ erg. The orange point illustrates the prediction of \citet[][S18]{sluder2016} adopting MNT for a 25 $M_\odot$ with explosion energy of $2.3 \times 10^{51}$ erg. The pink triangles show the values obtained by \citet[][B22]{brooker2022} applying KNT to three SN explosion models, selected from their grid: model M15bE0.92 for the 15 $M_\odot$ progenitor (from their Table 5), and their reference model for the 20 and 25 $M_\odot$ progenitors (from their Table 1). Violet points indicate the carbon dust mass predicted by \citet[][LH16]{lazzati2016} for their 15 $M_\odot$ for mixed (lower point) and unmixed (upper point) ejecta, adopting  their hybrid approach (see text). Here they assume a SN model from \citet{rauscher2002} with $1.2 \times 10^{51}$ erg. Except for \citet{sluder2016}, where the stars is assumed to have an initial metallicity of $Z = 0.007$, all the other models assume progenitor stars with solar metallicity.}
\label{fig:sn_comp1}
\end{figure}

A comparison between the predictions of all these theoretical models is shown in Fig.~\ref{fig:sn_comp1}. For each model, we report the total mass of dust predicted in core-collapse SN explosions with different initial progenitor masses (hereafter we refer to $m_{\rm star}$ as the zero-age main sequence stellar mass) and assuming that stars have initially a solar metallicity (except for the model by \citealt{sluder2016}). When reporting the results of each study, we attempted to select the SN models with explosion energies as close as possible to $10^{51}$ erg. The predicted dust masses are scattered between $\sim 0.03 \, M_\odot$ and $\sim 1 - 2 \, M_\odot$, with no clear coherent trend. In general, at least for the few stellar progenitors where the comparison is possible, models based on CNT \citep{bianchi2007, marassi2019} (represented with the blue color palette) tend to predict larger dust masses compared to models based
on MNT \citep[][red colour palette]{sarangi2015} or on KNT \citep[][pink and violet]{lazzati2016, brooker2022}. Note, however, that the results of \citet{brooker2022} and \citet{sluder2016} for the $25 M_\odot$ progenitor are very close to what expected on the basis on CNT by \citet{bianchi2007}.  At the same time, the comparison between the results of \citet{bianchi2007} and \citet{marassi2019} shows that -- even assuming a very similar approach to follow dust formation -- the resulting dust masses are sensitive to the adopted SN explosion models and to the assumed rotation rate of the progenitor star, at least for stars with initial masses $\le 25 \, M_\odot$. Similarly, assuming a clumpy rather than a homogeneous ejecta can increase the dust mass by almost 0.5 dex for the same progenitor mass (see the difference between the dark and light red points, corresponding to the clumpy and homogeneous ejecta model for a $19 \, M_\odot$ progenitor predicted by \citealt{sarangi2015}).
The figure also shows that the most efficient dust factories are SN explosions from low-mass rotating stellar progenitors ($13 - 15 \, M_\odot$), as a consequence of rotational mixing, which leads to more metal-enriched ejecta. This also causes stronger mass loss by stellar winds in the pre-SN evolution of more massive progenitors ($\gtrsim 20 \, M_\odot$), reducing the mass of the ejecta and of the newly synthesized dust compared to non rotating models. Finally, above $\sim 30 \, M_\odot$, the strong fallback experienced during the SN explosion is the main limiting factor to dust production, at least in models based on CNT.

\begin{figure}
\centerline{\includegraphics[width=12cm]{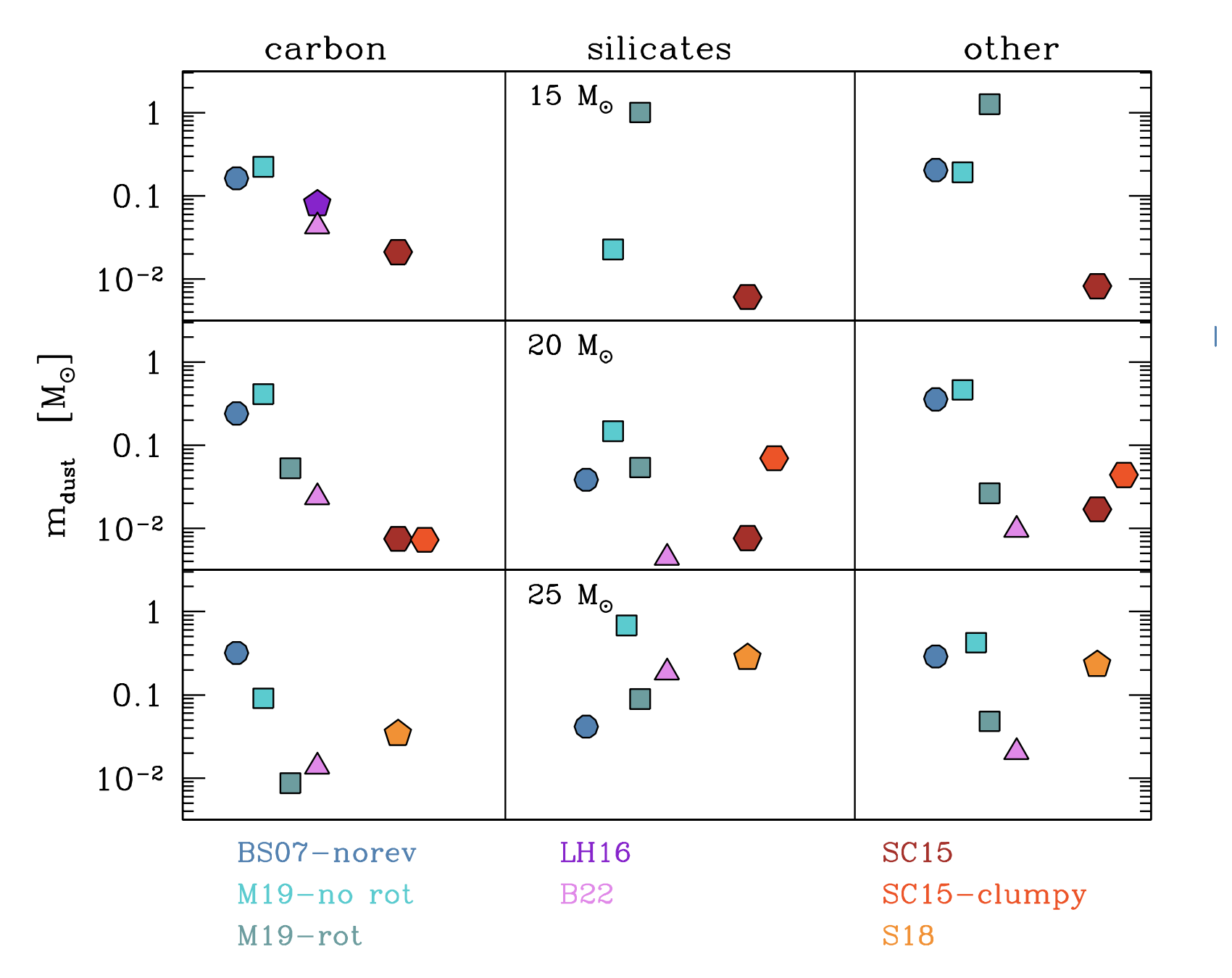}}
\caption{The mass of carbon (left), silicates (middle), and other grain species (right, see text) formed in SN explosions for three initial progenitor masses, 15 $M_\odot$ (top row), 19-20 $M_\odot$ (middle row), and $25 M_\odot$ (bottom row). The symbols and color-coding of theoretical models is the same as the one adopted in Fig.~\ref{fig:sn_comp1}, with BS07-norev, M19-no rot and M19-rot models based on CNT, LH16 and B22 based on KNT, and SC15, SC15-clumpy, and S18 based on MNT, the latter model including both grain coagulation and growth (see text). In the middle panels we show the results of \citet{bianchi2007} for a 22 $M_\odot$ SN
progenitor, and of \citet{sarangi2015} for a $19 \,M_\odot$ SN progenitor.
Note that the horizontal axis has no physical meaning and for each stellar progenitor mass and grain type, the models have been displaced horizontally to improve the clarity of the plot. }
\label{fig:sn_comp2}
\end{figure}
In Fig.~\ref{fig:sn_comp2} we show a comparison of the grain composition predicted by different theoretical models when applied to SN explosions with three initial progenitor masses, $15 \, M_\odot$ (top row), $19-22 \, M_\odot$ (middle row), and $25 \, M_\odot$ (bottom row). The symbols and color-coding of theoretical models is the same as the one adopted in Fig.~\ref{fig:sn_comp1}. Here we have broadly classified the grain species into carbon grains, silicates (which comprise enstatite, MgSiO$_3$, forsterite, Mg$_2$SiO$_4$,
silicon dioxide SiO$_2$, pure silicon, Si, and silicon carbide, SiC), and other grain types, which comprise alumina (Al$_2$O$_3$), pure iron (Fe), iron sulfide (FeS), iron oxyde (FeO), magnetite (Fe$_3$O$_4$), pure magnesium (Mg), and magnesia (MgO). The figure shows that a large variety of grain species are predicted to form. For the SN model with $15 M_\odot$ progenitor, all the non-rotating models predict the formation of more carbon grains than silicates, although the mass of carbon dust depends on the dust formation scheme adopted, being larger for models based on CNT \citep{bianchi2007, marassi2019}, and becoming progressively smaller for models based on KNT \citep{lazzati2016, brooker2022} and MNT \citep{sarangi2015}. For rotating models, instead, the abundance of heavier and more internal elements is very sensitive to rotational mixing, and the dominant grain species in the $15 M_\odot$ model are predicted to be magnetite and forsterite. 
Even for non-rotating models, silicate formation by the $15 \, M_\odot$ progenitor depends on the adopted SN explosion model, being negligible for \citet{bianchi2007} and \citet{brooker2022}, and small but not negligible for \citet{sarangi2015} and \citet{marassi2019}, despite the different microphysical approach to dust nucleation adopted in the latter models. Similar considerations apply for the $20$ and $25 M_\odot$ progenitors\footnote{Note that in the middle panels we
show the results of \citet{bianchi2007} for a $22 \, M_\odot$ SN progenitor, and of \citet{sarangi2015} for a $19 \, M_\odot$ SN progenitor.}. All the models predict the formation of carbon, silicates and other grains, with masses that are larger when CNT is adopted. 

\subsection{The case of SN 1987A} 
\label{sec:sn1987A}

When comparing the predictions of different SN dust models, it is important to consider the grain size distributions expected for different grain species. In fact, depending on the properties of the ejecta and on the timing of dust nucleation, the condensation phase via coagulation and/or accretion may lead to very different predictions regarding the characteristic grain sizes. This aspect is important when comparing with observational indications of the time evolution of dust formation in young SN remnants, 
and to estimate the fraction of newly formed dust that will be able to survive the passage of the reverse shock, with larger grains generally being more resistant to destruction (see Fig.~20 in \citealt{kirchschlager2023} for a discussion on the impact of gas density and magnetic field on the survival fraction of grains as a function of their sizes).

To this aim, we selected a few studies where the supernova model (progenitor mass, metallicity, explosion energy) has been chosen to provide a fair counterpart to SN 1987A \citep{sarangi2015, bocchio2016, sluder2016, brooker2022}. Depending on the model, both the time evolution of dust formation and the final dust mass, composition and sizes, can vary significantly. \citet{bocchio2016} select a $20 \, M_\odot$ progenitor exploding with an energy of $10^{51}$ erg, and releasing a $^{56}$Ni mass of $0.075 \, M_\odot$. Using CNT, they find that $0.84 \, M_\odot$ of dust forms in the ejecta (see their Table 2), mostly composed by Mg$_2$SiO$_4$ ($0.43 \, M_\odot$), SiO$_2$ ($0.19 \, M_\odot$), Fe$_3$O$_4$ ($0.11 \, M_\odot$), and carbon grains ($0.07 \, M_\odot$). The grain species follow a log-normal-like size distribution function, with central (peak) grain size which depends on the grain species, and which is larger for carbon grains (90.4 nm), Mg$_2$SiO$_4$ (68.9 nm) and SiO$_2$ (55.5 nm), and smaller for Fe$_3$O$_4$ (9.3 nm), reflecting the ejecta initial composition, and the timing of dust nucleation.

A similar SN model was considered by \citet{brooker2022} (see their M20cE1.00 model). Using KNT they find that $0.0378 \, M_\odot$ forms, mostly in the form of carbon  ($0.0237 \, M_\odot$), forsterite ($4.44 \times 10^{-3}\, M_\odot$) and alumina grains ($9.61 \times 10^{-3} \, M_\odot$). They do not show the time evolution of the dust mass and the final grain size distribution for this specific model, but based on the results of other $20 \, M_\odot$ SN models with the closest explosion energies, they predict silicate (carbon) grains to form $\sim 400$ ($\sim 900$) days after the explosion, and average grain sizes which range from $\sim 2 - 7 \mu$m ($\sim 0.8 - 3 \mu$m) for forsterite (alumina) grains, to $\sim 8 - 10 \mu$m for carbon grains. Hence, not only the total dust mass and composition is different, but also the average grain sizes are considerably larger compared to \citet{bocchio2016}.

\begin{figure*}
\centerline{\includegraphics[width=\textwidth]{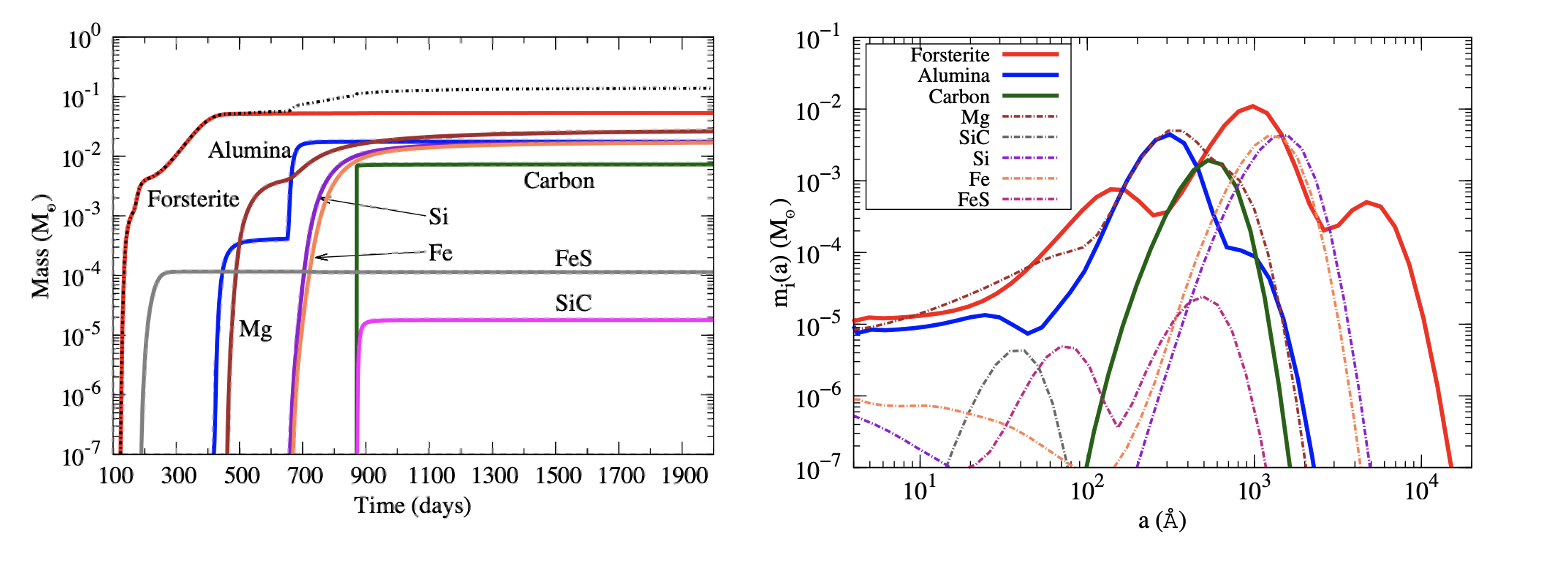}}
\caption{Properties of dust formed in the ejecta of the $19 \, M_\odot$ clumpy SN model investigated by \citet{sarangi2015} and considered to be their best SN 1987A analogue. {\bf Left panel:} mass of different grain species as a function of the post-explosion time. {\bf Right panel:} resulting dust mass as a function of the logarithm of the radius at 2000 days after the explosion. The figure has been adapted from \citet{sarangi2015}.}
\label{fig:sarangi}
\end{figure*}

\begin{figure*}
\centerline{\includegraphics[width=\textwidth]{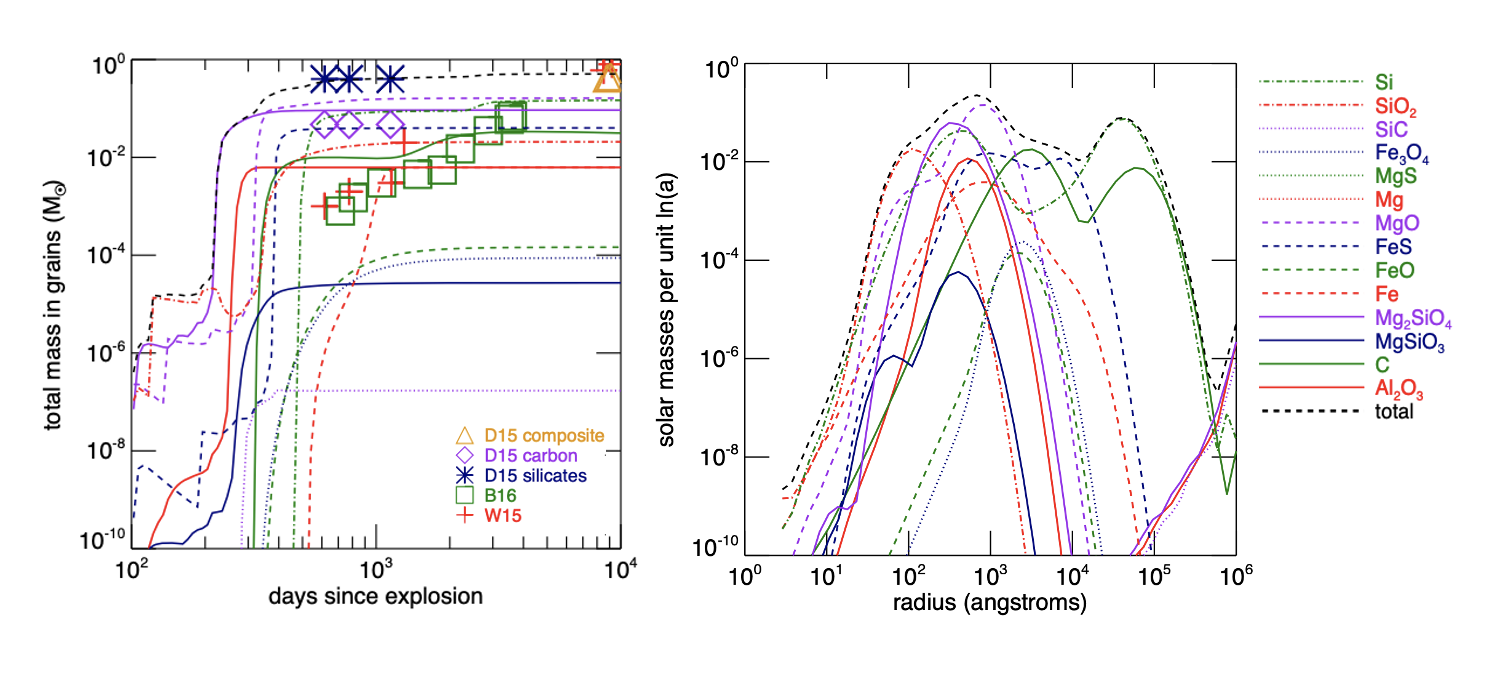}}
\caption{Properties of dust formed in the ejecta of the SN 1987A - model investigated by \citet{sluder2016}. {\bf Left panel:} mass of different grain species, as indicated in the legend, as a function of the post-explosion time. {\bf Right panel:} resulting dust mass as a function of the logarithm of the radius at $10^4$ days after the explosion. In the left panel, the data points refer to the dust mass observationally estimated by fitting the SED at different epochs by \citet{wesson2015} (W15), \citet{dwek2015} (D15), and \citet{bevan2016} (B16). The figure has been adapted from \citet{sluder2016}.}
\label{fig:sluder}
\end{figure*}

The results of SN 1987A models based on MNT have been discussed by \citet{sluder2016} (see their Sects.~7.1 and 7.2), who compare their 20 $M_\odot$ SN model with the 19 $M_\odot$ clumpy SN model considered by \citet{sarangi2015}. In Figs. \ref{fig:sarangi} and \ref{fig:sluder} we show the mass evolution as a function of the post-explosion time for different grain species as predicted by \citet{sarangi2015} and \citet{sluder2016}, respectively. In the same figures, we also show the dust mass per logarithm of the radius ($dM/d{\rm ln}\,a$) for different species at the end of the simulations. This corresponds, respectively, to $t = 2000$ days and $t = 10^4$ days after the explosion. 


At 2000 days after the explosion, the total dust mass predicted by \citet{sluder2016} is 0.44 $M_\odot$, while it is 0.14 $M_\odot$ in the model by \citet{sarangi2015}. This difference is attributed to the effect of grain growth by accretion, which is not considered by \citet{sarangi2015}. The final dust composition predicted by \citet{sarangi2015} is dominated by forsterite ($5.3 \times 10^{-2} M_\odot$), pure magnesium ($2.6 \times 10^{-2} M_\odot$), alumina ($1.8 \times 10^{-2} M_\odot$), pure silicon ($1.2 \times 10^{-2} M_\odot$) and iron ($1.2 \times 10^{-2} M_\odot$). Carbon grains represent only $\sim 5.3 \%$ of the total dust mass ($7.3 \times 10^{-3} M_\odot$). 

In the model by \citet{sluder2016}, the dust mass at the end of the simulation ($10^4$ days) is 0.51 $M_\odot$, mostly composed by magnesia (0.16 $M_\odot$), pure silicon ($0.15 \, M_\odot$), forsterite ($9 \times 10^{-2} \, M_\odot$), iron sulfide ($3.8 \times 10^{-2} \, M_\odot$), carbon ($3 \times 10^{-2} \, M_\odot$), and silicon dioxyde ($2.2 \times 10^{-2} \, M_\odot$). 

In both models, dust formation starts at 100--200 days after the explosion, but for most species appears to be more gradual in \citet{sluder2016}, as a consequence of the lower densities in the ejecta compared to \citet{sarangi2015}. The overall evolution of forsterite, alumina, iron sulfide, pure iron and silicon grains appears similar, despite the resulting masses are different. A striking difference is that C and SiC grains start to form at $\sim 300$ days in \citet{sluder2016} and only at $\sim 900$ days in \citet{sarangi2015}, and that magnesium grains do not form in \citet{sluder2016} due to the rapid formation of magnesia grains. These differences may be due to the different SN model considered, as well as to the inclusion of additional physical processes in \citet{sluder2016}, such as accretion of the grains, grain sublimation, grain charge (which may affect the coagulation rate, see \citealt{sluder2016} for more details). 

If we compare the final dust mass distribution as a function of the grains radii, we find that the peak radii agree to within a factor of a few for some grain species (forsterite, carbon, alumina, iron, and iron sulfide), while they differ significantly for others (silicon). In general, the bulk of the grains are found to have radii ranging between $\sim 10^2$ to $\sim 5 \times 10^4$ \AA \, in \citet{sluder2016}, and between $\sim 10^2$ to $\sim 5 \times 10^3$ \AA \, in \citet{sarangi2015}. These figures extend to significantly larger radii compared to the predictions of \citet{bocchio2016}, but are at the lower end of the range of grain sizes obtained by \citet{brooker2022}. It is hard to discriminate to what extent these differences can be attributed to the different microphysical processes implemented in the various models, and to what extent these depend on the adopted  physical properties of the expanding SN ejecta. Whatever the cause, these differences have important consequences for grain survival and ejection in the ISM.

It is important to comment on the comparison between model predictions and observations of dust formation in SN 1987A. This can be done by looking at the left panel of Fig.~\ref{fig:sluder}, where observationally estimated dust masses are reported by the coloured data points, 
as explained in the legend. These values have been obtained by fitting the observed spectral energy distribution (SED) at different epochs, as derived from observations made by the Kuiper Airborne Observatory (KAO) at $t = 60$, 250, 415, 615, and 775 days after the explosion \citep{wooden1993}, and at later time by Spitzer ($t \gtrsim 5800$ days, \citealt{dwek2010}), Herschel ($t \gtrsim 8000$ days, \citealt{matsuura2011, matsuura2015}), and ALMA ($t \gtrsim 9000$ days, \citealt{indebetouw2014, cigan2019}). 

The analyses have been made under different assumptions and using different methodologies. \citet{wesson2015} use a 3D radiative transfer model to fit the SED, finding a gradual increase in the dust mass, from 0.001 $M_\odot$ at 615 days, $0.02 M_\odot$ at 1300 days, $0.6 M_\odot$ at 8515 days, to $0.8 M_\odot$ at 9200 days (see the red crosses in the left panel of Fig.~\ref{fig:sluder} indicated by W15 in the legend). This gradual increase has been confirmed by \citet{bevan2016} (green open squares, B16), who used a 3D Monte Carlo model to estimate the dust mass from the observed blueshifting of the emission lines. 

None of the models that we have discussed above predicts a sufficiently slow gradual increase of the dust mass to be in agreement with these findings. However, the above interpretation has been questioned by \citet{dwek2016, dwek2019}, who argued that dust grains could have formed promptly, but could be hidden in optically thick clumps. By using a simple analytic approach to estimate the probability that a photon can escape a dusty sphere, they estimated that at 615 days the ejecta already contains $0.4 M_\odot$ of enstatite and $0.047 M_\odot$ of carbon dust, but the clumps are optically thick, and remain so until -- at 8815 days -- they become optically thin and enstatite and carbon grains have coagulated to form composite grains with masses $0.42 M_\odot$ at 8815 days and $0.45 M_\odot$ at 9090 days. Yet, studies based on radiative transfer modelling find it hard to hide early dust formation in clumps while at the same time reproducing the observed spectral energy distribution (SED) and emission line profiles of SN 1987A \citep{ercolano2007, wesson2015, bevan2016}. After a large parameter exploration of dust models with pure composition and a variety of spatial configurations, \citet{wesson2021} show that -- at an epoch of $\sim 800$ days, a carbon dust mass of $\sim 2 \times 10^{-3} M_\odot$, a clump volume filling factor of $f = 0.05$, and grain radius $a = 0.4 \mu$m is the only parameter set accounting for both the observed constraints on the SED and emission line profiles. Even if assuming carbon-silicate mixture would be consistent with a slightly higher dust mass, these constraints are still a factor of 50--100 below the masses estimated using the most recent observations of SN 1987A with Herschel \citep{matsuura2011, matsuura2015} and ALMA \citep{indebetouw2014}. 
Hence, these studies support a scenario in which dust formation in SN 1987A continues for many years after the 
supernova explosion and it is largely dominated by carbon grains\footnote{Note that, according to \citet{dwek2019}, if  
most of the dust forms within two years after the explosion, and the IR emission from the dust is initially self
absorbed, the lack of the 9.7 and 18 $\mu$m silicate emission features in
the spectra of SN 1987A is not evidence for the absence of silicate dust, but due to the large optical depth of
the ejecta \citep{dwek2015}.}, at odds with most (if not all) the theoretical models. 
Note, however, that these results assume spherically symmetric ejecta, while the 3D distribution of H, He, O, Mg, Si, Ca, and Fe has been found by \citet{larsson2016} to be sufficiently anisotropic at $10^4$ yrs after the explosion to 
explain on its own the spectral line asymmetries that are generally attributed to dust \citep{bevan2016, wesson2021}.

A more general discussion on observations of SN remnants is presented in Sect.~\ref{sec:snobservations}.

\subsection{Dust processing and survival in supernova remnants}
\label{sec:snrevshock}

It has been known since many years that not all the dust newly formed in SN ejecta will be able to enrich the ISM. On longer timescales, compared to the ones discussed above, the ejecta where dust resides 
is crossed by the reverse shock generated by the interaction between the expanding SN blast wave and the ISM. Depending on the grain properties (compositions and sizes) and on its spatial distribution, the processing by the reverse shock can lead to significant dust destruction. The \emph{effective}  SN dust yield (the dust mass that survives the passage of the reverse shock) is expected to have a different total mass, composition and grain size distribution compared to the newly formed grains that we have discussed above. 

The processing and survival of dust formation in SN remnants have been recently reviewed by \citet{micelotta2018}, 
where an extensive description of the observational evidences and theoretical models can be found. Here we provide a critical discussion of the main findings with the aim of providing a synthetic picture of our current understanding of the \emph{effective} SN yield. 

\subsubsection{Physical processes at work}
When dust grains are invested by the reverse shock, their interaction with gas particles and with other grains is mediated by different physical processes, such as: \emph{sputtering} (grain collision with high-velocity atoms and ions which leads to the erosion of the grains via ejection of atoms from its surface), \emph{sublimation} due to collisional heating to high temperatures, \emph{shattering} (grain--grain collisions that lead to fragmentation in smaller grains), and \emph{vaporization} (due to the intense heating generated during grain--grain collisions, that leads to partial or complete return of grain constituents to the gas phase). Sputtering is defined as \emph{kinetic} when the collision velocities are determined by the
relative motion between the grains and the gas (when the grain-gas relative velocity is much
larger than the gas thermal speed, generally in cold/warm gas phase, with $T \lesssim 10^4$ K), 
and as \emph{thermal} when the collision velocities arise from the thermal motion of the gas 
(when the gas thermal speed is much larger than the grain-gas relative velocity, generally in the hot gas phase, with $T \gtrsim 10^6$ K). Dust grains in the ionized shocked gas are heated mainly by collisions with electrons. If the grains are small, heating is stochastic and an equilibrium temperature does not exist. Instead, a broad temperature distribution establishes, but only a negligible fraction of the grains is found to exceed the sublimation temperatures \citep{bianchi2007}.

\begin{figure*}
\centerline{\includegraphics[width=10cm]{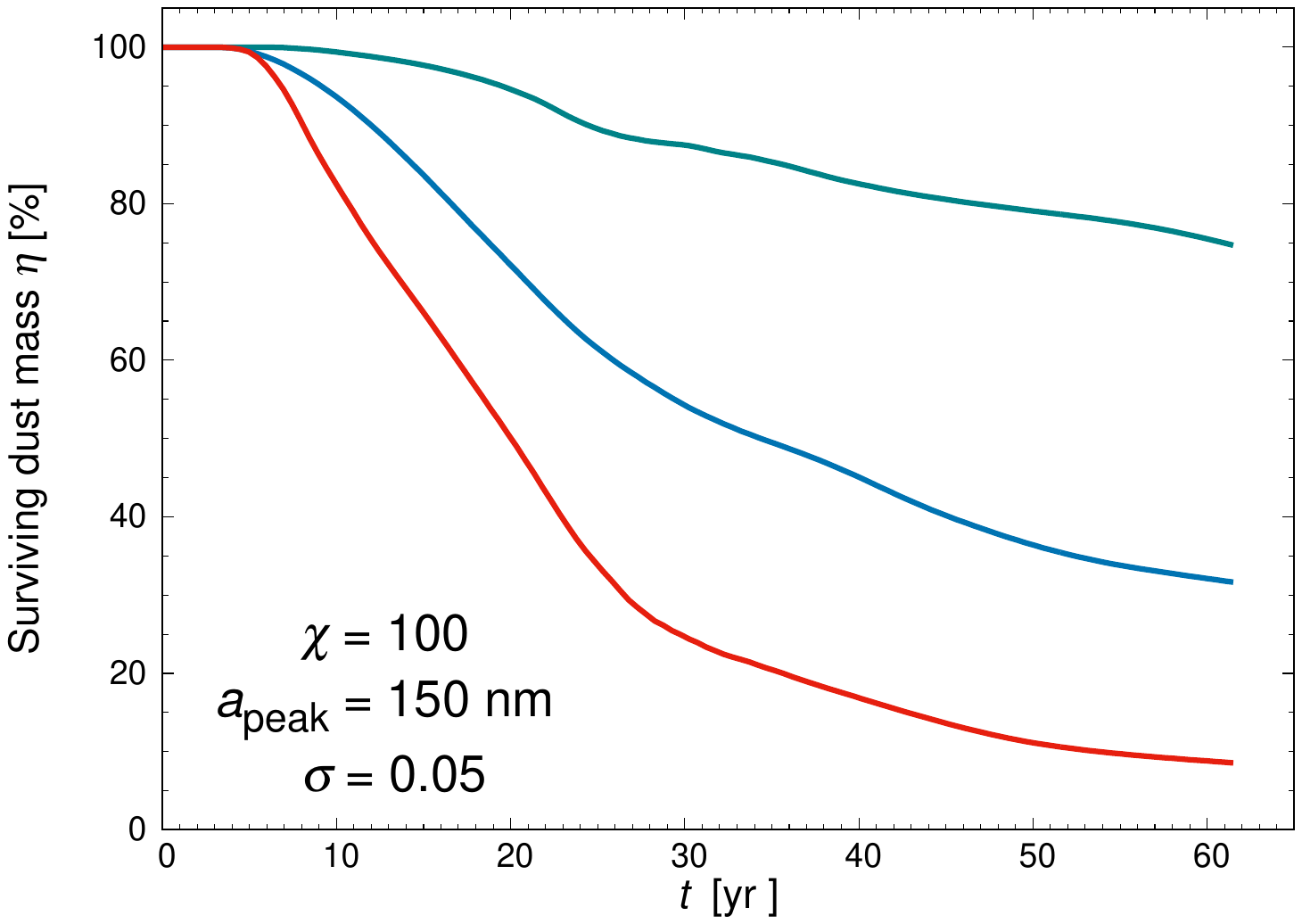}}
\caption{Result of a 2D hydrodynamical simulation showing the surviving dust mass as a function of time assuming only sputtering (blue), grain--grain collisions (green), and their combined effects (red). Here the initial grain population is assumed to be made by carbon grains with a log-normal-like size distribution peaked at $a_{\rm peak} = 3000$ \AA \, with a width $\sigma = 0.05$, initially distributed within clumps with 100 times larger density with respect to the smooth ejecta. Image reproduced with permission from \citet{kirchschlager2019}, copyright by the author(s).}
\label{fig:grainprocess}
\end{figure*}
The relative importance of these physical processes in SN remnants depends on the assumed initial dust spatial distribution: assuming a smooth, uniform distribution within the ejecta,
\citet{bocchio2016} find that due to the low dust density, grain-grain collisions are expected to be rare, and shattering and vapourisation lead to minor processing with respect to sputtering. 
Conversely, if the dust is initially located in overdense clumps within the ejecta, the increased
grain number density enhances grain-grain collision probabilities, while the grains are
sheltered in the clumps from the high gas velocities caused by
the shock and from the high gas temperatures in the inter-clump
medium, reducing the sputtering rates \citep{kirchschlager2019}. 
It is important to consider that grain--grain collisions and sputtering can be synergistic
processes since sputtering of the grain fragments resulting from collisions can be eroded in
a more efficient way than the larger colliding grains, as shown by \citet{kirchschlager2019}. In Fig.~\ref{fig:grainprocess}, we report the fraction of surviving dust mass as a function of time
obtained from their 2D hydrodynamical simulation of a shock wave interacting with a clumpy SN ejecta, where the clumps are assumed to be 100 times denser than the surrounding gas. The initial grain population in this particular case is made by carbon grains with a log-normal-like size distribution peaked at $3000$ \AA \, and with a width $\sigma = 0.05$. The green, blue and red lines show the results obtained considering, respectively, the effects of grain--grain collisions, sputtering, and the two processes acting together. It is evident that the
total dust destruction rate by sputtering and grain-grain collisions
can be significantly higher than their individual contributions acting
alone. Interestingly, \citet{kirchschlager2020} have shown that -- in suitable environments -- 
heavy ions that impact the grains can penetrate deep enough to be trapped, leading to grain growth and to an increase 
of the dust mass. Using the same set-up adopted by \citet{kirchschlager2019}, they show that grain growth can 
partly counteract destructive processes, increasing the fraction of surviving dust mass by 
by factors of up to two to four, depending on initial grain radii.

\begin{figure*}
\centerline{\includegraphics[width=8cm]{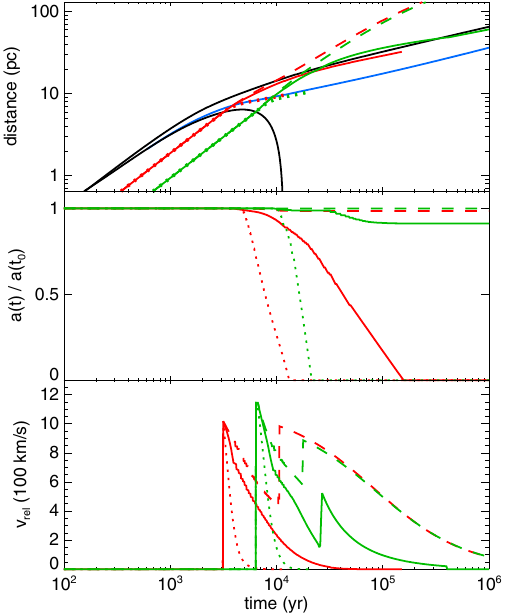}}
\caption{Examples of the dynamics of Mg$_2$SiO$_4$ grains with initial sizes of 10$^2$ (dotted lines), 10$^3$ (solid lines), and 10$^4$ \AA \, (dashed lines) in a homogeneous SN ejecta investigated by \citet{bocchio2016}. \textbf{Top panel:} Particles trajectories for grains initially located at one fourth (green) and one half (red) of the ejecta radius. The two black solid lines show the position of the forward and reverse shocks, while the blue line marks the transition from the ejecta to the ISM. \textbf{Middle panel:} time evolution of the grain sizes. \textbf{Bottom panel:} time evolution of the grain velocities relative to the gas. Image reproduce with permission from \citet{bocchio2016}, copyright by ESO.}
\label{fig:graindynamicsRS}
\end{figure*}

\subsubsection{Grain dynamics}
The efficiency of kinetic sputtering depends also on the dynamics of the grains. When invested by the reverse shock, the grains, which are initially coupled with the gas, have a different inertia with respect to the shocked gas, and start to move with respect to the gas with a velocity proportional to the velocity of the shock. Depending on the gas conditions and grain size, dust grains are slowed by drag forces and processed by collisions with gas particles. Small grains are quickly stopped and destroyed within the ejecta, while larger grains are eroded to a lower extent. However, if they are initially placed in the innermost part of the ejecta, the grains may not have enough inertia to cross the forward shock, and may be stopped and destroyed, or they may be reached by the faster forward shock, crossing the shock front a second time. 

Figure~\ref{fig:graindynamicsRS} exemplifies the time evolution of the position (top panel), size (middle panel), and velocity relative to the gas (bottom panel) of forsterite grains, adopting initial sizes of 10$^2$ (dotted lines), 10$^3$ (solid lines), and 10$^4$ \AA \, (dashed lines), as predicted by \citet{bocchio2016}. In this study, they adopt the self-similar analytical approximation of the ejecta dynamics by \citet{truelove1999}. The green and red lines refer to grains placed initially at one fourth and one half of the ejecta radius, respectively. As a reference, in the top panel the positions of forward and reverse shocks are indicated by black lines, while the position of the boundary between the ejecta and the ISM are plotted by the blue line. It is clear that 10 nm grains are quickly destroyed (on timescales of $\sim 10^4$ yr), while the fate of 100 nm grains depends on their initial position. The grains that are initially at one-fourth of the ejecta radius (green lines) have enough inertia to cross the forward shock, while grains that are initially at one-half of the ejecta radius (red lines) are stopped and destroyed on timescales of $\sim 10^5$ yr. Finally, the larger grains stream through the reverse and forward shock without suffering significant erosion, and they reach the ISM on timescales $\sim 10^4$ yr, gradually reducing their velocity until they get at rest with the surrounding gas ($t \sim 1$ Myr). Hence, the extent of grain destruction through sputtering depends on the initial size of the grains and on its initial position \citep{bocchio2016}. 
In addition, escaping grains are further destroyed while being slowed in the ISM. 
\begin{figure*}
\centerline{
\includegraphics[height=4.3cm]{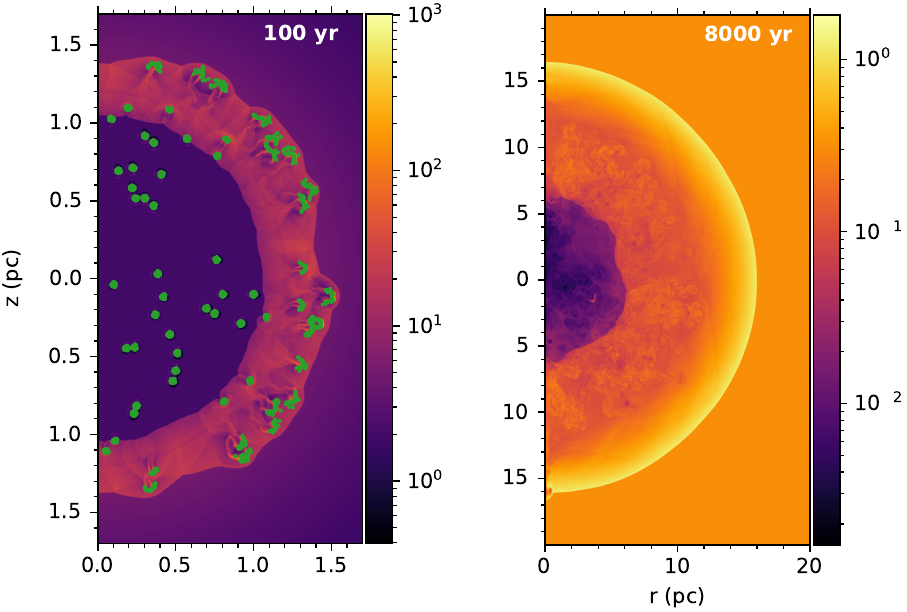}
\includegraphics[height=4.3cm]{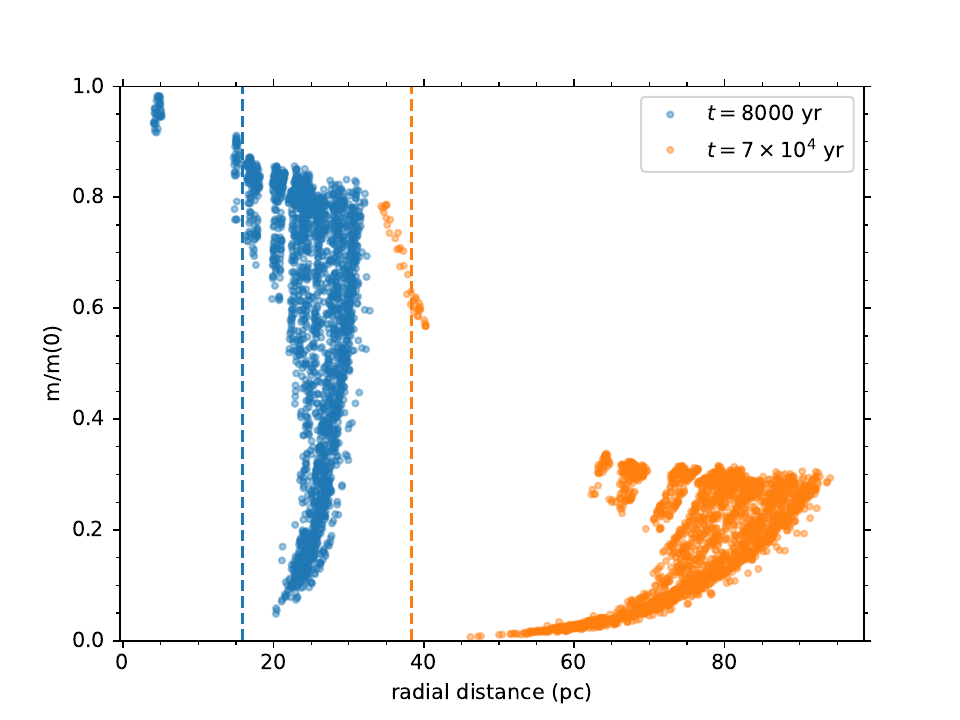}
}
\caption{The time evolution of grains in the expanding SN ejecta and the ISM. The \textbf{left} and \textbf{middle} panels show the gas density per cubic centimeter (as indicated by the color bars) and grain locations (green dots) at $t = 100$ and 8000 yr, respectively, for the 2D simulation of a clumpy SN ejecta by \citet{slavin2020} that includes silicate grains initialized with a radius of 10$^4$ \AA. 
Note that the spatial and the density scales vary by as much as an order of magnitude between panels and that the number of grains is the same in each panel, though at early times there is a lot of overlap of grains in gas clumps, which makes it appear that there are fewer grains. The \textbf{right} panel shows the ratio of grain mass at $t = 8000$ yr (blue dots) and $t = 7 \times 10^4$ yr (orange dots) as a function of distance, for silicate grains with initial size of $2.5 \times 10^4$ \AA. The two vertical dashed lines show the position of the forward shock at the corresponding time (see text). Although a vast majority of the grains of this size have crossed the forward shock already at $t = 8000$ yr, with a fractional mass reduction ranging between $\sim 5$ to $\sim 85 \%$,
substantial mass loss continues to occur at later times, as the grains are slowed in
the ISM. Images reproduced with permission from \citet{slavin2020}, copyright by AAS.}
\label{fig:graindynamicsISM}
\end{figure*}
In Fig.~\ref{fig:graindynamicsISM} we show time evolution of silicate grains with sizes $(1 - 2.5) \times 10^4$ \AA \, at $100$, $8000$, and $7 \times 10^4$ yr as predicted by \citet{slavin2020}. The first two panels are based on a 2D hydrodynamical simulations of the expanding SN ejecta, assuming that equal-sized grains are formed in gas clumps which have a radius of $R_{\rm clump} = 10^{16}$ \, cm and a density that is 100 times larger than that of the smooth ejecta ($\rho_{\rm sm} = 3.831 \times 10^{-23}$\,g\,cm$^{-3}$) with initial radius of $R_{\rm sm} = 3 \times 10^{19}$\, cm.  The color bars in the first two panels show the number density of the ejecta, and the green points show the position of the silicate grains, assuming that grain particles are initially randomly scattered inside the clumps (with 40 grains per clump), and have an initial size of $10^4$ \AA. At $t = 100$ yr, most of the grains are still in their birth clumps, but at 8000 yr, a significant fraction of the grains have streamed out of ejecta clumps, crossing the forward shock. In the right panel, the blue points indicate the fractional mass reduction at $t = 8000$ yr for silicate grains with initial size of $2.5 \times 10^4$ \AA, and vertical blue dashed line shows the position of the forward shock at the same time. Nearly all the grains have crossed the forward shock, with a fractional mass reduction ranging from $\sim 5$ to $\sim 85\%$. However, these grains are moving at substantial speed when they cross the forward shock (which has a speed of $ \sim 870$ km/s at 8000 yr). Hence, they will suffer further sputtering as they get slowed down, finally reaching a velocity of $\sim 10 - 20$ km/s, when the sputtering is no longer effective. To follow this additional evolution, \citet{slavin2020} used the radially averaged profiles of pressure, density and radial velocity from the final time step of the 2D simulation
to initiate a 1D, spherically symmetric simulation, adopting a Sedov--Taylor type similarity solution up to $t = 4 \times 10^4$ yr. The effects of the subsequent evolution of the grains in the ISM are represented by the orange points in Fig.~\ref{fig:graindynamicsISM} and show that substantial mass loss occurs between 8000 yr and $4 \times 10^4$ yr and that most of the initially $2.5 \times 10^4$ \AA \, silicate grains have a final mass that is $\sim$1 --30\% smaller than their initial mass.

\subsubsection{Effects of magnetic field}
It is important to note that the above studies do not consider the impact of magnetic fields on the dynamics of charged grains. As a consequence of Lorentz force, charged grains may gyrate around magnetic field lines, and the betatron acceleration can cause kinetic decoupling between gas and dust, enhancing grain sputtering \citep{slavin2015}. The importance of
magnetic fields in SN remnants depends on the strength and orientation of the magnetic field lines. Due to flux freezing, the stellar magnetic field is expected to be extremely small in the expanding SN ejecta. In the ISM, 
the magnetic fields may have typical magnitudes of several $\mu$G, but their importance for the trajectories of
the grains depends on the morphology of the field. A uniform field could allow for reflection of the grains back into the remnant \citep{fry2020}, while a field with a turbulent component could allow for diffusion of the grains
through the ISM \citep{slavin2020}. Given the uncertainties in the charging of the grains and in the ISM fields, the effects of magnetic fields on grain trajectories have been neglected in most studies \citep{bianchi2007, bocchio2016, micelotta2016,martinez2019,slavin2020}. Recent work by \citet{fry2020} show that charged Fe grains created in a unmagnetized SN can suffer large deflections when encountering the shocked ISM, in which the pre-existing turbulent magnetic fields have been amplified by shock compression. Due to magnetic trapping and mirroring, occurring at the interface between the SN
ejecta and the shocked ISM, the reflected particle moves back into the SN ejecta, traversing the SNR until it encounters the ejecta/ISM interface and is reflected again, in a sort of pinball behaviour within the SNR.

\begin{figure*}
\resizebox{\hsize}{!}{
   \includegraphics[trim=3.0cm 2.5cm 0.9cm 2.0cm, clip=true, page=3]{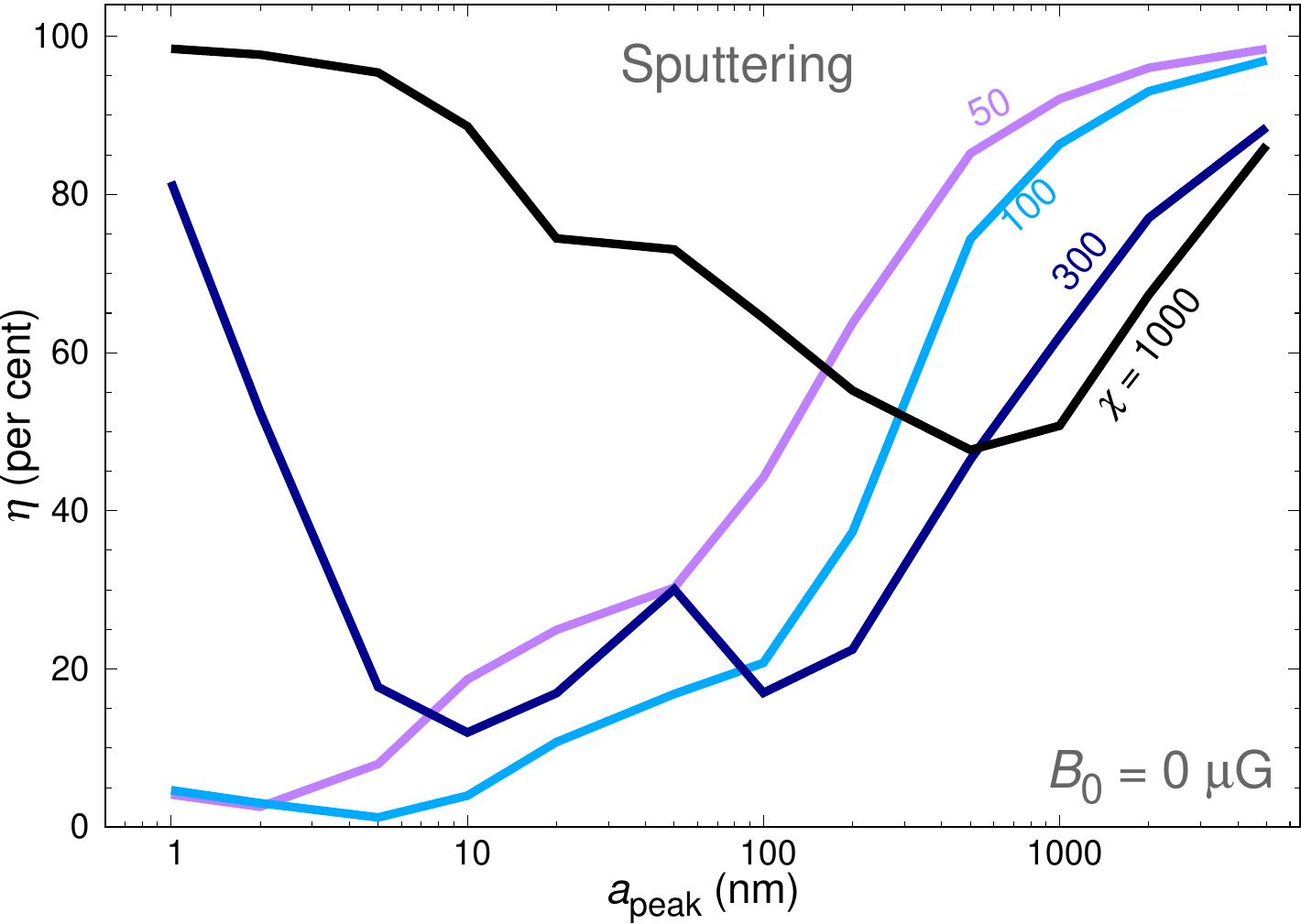}\hspace*{0.38cm}
   \includegraphics[trim=3.0cm 2.5cm 0.9cm 2.0cm, clip=true, page=2]{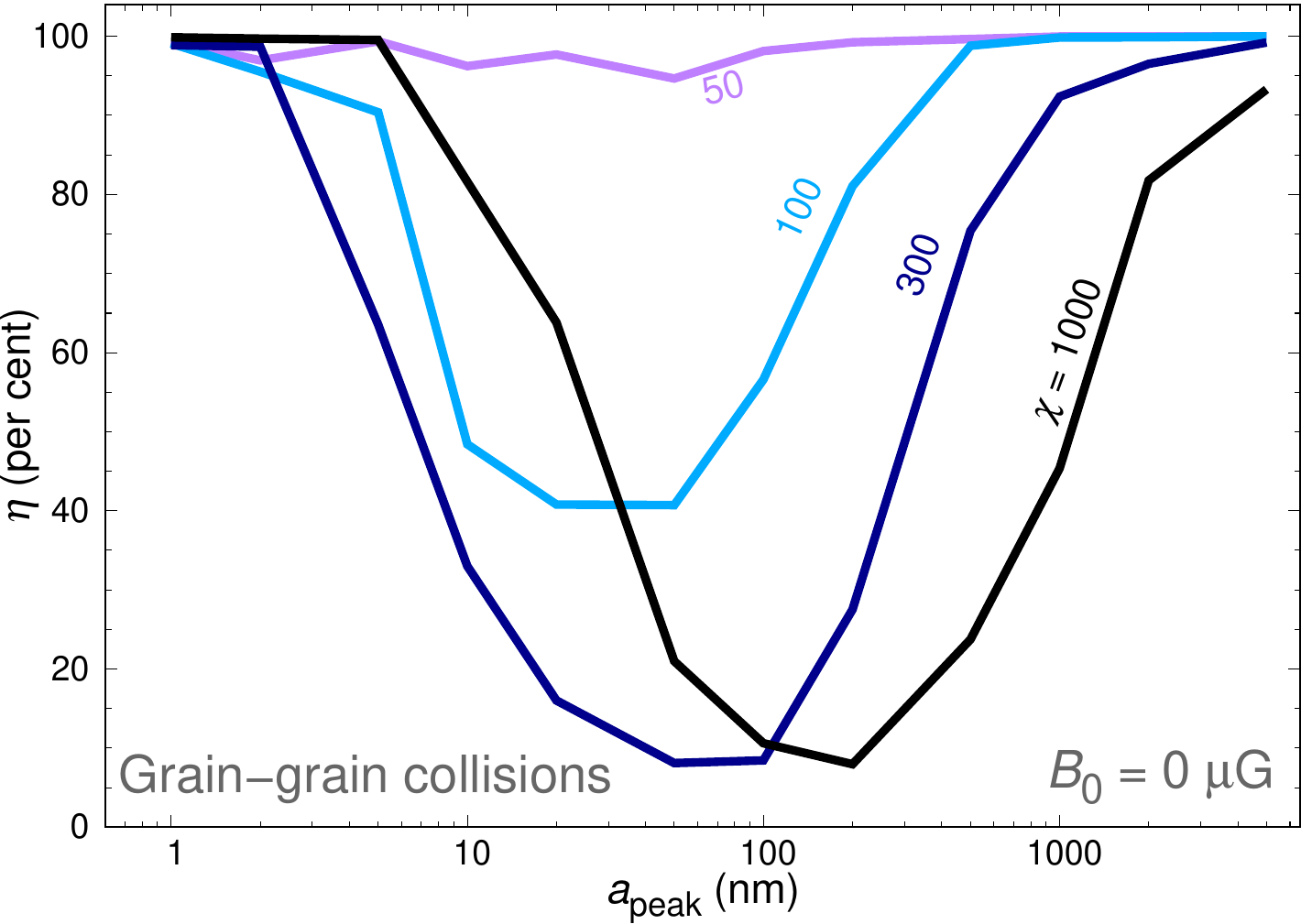}}\\
  \resizebox{\hsize}{!}{    
   \includegraphics[trim=3.0cm 2.5cm 0.9cm 2.0cm, clip=true, page=5]{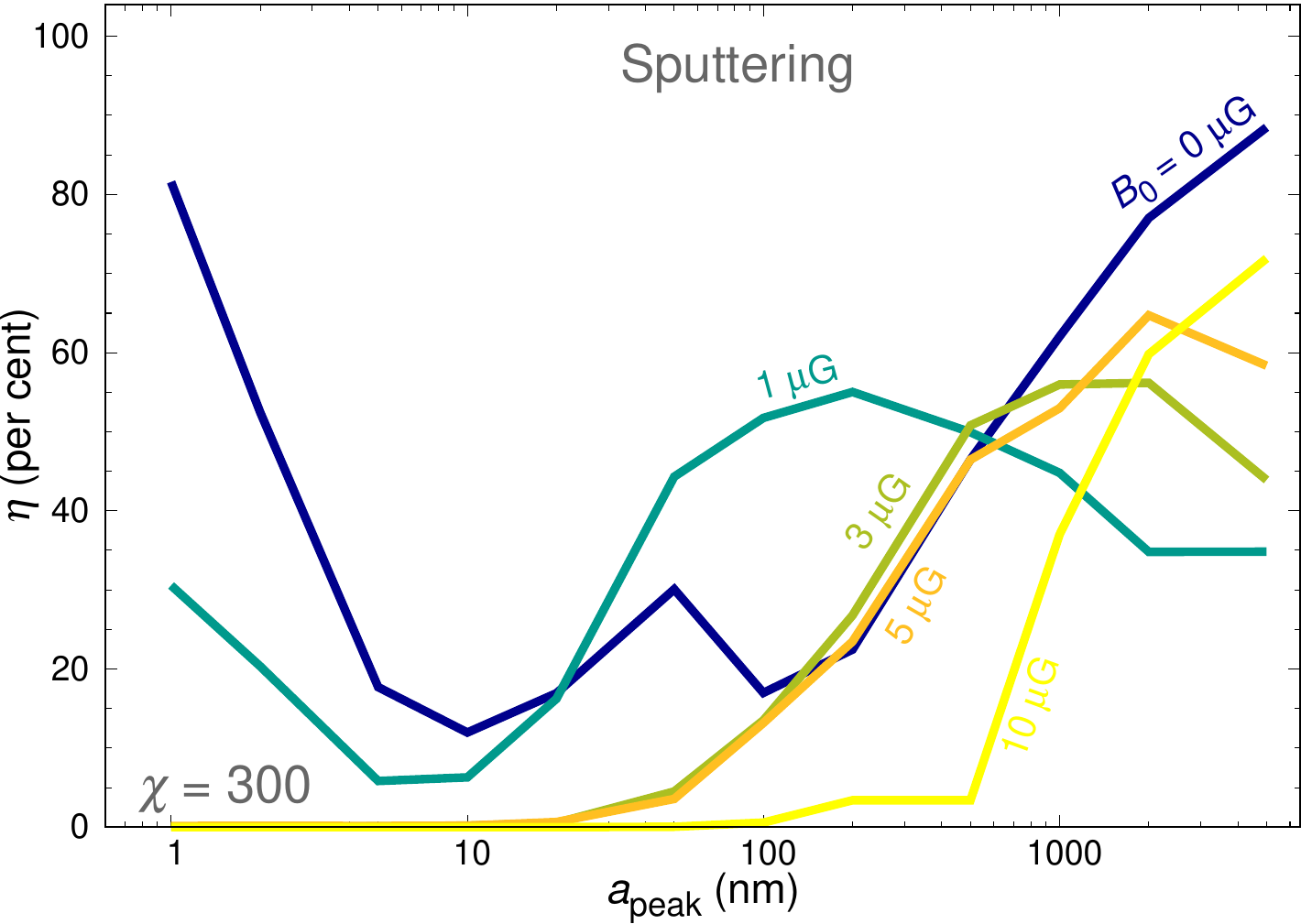}\hspace*{0.38cm}
   \includegraphics[trim=3.0cm 2.5cm 0.9cm 2.0cm, clip=true, page=4]{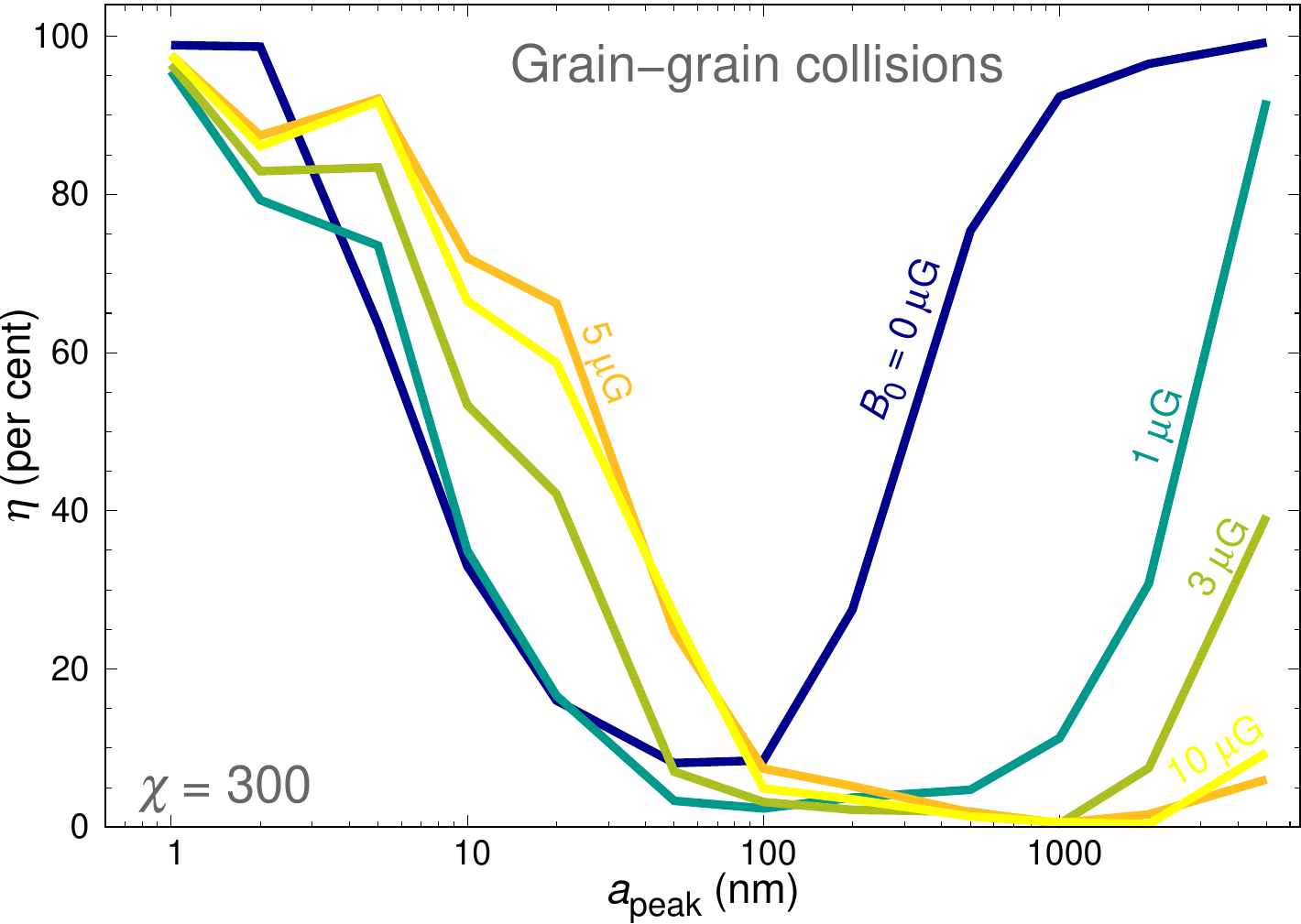}  
   }
\caption{Silicate dust survival fraction as a function of grain sizes from the magnetic hydrodynamic simulations of \citet{kirchschlager2023} of the Cas A SNR. The left and right panels show the effects of sputtering and grain-grain collisions, respectively. Upper and lower panels illustrate the dependence on the density contrast between the clumps and the ejecta $\chi$, assuming $B_0 = 0$, and on the magnetic field strength, $B_0$, assuming $\chi = 300$. Image reproduced with permission from \citet{kirchschlager2023}, copyright by the author(s).}
\label{fig:magnetic_grains}
\end{figure*}

Using a magneto-hydrodynamic simulation of a plane parallel shock investing a single clump embedded in a lower density magnetized medium (cloud-crushing problem), \citet{kirchschlager2022} find a significantly lower dust survival rate when magnetic fields are aligned perpendicular to the shock direction compared to the non-magnetic case. The grain survival fractions depend sensitively on the magnetic field strength, $B_0$, on the gas density contrast between the clump and the ambient medium, $\chi$, and on the grain sizes. A schematic summary of their findings is shown in Fig.~\ref{fig:magnetic_grains}. Three different grain size intervals can be identified: 
small grains, with radii $a \lesssim 10$\,nm, are mostly affected by sputtering and their survival fraction is very high for sufficiently large density contrasts and low magnetic field strengths, as the grains are effectively confined and shielded in ejecta clumps. These grains are completely destroyed when $\chi \lesssim 100$ and/or $B_0 \gtrsim 5 \mu$G. Large grains, with radii $a \gtrsim 100$\,nm are mostly affected by grain-grain collisions. Their survival fraction is almost 100 \% for very low density contrast ($\chi = 50$), but decreases with increasing $\chi$ and $B_0$, as these enhance the number density and collision velocities of the grains. Medium-sized grains, with $10 \, {\rm nm} \lesssim a \lesssim 100$\,nm experience a mixture of the effects described above: for low-density contrasts ($\chi \lesssim 100$) sputtering dominates and their survival fraction increase with grain size between a few \% to $\sim 40 \%$ when $B_0 = 0$, but magnetic fields reduce the surviving dust mass. For higher density contrasts, both sputtering and grain-grain collisions operate, and the survival fractions strongly decrease, with or without magnetic fields. Similar conclusions are found when carbon grains are considered. Although limited to a cloud-crushing set-up and therefore missing the global evolution of the SNR, this study shows that magnetic field strengths of a few $\mu$G may be able to destroy significant amounts of grains therefore limiting the amount of dust that will be injected in the ISM \citep{kirchschlager2023}.

\subsubsection{Models comparison}
Estimating the \emph{effective} SN dust yield requires to integrate the effects of sputtering on individual grains onto a grain size distribution. The results, which are generally expressed as the surviving dust mass fraction, depend on a number of assumptions, such as the composition, size, and spatial distribution of newly formed SN dust, the properties of the SN explosion and of the dynamical evolution of the SN remnant, the physical processes implemented in the models, the late-time evolution of the grains when they cross the forward shock and are slowed down in the ISM. \citet{micelotta2018} present a detailed description of the different assumptions made by different studies, warning that this prevents the possibility of making direct comparisons between different model results. While we agree with this concern, we believe that their Table 2 -- which summarizes the theoretically calculated fractions of surviving dust mass -- provides an important indication on the persisting uncertainties affecting the \emph{effective} SN dust yields. 
For this reason, in Table \ref{table:RSmassfrac} we update their original table with more recent findings and supplementary information. In particular, we provide the dust survival mass fraction ($\eta$), the timescale at which $\eta$ is estimated, the clumpy ejecta overdensity $\chi$ ($\chi = 1$ for uniform ejecta models), the physical processes that have been considered (sp = sputtering, th spu = thermal sputtering, sub = sublimation, gg = grain-grain collisions, B = magnetic field), the range of grain sizes before dust destruction (in nm), the SN progenitor/explosion properties, and the ambient medium density, $n_{\rm ISM} \,\, [\rm cm^{-3}]$. 

The lesson learnt from this model comparison is that there exist physical conditions for which a moderate to large fraction ($> 10 - 30$ \%) of SN dust is able to survive in the SNR phase, enriching the ISM. These are more easily met when clumpy ejecta, with moderate to high over-densities, produce grains with initial sizes $\gtrsim 100$ nm, and/or explode in a very tenuous ambient medium, and when the magnetic field is very low or absent.

More specific notes on individual studies reported in Table \ref{table:RSmassfrac} are given below: 

\begin{description}
\item[\bf a:] grid of core-collapse SN progenitors with $[12 - 40] \, M_\odot$, $Z = Z_\odot$, and explosion energy $E_{\rm exp} = 1.2 \times 10^{51}$ erg.

\item[\bf b:] core-collapse SN progenitors with $[13 - 30] \, M_\odot$, $Z = 0$, and explosion energy $E_{\rm exp} = 10^{51}$ erg with mixed ejecta.

\item[\bf c:] same as ${\bf b}$ but with stratified (unmixed) ejecta.

\item[\bf d:] Pair Instability Supernovae with progenitor masses $170$ and $200 \, M_\odot$, initial metallicity $Z = 0$, and explosion energy $E_{\rm exp} = [2 - 2.8] \times 10^{52}$ erg with mixed/unmixed ejecta.

\item[\bf e:] calculation performed for a core-collapse SN with ejecta mass $\sim 5 \, M_\odot$, explosion energy of $10^{51}$ erg, and assuming a uniform density core and a power-law density envelope. Graphite and silicate grains are considered, with power-law grain size distribution ($dn/da \propto a^{\alpha}$ with $\alpha = -3.5$ and -2.5). Note that we quote this estimate as an upper limit as the authors themselves warn that their analytical formalism ignores further sputtering of grains in hot plasma between the forward and reverse shocks.

\item[\bf f:] cloud-crushing set of simulations varying clump over-density, shock velocity (1000--5000 km/s), cooling timescale. The initial grain composition and size distribution is the same as the $20 \, M_\odot$ Pop III core-collapse SN progenitor of \citet{nozawa2003}, for the unmixed case, and the ejecta metallicity is $Z = Z_\odot$. We quote the range of surviving dust mass fractions of Si and Mg$_2$SiO$_4$ and FeS, which are the three dominant dust species.

\item[\bf g:] same as ${\bf f}$, but varying the ejecta metallicity from $Z = 1 Z_\odot$ to $100 Z_\odot$, and exploring shock velocities up to $10^4$ km/s.

\item[\bf h:] grid of Pop III core-collapse SNe with progenitor masses $[13 - 80] \, M_\odot$, initial metallicity $Z = 0$, and explosion energy $10^{51}$ erg.

\item[\bf i:] same as ${\bf h}$ but for faint SN explosion, where little mixing and strong fallback allows for carbon-rich ejecta.

\item[\bf l:] 4 different core-collapse SN models, tailored to reproduced the observed properties of Crab, Cas A, N49, and SN 1987 A, with progenitor masses 
$13$ and $20 \, M_\odot$, explosion energies $[0.5 - 1.5] \times 10^{51}$ erg, and metallicities $Z = 0.4$ and 1 $Z_\odot$. We report the survival fractions of the currently observed dust mass at the end of the simulation.

\item[\bf m:] clumpy ejecta of a type-IIb SN with progenitor mass $19 \, M_\odot$ appropriate to describe Cas A.

\item[\bf n:] same as the previous explosion model, but assuming a homogeneous ejecta with 2000 times larger density, appropriate to describe the ejecta of type-IIp SNe.

\item[\bf o:] clumpy ejecta of a type-IIb SN with progenitor mass $19 \, M_\odot$ and explosion energy of $2.2 \times 10^{51}$ \, erg, appropriate to describe Cas A. The initial grains (AC and MgSiO$_3$) are assumed to follow a power-law size distribution with $dn/da \propto a^{-3.5}$.

\item[\bf p:] 3D hydro-simulation of a type-IIp SN explosion with progenitor mass $60 \, M_\odot$, explosion energy $9.12 \times 10^{50}$\, erg, an equal amount of carbonaceous and silicate grains with a log-normal size distribution with $a_{\rm peak} = 100$\, nm, width $\sigma = 0.7$, and minimum/maximum sizes as reported in the table. In the first line we show $\eta$ when the explosion takes place in a uniform medium with density 1 and 1000 cm$^{-3}$. In the second line, we show $\eta$ when the explosion takes place after the wind-driven shell has excavated a very tenuous region, with density $\sim 10^{-3}$\, cm$^{-3}$.

\item[\bf q:] 2D hydro simulations with cloud-crushing set-up applied to Cas A, with shock velocity 1600 km/s and a range of cloud-overdensities, $\chi = 100, 200, 300, 400, 600, 1000$. The initial size distribution is assumed to be log-normal, with peak radii $ 10 \, {\rm nm} \le a_{\rm peak} \le 7000 \, {\rm nm}$ and width $ 0.02 \le \sigma \le 2.2$, and the dust survival rate $\eta$ is computed for carbon and silicate dust grains. Here we report the results for $a_{\rm peak} = 20, 100, 1000$\, nm with $\sigma = 0.02$ and $\chi = 100, 300, 1000$.

\item[\bf r:] same set-up as above, but computed using 3D simulations and assuming $\chi = 100$. The initial size distribution is assumed to be log-normal in the range 0.6 nm--10 $\mu$m, with peak radii $a_{\rm peak} = 0.01, 0.1, 1$ $\mu$m, and width $\sigma = 0.1$, and the dust survival rate $\eta$ is computed for silicate dust grains.

\item[\bf s:] clumpy ejecta of a type-IIb SN with progenitor mass $19 \, M_\odot$ and explosion energy of $1.5 \times 10^{51}$ \, erg, appropriate to describe Cas A. We report the $\eta$ for individual grain sizes and for two adopted initial size distribution, a log-normal with peak size 100 nm and width 0.1 (LN1) and a power-law with index $-3.5$ in the range [5--250] nm (PL1). For all these cases, lower (upper) values of $\eta$ refer to silicate (carbonaceous) grains.

\item[\bf t:] same as ${\bf q}$, but including the effects of magnetic fields. Here we report $\eta$ adopting a fixed overdensity $\chi = 300$ and varying the magnetic field strength from 0 to 10 $\mu$G. The results for carbonaceous and silicate grains are very similar.
\end{description}


\begin{landscape}
\begin{table}
\caption{Mass fraction of SN dust that survives the passage of the reverse shock as predicted by different theoretical studies. This table updates Table~2 of \citet{micelotta2018} with more recent studies and provides supplementary information.
For each study, we report the reference paper, the surviving dust mass fraction $\eta$, the post-explosion time at which $\eta$ was evaluated $t$ [$10^3$ yr], the adopted ejecta clumps overdensity $\chi$, the physical processes that have been considered (sp = sputtering, th spu = thermal sputtering, sub = sublimation, gg = grain-grain collisions, ion trp = ion-trapping, B = magnetic field, the range of grain sizes before dust destruction (in nm), the SN progenitor/explosion properties, circumstellar medium density, $n_{\rm ISM} \,\, [\rm cm^{-3}]$. For the notes on the Table, see main text.}
\label{table:RSmassfrac}
\begin{tabular}{llllllll}
\hline\noalign{\smallskip}
paper & $\eta$ \, [\%] & $t$ \, [$10^3$ yr] & $\chi$ & processes & $a_{\rm grain}$ [nm] & SN type & $n_{\rm ISM} \, [\rm cm^{-3}]$\\
\noalign{\smallskip}\hline\noalign{\smallskip}
\citet{bianchi2007} & 2\,--\,20 & [40\,--\,80]  & 1 & sp, sub & [2\,--\,60] & cc SNe$^a$ & 0.06, 0.6, 6  \\
\citet{nozawa2007} & 0\,--\,0.4 & [300\,--\,2000] & 1 & sp, sub & [0.2\,--\,100] & Pop III cc SNe$^b$ & 0.1, 1, 10\\
                   & 0.004\,--\,0.8 & [300\,--\,2000] & 1 & sp, sub & [0.2\,--\,500] & Pop III cc SNe$^c$ & 0.1, 1, 10\\
                    & 0\,--\,0.45 & [700\,--\,5000] & 1 & sp, sub & [0.2\,--\,300] & Pop III PISN$^d$ & 0.1, 1, 10\\
\citet{nath2008} & $<$ 0.8\,--\,1 & [1\,--\,4] & 1 & sp & [0.1\,--\,300] & cc SN$^e$ & 0.6 \\ 
\citet{silvia2010} & 0.05\,--\,0.89 & $\gtrsim 1$ & 100, 1000 & th sp & [0.2\,--\,500] & cloud-crush$^f$ & -- \\ 
\citet{silvia2012} & 0.02\,--\,1 & $\gtrsim 1$ & 1000 & th sp & [0.2\,--\,500] & cloud-crush$^g$ & -- \\ 
\citet{marassi2015} & 3\,--\,50 & $\sim 10$  & 1 & sp, sub & [1\,--\,500] & Pop III cc SNe$^h$ & 0.06, 0.6, 6  \\
                     & 10\,--\,80 & $\sim 10$  & 1 & sp, sub & [30\,--\,220] & Pop III faint SNe$^i$ & 0.06, 0.6, 6  \\
\citet{bocchio2016} & 1\,--\,8 & $\sim 10^3$  & 1 & sp, sub, gg & [2\,--\,400] & cc SNe$^l$ & 0.9\,--\,1.9  \\
\citet{biscaro2016} & 6\,--\,11 & $ \sim 40$   & 200 & sp & [0.1\,--\,20] & SN-IIb$^m$ & --  \\
                     & 14\,--\,45 & $ \sim 40$ & 2000 & sp & [0.1\,--\,50] & SN-IIp$^n$ & --  \\
\citet{micelotta2016} & 9\,--\,16 & $ 8 $ & 100 & sp & [5\,--\,250] & SN-IIb$^o$ & 2 \\
\citet{martinez2019} & 60, 0 & $6.5, 0.4$ & 1 & th sp & [10\,--\,500] & SN-IIp$^p$ & 1, 1000 \\
                     & 96 & $12, 100$ & 1 & th sp & [10\,--\,500] & SN-IIp$^p$ & $10^{-3}$ \\
\citet{kirchschlager2019} & 0, 5, 30 & $\sim 0.0615, 0.1, 0.2$ & 100, 300, 1000 & sp, gg & 20 [0.02] carb & cloud-crush$^q$ & 1 \\
                    & 6, 2, 4 & $\sim 0.0615, 0.1, 0.2$ & 100, 300, 1000 & sp, gg & 100 [0.02] carb & cloud-crush$^q$ & 1 \\
                    & 13, 28, $\lesssim 1$ & $\sim 0.0615, 0.1, 0.2$ & 100, 300, 1000 & sp, gg & 1000 [0.02] carb & cloud-crush$^q$ & 1 \\
                    & 6, 15, 37 & $\sim 0.0615, 0.1, 0.2$ & 100, 300, 1000 & sp, gg & 20 [0.02] sil & cloud-crush$^q$ & 1 \\
                    & 8, 3, 2 & $\sim 0.0615, 0.1, 0.2$ & 100, 300, 1000 & sp, gg & 100 [0.02] sil & cloud-crush$^q$ & 1 \\
                    & $\lesssim 1$ & $\sim 0.0615, 0.1, 0.2$ & 100, 300, 1000 & sp, gg & 1000 [0.02] sil & cloud-crush$^q$ & 1 \\  
\citet{kirchschlager2020} & 5\,--\,22 & $\sim 0.0615$ & 100 & sp, gg, ion trp & 10\,--\,1000 [0.1] sil& cloud-crush$^r$ & 1 \\        \citet{slavin2020} & $0.4, 2$ & 70 & 100 & sp & 40 sil, carb& SN-IIb$^s$ & 0.3\,--\,1.5 \\
                   & $3, 35$ & 70 & 100 & sp & 100 sil, carb & SN-IIb$^s$ & 0.3\,--\,1.5 \\
                   & $20, 70$ & 70 & 100 & sp & 250 sil, carb & SN-IIb$^s$ & 0.3\,--\,1.5 \\
                   & $40, 82$ & 70 & 100 & sp & 395 sil, carb& SN-IIb$^s$ & 0.3\,--\,1.5 \\
                   & $60, 90$ & 70 & 100 & sp & 625 sil, carb& SN-IIb$^s$ & 0.3\,--\,1.5 \\
                   & $3, 35$ & 70 & 100 & sp & LN1 sil, carb& SN-IIb$^s$ & 0.3\,--\,1.5 \\
                   & $4, 32$ & 70 & 100 & sp & PL1 sil, carb& SN-IIb$^s$ & 0.3\,--\,1.5 \\
\citet{kirchschlager2023}  & 0\,--\,10 & 0.1 & 300 & sp, gg, B & 20 [0.02] & cloud-crush$^t$ & 1 \\
                           & 0\,--\,10 & 0.1 & 300 & sp, gg, B & 100 [0.02] & cloud-crush$^t$ & 1 \\
                           & 0\,--\,70 & 0.1 & 300 & sp, gg, B & 1000 [0.02] & cloud-crush$^t$ & 1 \\
\noalign{\smallskip}\hline
\end{tabular}
\end{table}
\end{landscape}

\subsection{Confronting models with observations of SNRs}
\label{sec:snobservations}

While some simulations show that dust sputtering continues on timescales $> 10^3$ yr \citep{nozawa2007, bocchio2016, slavin2020}, important constraints on model predictions are provided by observations of SN remnants. 

Dust mass estimates have been derived through modelling dust thermal emission and/or the effects of dust absorption and scattering on the optical line emission profiles (see also Sect.~\ref{sec:sn1987A} for the application of these observational techniques to the specific case of SN1987A). A first collection of observations, mostly obtained with the \textit{Spitzer Space Telescope}, of warm dust emission in SNe and SNRs was reported by \citet{gall2011} (see, in particular, their Tables 3 and 4). These observations showed the unambiguous presence of hot dust, with temperatures ranging from  $T_{\rm d} \sim 200-300$\,K up to $\sim 1000$\, K, at early epochs (post-explosion time 
$t_{\rm pe} \lesssim 2000$ \, days), with a maximum inferred dust mass of $m_{\rm dust} < 3 \times 10^{-3} \, M_\odot$. 
At later epochs ($t_{\rm pe} \lesssim 5000$ \, days), studies based on observations of SNRs provided a large dispersion 
in the inferred dust masses (from $10^{-7} \, M_\odot$ to $\sim 1 \, M_\odot$), with higher inferred dust masses related to cold dust.

Since then, and starting from the work of \citet{matsuura2011, indebetouw2014, matsuura2015,cigan2019} on SN1987A, FIR and sub-mm observations of several Galactic SNRs \citep{barlow2010, gomez2012, arendt2014, delooze2017, tenim2017, rho2018, delooze2019, chawner2019, chawner2020, chastenet2022} using the \textit{Herschel Space Observatory} and the \textit{Atacama Large Millimeter Array} (ALMA) have been obtained, reporting up to $\sim 0.7 M_\odot$ of cold dust ($T_{\rm d} \sim 10 - 40$\,K) in several of these. Similarly large dust masses have been inferred by modelling the red-blue asymmetries of optical line emission profiles for a sample of (mostly) extragalactic SNRs \citep{bevan2016, bevan2017, bevan2019, bevan2020, wesson2021, niculescu2022, wesson2023}. Very recently \cite{Shahbandeh2023} has obtained the first mid-IR detections with JWST of two SNR (5--18 years old), revealing masses of warm ($T_{\rm d} \sim 100 - 150$\,K) dust higher than $\rm 10^{-2}~M_\odot$ in the older of the two systems (although they warn that this is likely a lower limit), second only to SN1987A.

\begin{figure*}
\includegraphics[width=.49\textwidth]{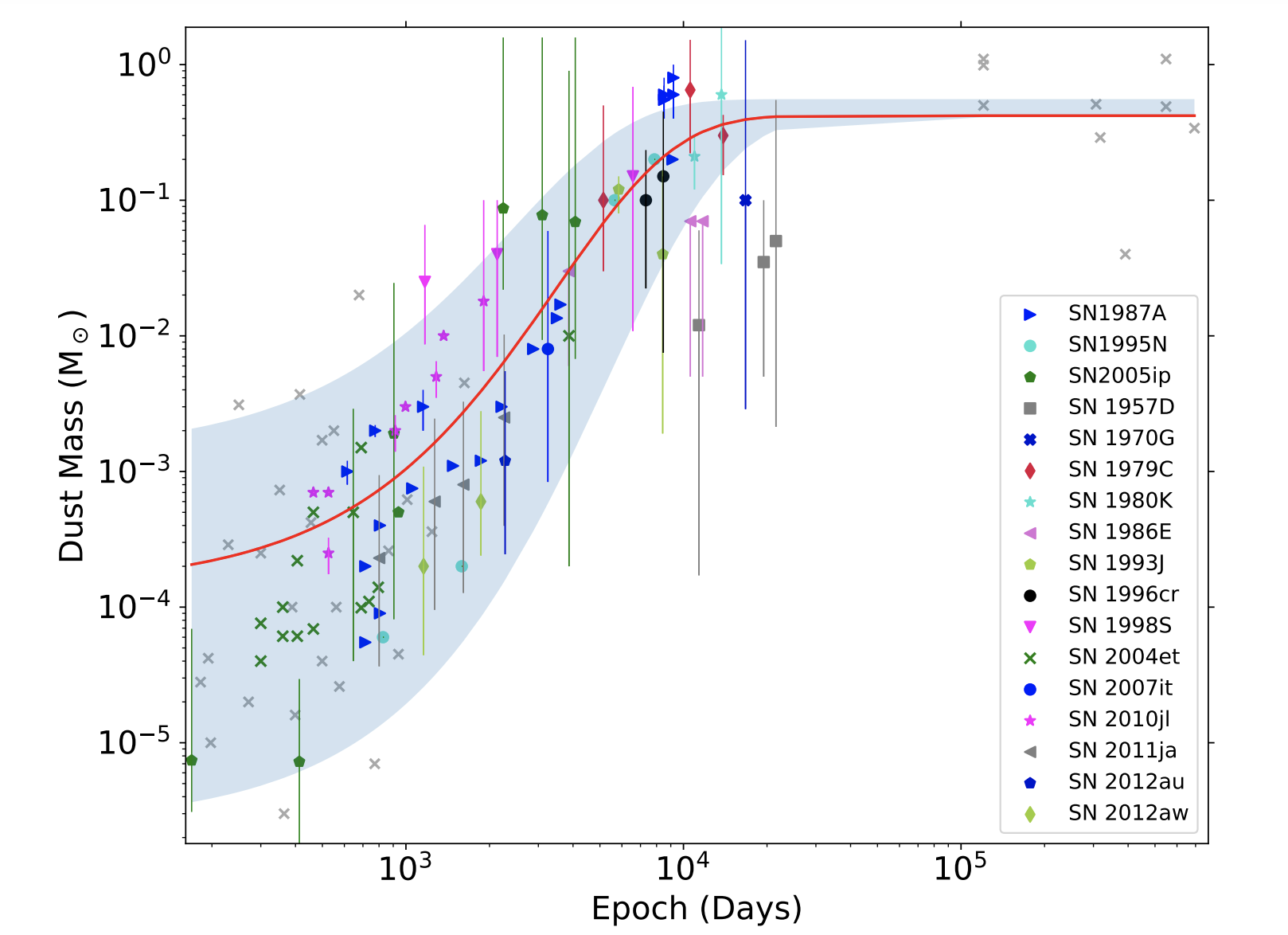} 
\includegraphics[width=.45\textwidth]{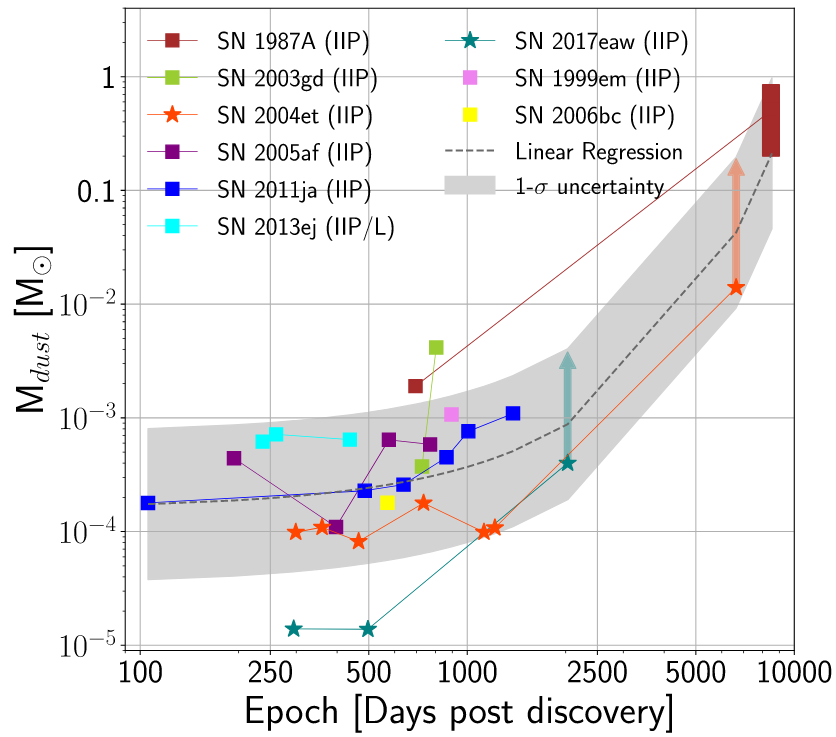}
\caption{\textbf{Left:} Dust masses inferred by \citet{niculescu2022} by modeling the red-blue asymmetries of optical emission lines of SNRs (coloured points, see also their Table~IX), complemented by additional sources from independent studies (grey crosses, see their Table~A1), as a function of the post-explosion times. 
The red solid line shows a best-fit and the light blue band encloses the error region on the best fit.
As a function of increasing age, the six supernova remnants for which dust masses are plotted are
Cas A, G29.7-0.3, G21.5-0.9, the Crab Nebula, G54.1+0.3 and G11.2-0.3, with ages $\sim $330, 850, 880, 1100, 1500, 1900 years respectively. \textbf{ Right:} Dust masses collected by \cite{Shahbandeh2023} as a function of time after explosion, based on mid-IR observations, including their recent JWST detections for SN2004et and SN2017eaw. Images reproduced with permission, copyright by the author(s).}
\label{fig:snr_obsdustmasses}
\end{figure*}

A collection of dust mass determinations as a function of the estimated post-explosion time is shown in Fig.~\ref{fig:snr_obsdustmasses}, from \cite{niculescu2022}, based on the modelling of red-blue asymmetries of optical emission lines (left), and from \cite{Shahbandeh2023}, based on mid-IR observations (right). In the left panel the red solid line represents a best fit to the observations, with the grey band representing its uncertainty. Taken at face value, these results suggest a dust mass growth with time that can be fit by a sigmoid \citep{wesson2015},

\begin{equation}
m_{\rm dust} (t) = a \, {\rm e}^{b \, {\rm e}^{c\, t}},
\end{equation}
\noindent
with $a = 0.42^{+0.09}_{-0.05} \, M_\odot$, $b = -8.0^{+4.0}_{-2.0}$, and $c = -2.88^{+1.03}_{-1.27} \, \times 10^{-4}$\, days$^{-1}$, which implies a saturation around a value of $\sim 0.42 \, M_\odot$ at around 55 years after the explosion (see however the discussion on SN1987A in Sect.~\ref{sec:sn1987A} for a different interpretation of this time sequence). Right panel, which includes recent JWST detections, show a similar increase of dust mass, although \cite{Shahbandeh2023} warn that the observed trend may actually trace a variation in optical thickness, and also warn about inhomogeneity of the data and in the way dust masses have been inferred in different studies.

One important point to keep in mind is that a considerable spread in the dust mass determination at given epochs is found when the same set of data is analysed by independent studies (see, for example, the case of SN1987A represented in Fig.~\ref{fig:snr_obsdustmasses} with blue horizontal triangles). Indeed, as already emphasized by \citet{gall2011} and thoroughly discussed by \citet{micelotta2018}, dust mass determinations are affected by the adopted dust composition, optical constants, and grain sizes\footnote{The dust masses are $\propto a^3$, so even a relatively small variation in the adopted grain sizes can lead to a significant variation in the inferred dust mass.}.
For most of the SNRs considered by \citet{niculescu2022}, the best-fitting line profiles require either 100\% silicate composition or 50\% silicates and 50\% amorphous carbon grains, with grain radii between 100 and 500 nm. While JWST observations promise to significantly advance our understanding of dust formation and survival in SN ejecta, currently dust composition remains largely unconstrained for most of the sources, with the exception of those for which polarized dust emission has been detected, such as Cas~A \citep{rho2022}, and the Crab pulsar wind nebula \citep{chastenet2022}. In the latter case, the polarized signal suggests the presence of (50--100)\, nm grains, carbon-rich grain mass fraction that varies between 12 and 70\%, and temperatures that range from $\sim 40$ to $\sim 70$\, K ($\sim 30$ to $\sim 50$\, K) for carbonaceous (silicate) grains. 

Table~\ref{table:snr_observations} attempts to summarize dust mass determinations in Cas A and Crab SNRs obtained by different studies. For Cas A, we have collected the results from \citet{barlow2010}, \citet{arendt2014}, \citet{delooze2017}, \citet{bevan2017}, \citet{niculescu2021}, and \citet{priestley2022}. For the Crab, we have integrated the compilation recently presented by \citet{chastenet2022} (see their Table~2). The table shows that the most recent studies tend to converge towards values of the dust masses in the range $\sim [0.3 - 1] \, M_\odot$ for the younger Cas A, and in the range $\sim [0.026 - 0.049] \, M_\odot$ for the older Crab. 
\begin{landscape}
\begin{table}
\caption{Compilation of dust mass determinations for Cas A and Crab SNRs. We report here the cold dust mass component, which is believed to be associated to the interior of the SNR, where the reverse shock has not yet swept up and heated the ejecta.}
\label{table:snr_observations}
\begin{tabular}{llll}
\hline\noalign{\smallskip}
& Cas A & 340 [yr] &  \\
\hline\noalign{\smallskip}
Reference & $m_{\rm dust} [M_\odot]$ & $T_{\rm d} [\rm K]$  & Notes \\
\hline\noalign{\smallskip}
\citet{barlow2010} & $0.075$ & $35$ & silicates \\
\citet{arendt2014} & $\lesssim 0.1$ & cold & undetermined \\
\citet{delooze2017} & $[0.3 - 0.5]$ & [30 - 32] & silicates \\
                     & $[0.4 - 0.6]$ &              & 50\% silicates, 50\% carbonaceous grains \\
\citet{bevan2017}    &   $1.1$ &        & $50$nm 50\% silicates, 50\% carbonaceous grains \\
\citet{niculescu2021} & $0.99 \pm 0.1$ & & $[50 - 200]$nm 50\% silicates, 50\% carbonaceous grains \\
                       & $0.99 \pm 0.1$ & & $[200]$nm 75\% silicates, 25\% carbonaceous grains \\
\citet{priestley2022}  & $[0.6 - 0.8]$ & cold & $100$nm silicate grains \\
                       & $\sim 0.13$ & cold & $100$nm 50\% silicates, 50\% carbonaceous grains \\
\noalign{\smallskip}\hline
& Crab & 969 [yr] &  \\
\hline\noalign{\smallskip}
Reference & $m_{\rm dust} \, [M_\odot]$ & $T_{\rm d} \, [\rm K]$  & Notes \\
\hline\noalign{\smallskip}
\citet{gomez2012} & $0.24^{+0.32}_{-0.08}$ & [25 - 34] & silicate grains \\
                  & $0.11 \pm 0.01 $ & [32 - 36]  & carbonaceous grains \\
                  & $[0.14 + 0.08] $ & & silicate + carbonaceous grains \\
\citet{temim2013} &  $0.19^{+0.010}_{-0.003}$ & $56 \pm 2$  & $\le 100$nm carbonaceous grains \\
                  &  $0.13 \pm 0.01$ & [23 - 55] & $\le 5000$nm silicate grains \\
\citet{owen2015} &  $[0.18 - 0.27] $ & - & [50-700]nm clumped carbonaceous grains \\
                &  $[0.98 - 1.10] $ & - & [10-900]nm clumped silicates \\
                 &  $[0.38 - 0.47]$ + $[0.11 - 0.13]$ & - & [10-1000]nm silicates + carbonaceous grains \\
\citet{delooze2019} & $[0.032 - 0.049]$ & $41 \pm 3$  & 1000 nm carbonaceous grains \\
\citet{nehme2019} & $0.056 \pm 0.037$ & $42.06 \pm 1.14$ & \\
\citet{priestley2020} & $0.026 - 0.039$ & - & [1-1000]nm carbonaceous grains \\
                      & $0.076 - 0.218$ & - & [1-1000]nm silicate grains \\
\citet{chastenet2022} &   $< [0.0002 - 0.0036]$  & $\sim [40 - 70]$ & [100-5000]nm carbonaceous grains \\
                      &   $< [0.026 - 0.059]$  & $\sim [30 - 50]$  & [100-5000]nm silicate grains \\
\end{tabular}
\end{table}
\end{landscape}

How do theoretical models compare with these findings? 
The models by \citet{bianchi2007} and \citet{nozawa2010} predict that $0.05 \, M_\odot$ and $0.08 \, M_\odot$ of dust should be currently present in Cas A, corresponding to $\sim 50\%$ of the initial dust mass formed in the ejecta. These values are smaller than the most recent observational estimates reported in Table \ref{table:snr_observations}. Due to the relatively small grain sizes, most of this dust will be destroyed before being injected in the ISM (see Table \ref{table:RSmassfrac}). \citet{micelotta2016} predict a surviving dust mass fraction which ranges between $13.3 \, (10.4$) to $16.9 \, (13.4)\%$ for amorphous carbon (silicate) grains. Taking a reference value of $\sim 0.5 \, M_\odot$ for the currently observed dust mass in Cas A \citep{delooze2017}, with 50\% silicates and 50\% carbonaceous grains, these figures imply an unrealistically high dust condensation efficiency in SN ejecta, with an initial dust mass of $[1.48 - 1.88]\, M_\odot$ of carbonaceous and $[1.87 - 2.40] \, M_\odot$ silicate grains. These values are larger than those typically found by theoretical models (see Sect.~\ref{sec:snmodels}), and require very large dust-to-gas mass ratios, given the $\sim 3.47 \, M_\odot$ ejecta mass determination (of which $\sim 3 M_\odot$ have been already been through the reverse shock) obtained by \citet{laming2020} by modeling the IR emission spectrum of Cas~A. \citet{biscaro2016} model grain formation in the SN ejecta and its reprocessing by the reverse shock and predict the time evolution of the total grain mass, composition and size distribution. They find that of the initial $\sim 0.03 \, M_\odot$ of dust formed in the ejecta, between 30\% and 60\% is present today in Cas A, and only 6\% to 11\%  will survive in the SNR and be injected in the ISM. Similar to \citet{bianchi2007} and \citet{nozawa2007}, their selected progenitor model for Cas A appears to produce too little dust mass, being at least a factor of $\sim 3 - 10$ smaller than the currently observationally estimated dust mass. 
\begin{figure*}
\centerline{\includegraphics[width=10cm]{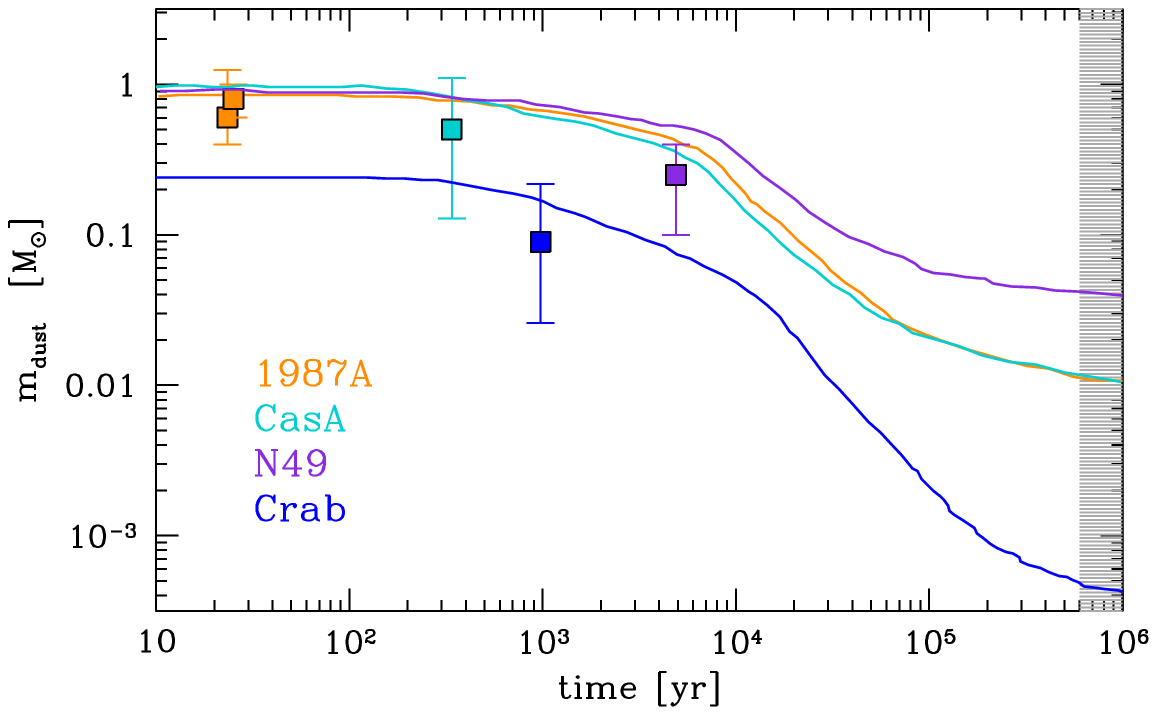}}
\caption{Time evolution of the total dust mass predicted by \citet{bocchio2016} (solid lines) for four simulated SNRs: SN1987A (orange), Cas A (turquoise), Crab (blue) and N49 (violet). Data points represent the observationally inferred dust masses, and the shaded region indicate the time interval when dust processing fades out. The data points for SN 1987A are taken from the best-fit values obtained by \citet{wesson2015} at two different ages (8515 and 9200 days). For Cas A, we considered the average value of 0.5 $M_\odot$ obtained by \citet{delooze2017} and the minimum and maximum values obtained by \citet{bevan2019} and \citet{priestley2022} assuming a 50\% silicates, 50\% carbonaceous grains mixture (see Table \ref{table:snr_observations}). For Crab and N49, we took the average value obtained assuming 100\% silicates and 100\% carbon grains by \citet{priestley2020} and \citet{otsuka2010}, respectively. Image adapted from \citet{bocchio2016}.}
\label{fig:bocchio}
\end{figure*}
Finally, \citet{bocchio2016} considered four different SN remnants (1987A, Cas A, Crab, N49), with ages ranging between 36 to 4800 years. They used observed/estimated physical properties of the four SNe to select the input parameters (progenitor mass, metallicity, explosion energy, and ambient gas density) of their simulations to model the time dependent dust mass in the SNRs. The results are shown in Fig.~\ref{fig:bocchio}. According to these model predictions, a dust mass of $[0.7 - 0.9] \, M_\odot$, such as the one observed in SN 1987A, can be indicative of the efficiency of dust production in massive stars, as the ejecta have not yet been invested by the reverse shock. This conclusion does not apply to the other three SNRs where, at their currently estimated age, between 10 to 40\% of the initial dust mass has already been destroyed. For Cas A and Crab, this translate into $0.83 \, M_\odot$ (15\% carbonaceous, $\sim 70\%$ silicates, and $\sim 16\%$ magnetite) and $0.17 \, M_\odot$ (46\% carbonaceous, $\sim 40\%$ silicates, and $\sim 14\%$ alumina) of dust grains, in reasonable agreement with the values estimated from the observations. We note that for Crab the estimated dust mass is smaller than for the other SNRs considered, as this could be the remnant of an electron-capture SN event from a lower mass progenitor of 8--10 $M_\odot$ \citep{smith2013}.

According to \citet{bocchio2016}, for none of the SNRs considered the reverse shock travelled to the center of the ejecta, and the simulations predict that the surviving dust mass that will be injected in the ISM (the effective dust yield), will be significantly smaller, ranging between $4.2 \times 10^{-4} \, M_\odot$ (for Crab) to $[1 - 4] \times 10^{-2} \, M_\odot$, for the other three SNRs, indicative of survival fractions of $\eta = 1 - 8 \%$ (see Table~\ref{table:RSmassfrac}). It is important to stress that these conclusions depend on the dynamical modeling of the reverse shock, as a faster moving reverse shock would have affected a larger fraction of the ejecta volume by the present time, implying that smaller additional dust destruction will take place before the SN dust will be injected in the ISM. Indeed, \citet{delooze2017} estimate that the reverse shock in Cas A has already travelled through $\sim 76\%$ of the ejecta volume (in agreement with the estimate by \citealt{laming2020} reported above), and that a large fraction of dust destruction ($\gtrsim$ 70--90\%) has already occurred, leading to an effective dust yield for Cas A of $\sim 0.05 - 0.30 \, M_\odot$ \citep{delooze2017, priestley2022}. However, as explained above, this would imply unrealistically large dust condensation efficiencies in the ejecta, and additional work is required to better understand the origin of this tension.

\section{Asymptotic giant branch stars (AGBs)}
Low- and intermediate-mass ($0.8 \, M_\odot \le m_{\rm star} \le 8 \, M_\odot$, where $m_{\rm star}$ is the zero-age main sequence mass) stars end their lives with a phase of strong mass loss and thermal pulses (TP) on the asymptotic giant branch (AGB), and are one of the main contributors to the chemical enrichment of the interstellar medium. The mechanisms responsible for driving the winds, the status of theoretical models and of high-resolution observations have been recently reviewed by \citet{hofner2018}. Here we present a summary of the most important results regarding dust formation in AGBs.

\subsection{Theoretical models} 
\label{sec:agbmodels}
The most promising scenario to explain the large mass loss rates observed in AGBs is based on a combination of atmospheric levitation by pulsation-induced shock waves and radiative acceleration of dust grains which form in the atmospheres. 
These models are generally referred to as 
Pulsation-Enhanced Dust-DRiven Outflow (PEDDRO) and their theoretical description requires full hydrodynamical computations with self-consistent dust formation and multi-wavelength radiative transfer (see \citealt{hofner2018} for a recent review on AGB mass loss). Extensive grids of C-stars \citep{mattson2010, eriksson2014} and M-stars\footnote{A star is classified as a C-star when carbon is more abundant than oxygen in its atmosphere. M-stars are oxygen-rich and S-stars are an intermediate class, when C/O $\sim 1$.} \citep{bladh2019} have been computed using the DARWIN code \citep{hofner2003, hofner2016}, which combines frequency-dependent radiative
transfer with nonequilibrium dust formation in 1D. Their results provide mass loss rates and other wind properties that well compare with observational data, but still rely on a parameterized description for treating convective 
energy transport, which is probably not adequate to account for the strongly nonlinear, large-scale convective 
motions that couple to AGB stellar pulsations \citep{ahmad2023}. Recently, \citet{freytag2023} have presented the first 3D radiation-hydrodynamical simulations of dust-driven winds of AGBs, exploring the interplay of convection, pulsation, atmospheric shocks, dust formation, and wind acceleration. In these simulations, computed with the CO5BOLD code, convection and pulsations emerge self-consistently and strong deviations from spherical symmetry lead to a patchy stellar atmospheric structure. As a result, dust grains can efficiently form closer to the star than spherical averages of the temperature would indicate, in dense regions with enhanced grain growth rates. This can lead to dust-driven outflows with low mass-loss rates in situations where 1D models with the same stellar parameters do not produce winds. In contrast, for stars where the overall conditions for dust formation and wind acceleration are favorable, it is not obvious whether the resulting mass-loss rates will be higher or lower than predicted by 1D models, as the increased efficiency of dust formation in high-density clumps may be set off by a lower filling factor of these regions \citep{freytag2023}. While these first exploratory 3D models are important to recover the complex 3D morphology and clumpy dust distribution observed in AGBs, they are time-consuming to run and analyse. 

For this reason, AGB dust yields have been generally computed assuming a stationary, spherically symmetric wind \citep{ferrarotti2006}. This approximation is not fully-consistent, as it decouples the properties of the wind from dust growth that,
as we have seen above, plays a major role in the wind dynamics.
However, at present this simplified method is the only feasible way
to couple dust formation with stellar evolution calculations.

The first pioneering works by \citet{ferrarotti2006, zhukovska2008,  gail2009} were based on synthetic models to describe the properties of the central star, and allowed to compute the dust yields for the three different evolutionary phases of AGBs, when the stars appear spectroscopically as M-, S- or C-stars. In all these models, it is assumed that the grains grow on some kind of seed nuclei, such as small TiO$_2$ clusters, but the results are shown to be independent of the adopted size and composition of these seeds. The grain growth rate is determined by the slowest reaction, which generally involve the least abundant chemical element (the key species). The evolution of grain size with time is determined by the competition between grain growth and grain destruction by thermal evaporation and/or chemical sputtering. 

As a lower limit to the stellar mass spectrum, they consider $1 \, M_\odot$ because the mass loss rate in lower mass stars that reach the thermally-pulsating AGB phase (down to $0.8 \, M_\odot$) is too small to significantly contribute to dust production. They explore stellar initial masses up to $7 \, M_\odot$ \, and vary the initial stellar metallicity from $Z=0.001$ ($Z = 0.07 \, Z_\odot$) to $Z=0.02$ ($Z = 1.4 \, Z_\odot$)\footnote{Here we have assumed a solar metallicity of $Z_\odot$=0.0142 \citep{asplund2009}.}. 

The grain properties depend on the elemental composition of the stellar atmospheres, which change during stellar evolution. Particularly important are the third dredge-up (TDU) episodes that follow each thermal pulse and that transport the products of the He-burning shell to the stellar surface, transforming the original O-star in a C-star.  

However, stars more massive than 3--4 $M_\odot$ \, experience hot bottom burning (HBB), i.e. proton-capture nucleosynthesis at
the base of the outer envelope, that favours the conversion of C to
N by the CN-cycle and the reconversion of the C-rich to an O-rich
atmosphere. 

Simple parametric formulae are used to describe TDU in the models by \citet{ferrarotti2006, zhukovska2008, gail2009}, where TDU parameters depend on the initial stellar mass, metallicity and on the number of pulses \citep{karakas2002} and are fixed by comparing the model results to the observed distribution of carbon stars in the Large Magellanic Cloud (LMC). 

Similarly, HBB is also accounted for in a simplified way, assuming that it occurs for all stars with masses above 4 $M_\odot$ as long as the envelope mass remains above a critical metallicity-dependent limit. When this occurs, all the C and N nuclei are immediately converted into their equilibrium abundance predicted by the CN-cycle. Hence, the models do not account for partial conversion of C into $^{14}$N for stellar masses at the lower mass limit for HBB, potentially overestimating the duration of the M-stars phase \citep{ferrarotti2006}.

\begin{figure}
\centerline{\includegraphics[width=10cm]{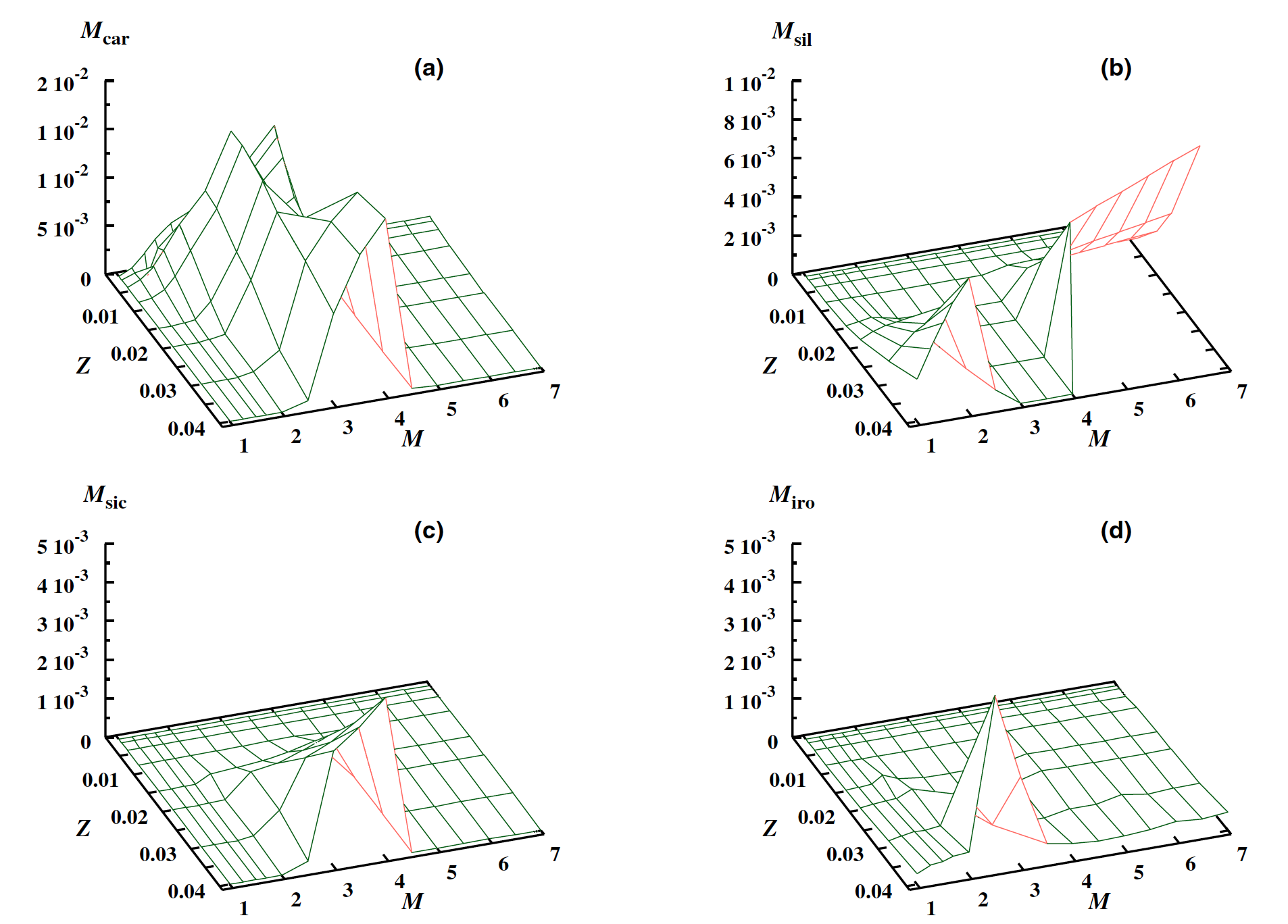}}
\caption{The metallicity-dependent dust mass produced by individual AGBs according to the models by \citet{ferrarotti2006,zhukovska2008,gail2009}. The four panels represent the results for the four main dust species: carbon (a), silicates (b), silicon carbide (c), and iron grains (d). From \citet{zhukovska2008}.}
\label{fig:zhukovska2008_fig10}
\end{figure}

The resulting dust yields (the total dust mass released by individual AGBs) for a finer metallicity grid have been presented by \citet{zhukovska2008, gail2009} and are shown in Fig.~\ref{fig:zhukovska2008_fig10}. The four panels illustrate the predicted mass of the four main dust species: carbon dust, silicate dust (that comprises forsterite, Mg$_2$SiO$_4$, fayalite, Fe$_2$SiO$_4$, enstatite, MgSiO$_3$, ferrosilite, FeSiO$_3$, and quarz SiO$_2$), iron dust, and silicon carbide\footnote{Part of the data shown in Fig.~\ref{fig:zhukovska2008_fig10} can be found as online material in \citet{ferrarotti2006}.}.

The production of carbon dust is essentially limited to stars from the mass range $2 \, M_\odot \leq m_{\rm star} \leq 4 - 5 \, M_\odot$ that become C-rich during the TP-AGB phase. It is largely independent of the initial stellar metallicity, although TDU becomes more efficient with decreasing metallicity and even lower mass stars produce carbon dust \citep{gail2009}. Silicate dust is formed only by low-mass stars, with $m_{\rm star} \le 2 \, M_\odot$, that enter the instability strip before they become C-stars, and by stars suffering HBB, $m_{\rm star} \geq 4-5 \, M_\odot$. 

Silicate dust production depends strongly on metallicity, as the refractory elements required for its formation (O, Si, Mg, Fe) are not synthesized by low and intermediate mass stars and have to be present in the material from which the stars formed. This requires a metallicity $Z > 0.1 \, Z_\odot$ ($Z > 0.001$). The production of silicon carbide and iron grains requires even higher metallicity and is less efficient than carbon or silicate dust production \citep{ferrarotti2006, zhukovska2008}. 

The results presented above are strongly dependent on the efficiency of TDU and HBB, which in turn depend on the modelling of convection and on the complex coupling between the outer region of the degenerate core and the inner region of the external mantle during the TP-AGB phase. 

Dust formation calculations based on improved models have been accomplished independently by \citet{nanni2013, nanni2014} and by \citet{ventura2012a, ventura2012b, dicriscienzo2013, dellagli2014, dellagli2017, dellagli2019}.

The models\footnote{The total dust yields of TP-AGB stellar evolution models as a function of the stellar mass and metallicity are available online at \url{https://ambrananni085.wixsite.com/ambrananni/online-data-1}.} by \citet{nanni2013, nanni2014} are based on a set of TP-AGB evolutionary tracks with initial mass $1 \, M_\odot \leq m_{\rm star} \leq 5 - 6 \, M_\odot$ and metallicity $0.001 \, Z_\odot \leq Z \leq 0.04 \, Z_\odot$, computed with the COLIBRI code \citep{marigo2013}. This integrates the stellar structure equations from the atmosphere down to the bottom of the hydrogen-burning shell, using the characteristic quantities at the first thermal pulse obtained from the PARSEC database of stellar models \citep{bressan2012}. Compared to purely synthetic models, this approach allows to accurately follow the changing envelope structure and the energetics and nucleosynthesis of HBB. The models uses specific prescriptions for mass loss and TDU, whose efficiencies are parametrized as a function of the current stellar mass and metallicity. The models are then calibrated against observations of AGBs in the Galaxy \citep{nanni2013} and in the Magellanic Clouds \citep{nanni2016, nanni2018}. 

Conversely, the models first presented by \citet{ventura2012a, ventura2012b} and later expanded to lower \citep{dicriscienzo2013, dellagli2019} and higher \citep{dellagli2017, ventura2018} initial stellar metallicity, are based on the ATON code \citep{ventura1998}, which integrates the evolution of the stars from their pre-main-sequence phase until the almost complete ejection of their external mantle.
This allows to compute the changes in the surface chemical composition due to TDU or HBB in a self-consistent way, overcoming some of the limitations of synthetic and semi-synthetic stellar models. Mass loss is computed as function of the current stellar mass and luminosity using empirical prescriptions, with parameters calibrated against observations \citep{ventura2012b}. 

The models are very similar for what concerns the wind description and dust formation model, as both \citet{nanni2013, nanni2014} and \citet{ventura2012a, ventura2012b} follow the method adopted by \citet{ferrarotti2006}: a spherically symmetric wind and a two-steps dust formation scheme. Yet, \citet{nanni2013} discuss two alternative cases, named low condensation temperature (LCT) and high condensation temperature (HCT) models, where different assumptions are made regarding the regimes where silicate and carbon grains are allowed to grow. Here we discuss only their more recent data \citep{nanni2014, nanni2015}, where the conditions for carbon dust growth are the same as in \citet{ferrarotti2006} and  \citet{ventura2012a}, but silicates are allowed to form when $T_{\rm dust} < 1400$K, as if the only grain destruction process were sublimation due to heating by the stellar radiation\footnote{In \citet{ferrarotti2006} dust species that can react with hydrogen molecules can be chemi-sputtered and this process inhibits dust condensation unless $T_{\rm dust} < 1000$K.}

\begin{figure}
\centerline{\includegraphics[width=12cm]{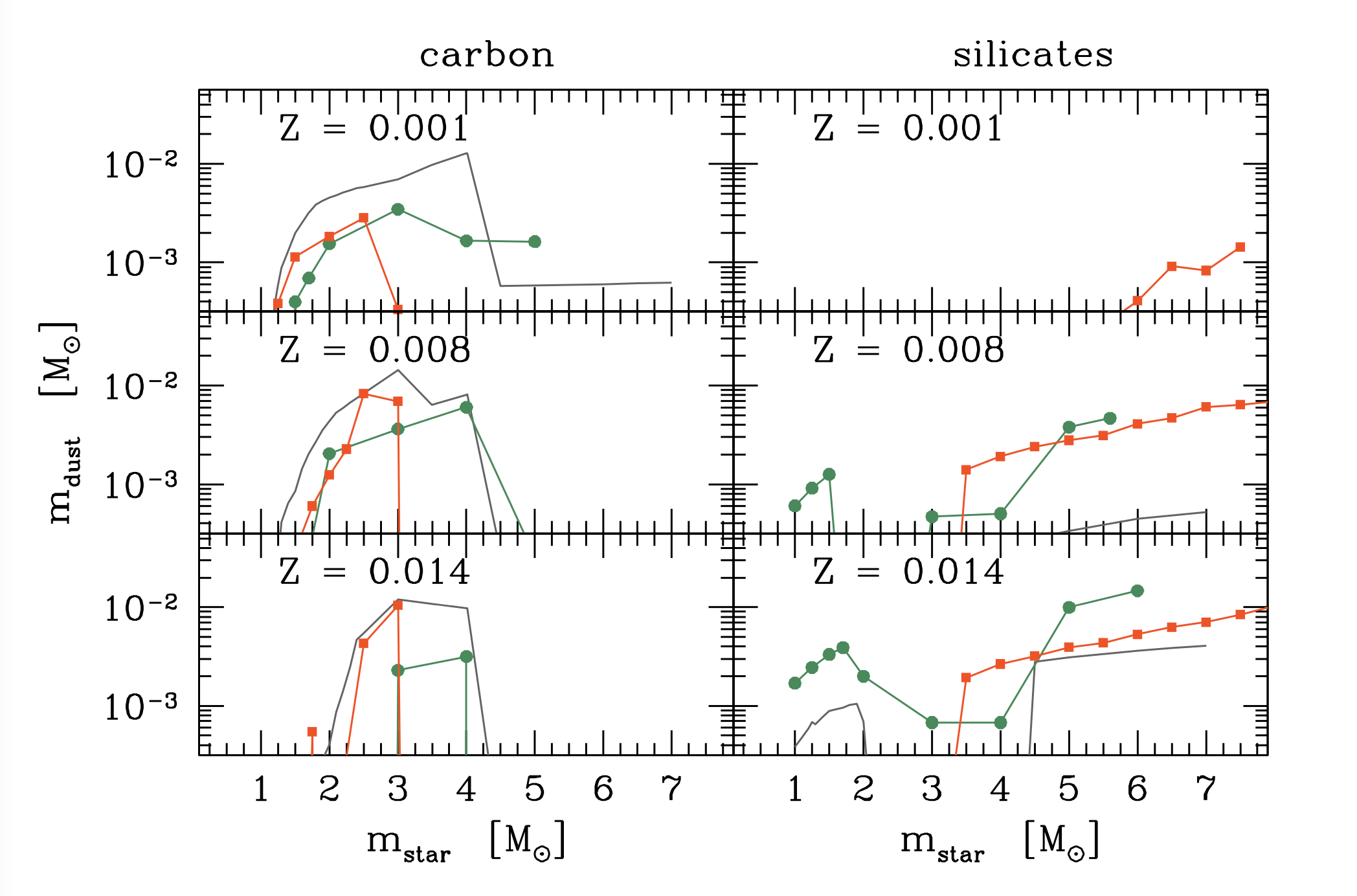}}
\caption{The mass of carbon ({\bf right panels}) and silicates ({\bf left panels}) produced by individual AGBs as a function of their initial masses assuming an initial metallicity of $Z = 0.001 \, (Z = 0.07 Z_\odot)$, $Z = 0.008 \, (Z = 0.57 Z_\odot)$, $Z = 0.014 \, (Z = Z_\odot)$. We compare the results of \citet{zhukovska2008} (gray solid lines), \citet{nanni2014, nanni2015} (green dots and lines), and \citet{ventura2012a, ventura2018} (red squares and lines). In the bottom panels, the tracks from \citet{zhukovska2008} and  \citet{nanni2015} correspond to $Z = 0.02$.}
\label{fig:agb_comp1}
\end{figure}
\begin{figure}
\centerline{\includegraphics[width=12cm]{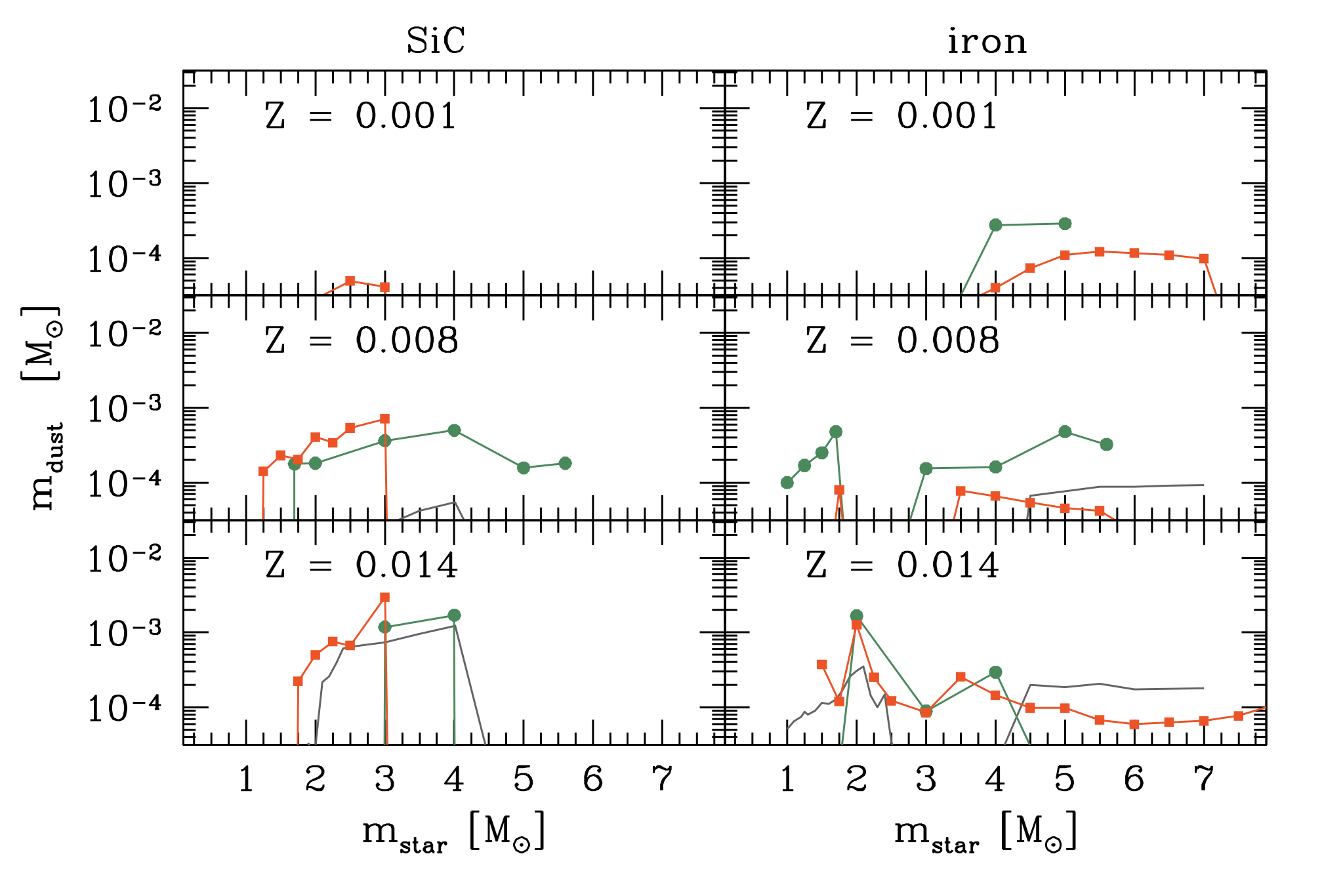}}
\caption{Same as Fig.\ref{fig:agb_comp1} but for SiC and iron grains.}
\label{fig:agb_comp2}
\end{figure}

\begin{figure}
\centerline{\includegraphics[width=12cm]{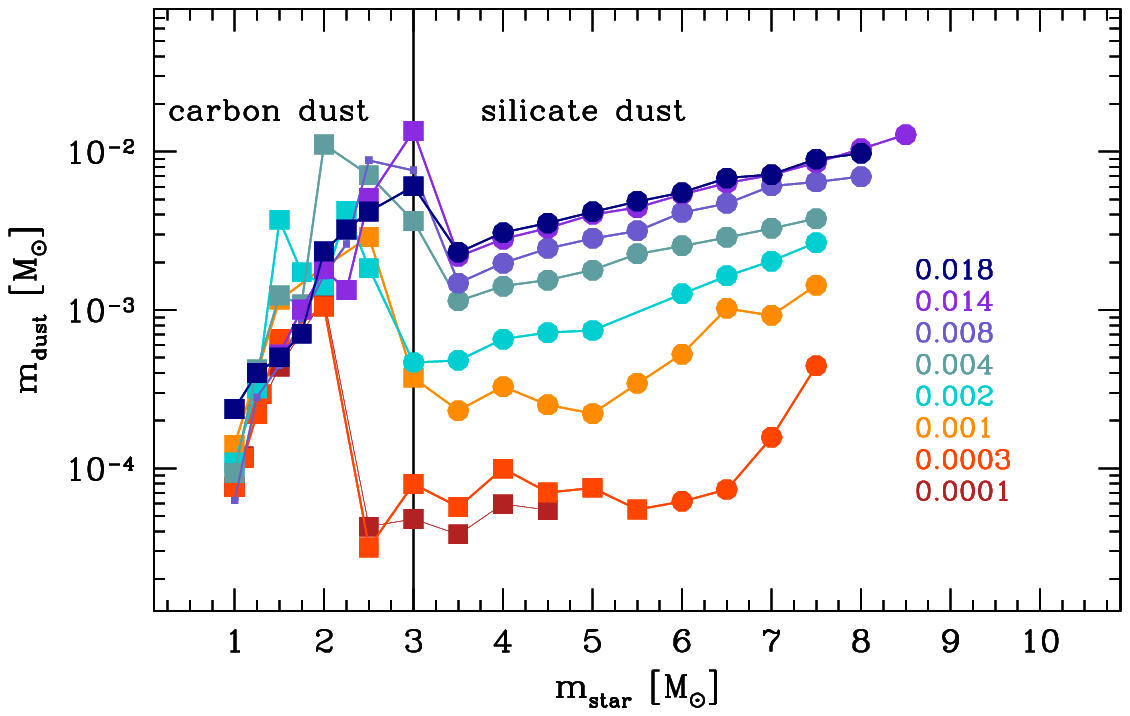}}
\caption{Total dust mass produced by individual AGBs as a function of the initial stellar mass predicted by ATON models for varying initial metallicity, as indicated by the different colours in the legend on the right. Squares and circles indicate models where the dominant dust species are, respectively, carbon and silicates. The vertical line marks the transition between carbon and silicate dust production (adapted from \citealt{dellagli2019}).}
\label{fig:agbdust_aton}
\end{figure}

A comparison between the AGB dust yields predicted by COLIBRI and ATON models is shown in Figs.\ref{fig:agb_comp1} and \ref{fig:agb_comp2},where  we also report the results obtained by \citet{zhukovska2008} for the same stellar initial metallicity. Some general trends are common to all the models: carbon and SiC dust are mostly formed by lower mass AGBs while silicates are more efficiently formed by higher mass AGBs. In addition, all the models show that the formation of silicates, SiC and iron dust strongly depends on metallicity. However, in some mass and metallicity ranges, the dust mass released by individual AGBs can greatly vary between models. These variations can be mostly ascribed to differences in the efficiency of TDU and HBB and in the duration of the TP-AGB phase \citep{nanni2013,ventura2014}. 

One peculiar feature of ATON models is the sharp transition from carbon dust to silicate dust production that occurs around $m_{\rm star} \sim 3 - 3.5 \, M_\odot$ \citep{ventura2012a}. Indeed, stars with lower masses never reach the conditions for HBB (the temperature at the base of the convective envelope must become $\geq 40$ MK to activate advanced proton capture nucleosynthesis), independently of the convective model adopted. However, the efficiency of TDU for stars with masses $m_{\rm star} \ge 3 - 3.5 \, M_\odot$, close to the limit for HBB, is very sensitive to the convection model adopted\footnote{All the ATON models are based on the full spectrum of turbulence and evolve at larger luminosities, on more expanded configurations, in comparison with their counterparts calculated with the traditional mixing length theory. This partially limits the efficiency of TDU for masses close to the limit for HBB \citep{ventura2014}.} \citep{ventura2014}. Fig.\ref{fig:agbdust_aton} shows that this trend is maintained for all ATON models down to $Z = 0.001 \, (0.07 \, Z_\odot$). At lower metallicity, the dust production efficiency in stars with $m_{\rm star} > 2 \, M_\odot$ drops by more than 1 dex, because HBB prevents the formation of carbon-type dust and their mass-loss rates and metallicity are too small to form silicates. For $Z = 0.0003$, higher mass models produce more dust, mostly in the form of silicates, because they evolve at
larger luminosities and experience larger mass loss rates \citep{dellagli2019}. 

One obvious implication of the above findings is that the contribution of AGBs to dust enrichment in young 
($< 300$ Myr) starbursts is mostly limited to silicate dust
and starts to be significant when $Z > 0.07 \, Z_\odot$. On longer
timescales, when $m_{\rm star} \leq 3 \, M_\odot$ reach their TP-AGB
phase, AGBs can contribute to carbon dust enrichment, independently of the initial stellar metallicity.

\subsection{Confronting models with observations}

Theoretical models of AGB winds based on the PEDDRO scenario provide information on the mass loss rate, wind velocity and dynamical structure that can be compared to observations. The agreement is generally good (see Figs.~15 and 18 in \citealt{hofner2018}) for both C-stars \citep{eriksson2014} and M-stars \citep{bladh2015, bladh2019}. The comparison is limited by the number of models, that often do not cover all the ranges of possible stellar parameters, and by the difficulty in inferring the stellar parameters from the observations. With this in mind, the comparison seems to suggest a scarcity of models with low outflow rates compared to the observations, pointing to possible deviations from the adopted spherical symmetry or to clumpy gas and dust distributions in the atmospheres \citep{hofner2018}.

Regarding specifically dust in AGB envelopes (and also in post-AGB  envelopes, which may provide a cleaner view of dust produced in the AGB phase),
extensive observations have been performed to characterize it at millimeter, infrared and optical wavelengths \citep[e.g.][]{Mauron2006,
Buemi2007,Groenewegen2011,
Matsuura2013,Gottlieb2022,Velilla-Prieto2023,Montarges2023}.
AGB dust formation models have  been used to interpret such observations  in our Galaxy \citep{ventura2018,Tosi2023,Dellagli2023}, in the Magellanic Clouds \citep{dellagli2014,dellagli2015,dellagli2015b,nanni2016,nanni2018,Matsuura2013} and in Local dwarf galaxies \citep{dellagli2016,dellagli2018, dellagli2019b}. By running radiative transfer calculations on time-dependent grids of AGBs, these studies allow to characterize individual sources as well as to derive global information on the galaxies, such as the AGB dust production rate (DPR) at the present time. Table~\ref{table:DPR} presents a summary of these results, indicating -- where possible -- the sources that currently dominate dust production and the observations used in the analysis.  

Table~\ref{table:DPR} shows that current DPRs strongly depend on the galaxy properties, particularly the total stellar mass and star formation history. Yet, for all the systems considered so far, the current DPR of AGBs is found to be largely dominated by carbon grains produced by $m_{\rm star} < 3 \, M_\odot$ C-stars. Very likely, the contribution of AGBs to silicate dust enrichment would be apparent in metal-rich young starbursts, with stellar populations younger than $300$\, Myr.

\begin{figure*}
\centering
\includegraphics[width=\textwidth]{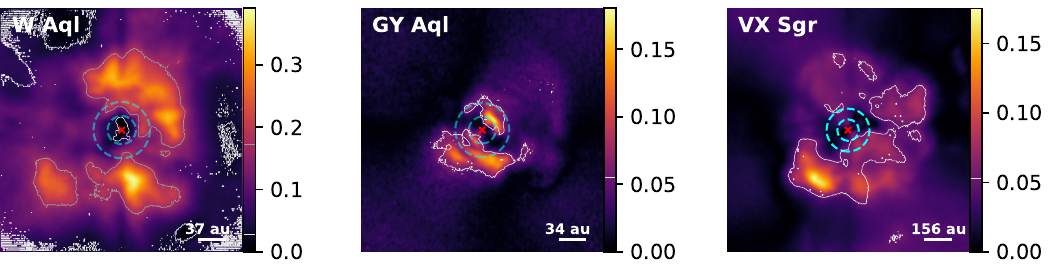} 
\caption{Maps of degree of linear polarization (DoLP) at optical wavelengths in three AGBs. The red cross shows the location of the star centre, while the green and white dashed circles have radii 10 and 20 times the stellar radius, respectively. Contours correspond to the 5$\sigma$ level of the DoLP in GY Aql and VX Sgr, and to the 30$\sigma$ level in W Aql. The maps illustrate the clumpiness and  inhomogeneous distribution of dust in the envelopes. Images reproduced with permission from \citet{Montarges2023}, copyright by the author(s).}
\label{fig:agb_polarization}
\end{figure*}

It should be noted that a major limitation of the vast majority of the models is that they assume spherical symmetry. Indeed, recent high resolution millimeter and optical observations of AGBs (Fig.~\ref{fig:agb_polarization}) have revealed a very clumpy and inhomogeneous structure of their dusty envelopes. Additionally, the distribution of dust is found to be decoupled from the distribution of molecular gas \citep{Montarges2023,Velilla-Prieto2023}. These observations are a clear warning about the suitability of simple, spherically symmetric models.
Additionally, the same authors report that the degree of polarisation and its dependence on wavelength is different for different AGBs in their sample, suggesting that both the dust chemical composition and grain size distribution is different among different AGBs. This could be an evolutionary effect and/or indicate a dependence on the metallicity and mass of the progenitor, as expected by theoretical models.

\begin{landscape}
\begin{table}
\begin{tabular}{llllllll}
\hline
\hline
Name & Log $M_{\rm star}/M_\odot$ & $Z/Z_\odot$ & SFR & DPR  & dominant species & relevant data & reference \\     
\hline
\hline
Sextan A & 6.6 & 0.07 & 0.006 & $6 \times 10^{-7}$ & 90\% $[0.1 - 0.2]\, \mu$m carbon  & WHIRC/WIYN$^c$, Spitzer$^b$ & \citet{dellagli2019b} \\
& & & & & 10\% $[0.05 - 0.07]\, \mu$m silicates & & \\
\hline
IC10 & 8.5 & 0.2 & 0.01 & $7 \times 10^{-6}$ & $\sim 0.15 \, \mu$m carbon & WFCAM/UKIRT$^a$, Spitzer$^b$ & \citet{dellagli2018} \\
\hline
IC1613 & 6.7 & 0.05 & 0.08 & $5 \times 10^{-5}$ & 80\% $[0.003-0.18] \, \mu$m carbon  & WFCAM/UKIRT$^d$, Spitzer$^b$ & \citet{dellagli2016} \\
& & & & & 20\% $[0.001 - 0.08]\, \mu$m silicates & & \\
\hline
LMC & 9.04 & 0.6 & 0.39 & $4.5 \times 10^{-5}$ & 85\% $[0.05 - 0.2]\,\mu$m carbon & Spitzer$^e$ & \citet{dellagli2015} \\
 & & & & & 15\% $0.1 \, \mu$m silicates & & \\
 \hline
 LMC & 9.04 & 0.6 & 0.39 & $1.77 \times 10^{-5}$ & carbon & mixed catalogues$^{f,g,h}$ & \citet{nanni2019} \\
 \hline
SMC & 8.5 & 0.2 & 0.03 &  &  & Spitzer$^i$ & \citet{dellagli2015b} \\
 & & & & &  & & \\
 \hline
 SMC & 8.5 & 0.2 & 0.03 & $2.56 \times 10^{-6}$ & carbon & mixed catalogue$^{h,l}$ & \citet{nanni2019} \\
 & & & & &  & & \\
\hline
\hline
\end{tabular}
\caption{Total dust production rates (DPR) in [$M_\odot/$yr] inferred by comparing AGB dust formation models with observations of local galaxies. For each system, we also report the total stellar mass, metallicity, current star formation rate (SFR) [$M_\odot/$yr], the dominant dust species, the reference study and the data sample used.
References for the observational samples: $a$: \citet{gerbrandt2015}; $b$: \citet{boyer2015}; $c$: \citet{jones2018}; $d$: \citet{sibbons2015}; $e$: \citet{meixner2006}; $f$: \citet{riebel2012}; $g$: \citet{jones2017}; $h$: \citet{groenewegen2018}; 
$i$: \citet{gordon2011}; 
$l$: \citet{srinivasan2016}.
}
\label{table:DPR}
\end{table}
\end{landscape}

\section{Additional stellar sources of dust}

\subsection{Red Super Giants}
\label{sec:rsg}

Red Super Giants (RSGs) are the evolved, He-burning descendants of stars with initial masses between about 12 and 30 $M_\odot$ \citep{levesque2006}. Stars that are less massive undergo second dredge-up and end their lives as asymptotic giant branch (AGB) stars \citep{eldridge2007}. Mass-loss rates increase with mass and stars that are more massive suffer enough mass loss to remove their hydrogen envelopes and become Wolf--Rayet stars. 

\citet{levesque2006} modeled the optical spectra of RSGs in the Milky Way showing that many of these stars suffer extinction beyond that of their neighbouring O and B stars. They attribute this excess to dust extinction in the circum-stellar shells of these ``smoky'' stars, in the sense that when the surface temperature of a red supergiant falls below about 5000 K, dust begins to condense out of the stellar wind at a distance of around 5--10 $r_{\rm star} \approx 1000 \, R_\odot$ \citep{massey2009}. This dust is thought to be partially responsible for driving the stellar wind via radiation pressure, and interferometry in the IR has demonstrated that for some RSGs the dust is found very close to the star itself (3--5 $r_{\rm star}$), while in other cases it is found at greater distances, suggesting that the production of substantial amounts of dust is episodic in nature, with timescales of a few decades \citep{danchi1994}. 

It might be expected that the amount of dust production would correlate with the mass loss, which in turn correlates with the luminosity because this is responsible for the stellar wind \citep{vanLoon2005}. \citet{massey2005} show that the RSG dust production rate (indicated by the 12 \mic \,  excess) is well correlated with bolometric luminosity:

\begin{equation}
{\rm log} \frac{dm_{\rm dust}}{dt} = - 0.43 \, M_{\rm bol} - 12.0 
\end{equation}
\noindent
for $M_{\rm bol} < - 5$, which roughly corresponds to masses $> 10 \, M_\odot$ and where the dust production rate is measured in $M_\odot$/yr. In \citet{levesque2005} they used the evolutionary tracks of \citet{meynet2003} to estimate that the masses of RSGs scale with luminosity as ${\rm log} \, (m_{\rm star}/M_\odot) = - 0.50 - 0.099 \, M_{\rm bol}$, so they expect:

\begin{equation}
{\rm log} \, \frac{dm_{\rm dust}}{dt} = 4.3 \, {\rm log} \, \frac{m_{\rm star}}{M_\odot} - 14.2
\end{equation}
\noindent
for RSGs with masses $10 < (m_{\rm star}/M_\odot) < 25$. From the evolutionary models of \citet{meynet2003}, they find that the RSG phase lasts 2 Myr (10 $M_\odot$) to 0.4 Myr (25 $M_\odot$), and they use the models to approximate:

\begin{equation} 
{\rm log} \, t_{\rm RSG} = 8.1 - 1.8 \, {\rm log}\, \frac{m_{\rm star}}{M_\odot}.
\end{equation}
\noindent
From the two above relations, one finds a scaling between the dust mass released by a RSG and its stellar mass:
\begin{equation} 
{\rm log} \, \frac{m_{\rm dust}}{M_\odot} = 2.5 \, {\rm log} \, \frac{m_{\rm star}}{M_\odot} - 6.1
\label{eq:RSGyields}
\end{equation}
\noindent
so that RSGs with masses $10 < m_{\rm star}/M_\odot < 25$ release a dust mass in the range $ -3.6 \leq {\rm log}\, m_{\rm dust}/M_\odot  \leq - 2.6$ during their evolution. These figures are comparable to the expected dust masses released by core-collapse SNe when accounting for reverse shock destruction.

It is noteworthy that extra intrinsic extinction close to the red supergiant progenitors would give reduced luminosities and lower predicted masses, since mass estimates are based on mass-luminosity relations. \citet{walmswell2012} show that even using a crude spherically symmetric model, with no metallicity variation, dust extinction and corrected mass determinations at the high-mass end provide a solution to the so-called RSG problem, i.e. the apparent lack of RSG progenitors with masses $> 17 \, M_\odot$ in the pre-explosion images of Type IIP SN \citep{smartt2009, smartt2015}. Indeed, evidence for circumstellar extinction around SN2017eaw progenitor star supports the conclusion that progenitor mass estimates can be low if circumstellar
extinction is not properly accounted for \citep{kilpatrick2018}. By modeling the optical-to-mid IR SED in the 5 months before the explosion, \citet{kilpatrick2018} show that the dust shell around SN2017eaw was compact, with a radius of approximately 4000 $R_\odot$ (five times the photospheric radius), and so it is likely that this dust was vapourized within the first few days after explosion. 
Hence, the question of whether such dust can survive the impact of high-velocity ejecta from the subsequent supernova and thereby go on to enrich the dust content of the host galaxy is an interesting one. 

\subsection{Wolf--Rayet stars}
\label{sec:wr}

Stars with initial masses greater than 30 $M_\odot$ loose their hydrogen envelope and become Wolf--Rayet (WR) stars.  Their different emission spectra suggest that WR stars experience three phases of evolution: the WN phase is characterised by hydrogen burning in the core, the more evolved WC phase corresponds to helium burning in the core, have no hydrogen left in their atmosphere, are rich in helium and carbon and have a varying amount of oxygen. Those WCs with the most oxygen are classified as WO stars.

Only WCs are known to produce dust, and even those may require a hydrogen-rich OB binary companion for dust ejection (see \citealt{crowther2003, lau2020, lau2021, lau2022}). Owing to the photospheric chemical composition, carbon-based dust is expected to form \citep{cherchneff2000}. The mixing of carbon-rich material from the WC star plus hydrogen from the OB companion, together with efficient shielding from their harsh ionizing fluxes via the shocked region of their colliding winds most likely provides the necessary ingredients for dust nucleation and growth to occur \citep{cherchneff2000,crowther2003}. Indeed, \citet{lau2020} performed a dust
spectral energy distribution analysis of 19 Galactic WC binaries, revealing a broad range of dust production rates, $\dot{m}_{\rm dust} = 10^{-10} - 10^{-6} \, M_\odot/{\rm yr}$, and carbon dust condensation fractions between 0.002\% and 40\%, consistent with predictions from theoretical models of dust formation in colliding-wind binaries. Recent observations with JWST reveal the spatial and spectral signatures of over 17 carbon-rich nested circumstellar dust shells around the WR binary WR 140, indicating their persistence in the luminous and hard radiation field of the central binary system for at least 130 yr after the initial dust-formation event \citep{lau2022}.

According to the evolutionary models of \citet{meynet2003}, WCs come from stars with masses greater than 40 $M_\odot$; the lifetime of the WC stage is independent of mass, with $t_{\rm WC} = 0.2$ Myr. The (total) mass-loss rates of WCs are roughly independent of mass, and are about $10^{-5} \, M_\odot$/yr \citep{nugis2000}. Assuming a dust nucleation efficiency of $f_{\rm cond} = 1 \%$ \citep{dwek1998}, the dust mass released by a WC star is expected to be:
\begin{equation}
{\rm log}\, \frac{m_{\rm dust}}{M_\odot}  =  - 1.7 + {\rm log}\, \frac{f_{\rm cond}}{0.01}
\label{eq:WRyields}
\end{equation}
\noindent
Similarly to RSGs, it is possible that a large fraction of this dust will be destroyed by the explosion of the SN Ib/c that ends the evolution of WR stars.

It is important to stress that for a standard stellar initial mass function, WC stars are very rare, such that they are expected to be minor contributors to interstellar dust. In addition, late-type WCs are known to be absent in low-metallicity galaxies (see \citealt{massey2003} and references therein), although \citet{lau2021} show that efficient dust-formation is feasible even at metallicities $Z \simeq 0.6 \, Z_\odot$, provided that the systems host an O-type companion with a high mass-loss rate ($\dot{m} \gtrsim 1.6 \times 10^{-6} \, M_\odot/{\rm yr}$). In more extreme environments (such as very metal-poor starbursts, or for most galaxies at early times), RSGs could play a more prominent role compared to WR stars.

\subsection{Classical Novae}

Classical Novae (CNae) are transient events driven by runaway thermonuclear reactions on the surfaces of accreting white dwarfs (WDs) in interacting binary systems 
(see \citealt{starrfield2016} for a recent review). Observations of the outburst show that a CN eject metal-enriched gas, and when the expanding gas has cooled to temperatures of 1500 K (50 to 200 days after the eruption), a dust condensation phase starts, characterized by declining visual light and rising IR emission \citep{gehrz1988}. 

Infrared observations have confirmed the formation of carbon, SiC and oxygen-rich silicate grains (\citealt{starrfield2016} and references therein). The IR emission continues to rise as the grains grow to a maximum radius of 0.2--0.55 $\mu$m within a few hundred days after their condensation, and then falls as the mature grains are dispersed by the outflow into the ISM \citep{gehrz1998}. 
The rate of decline of the IR radiation suggests that the grains begin to decrease in radius shortly after having grown to their maximum size, consistent with the hypothesis that they could be processed by evaporation or sputtering before eventually reaching the ISM. About $10^{-8} - 10^{-6} M_\odot$ of dust forms in each episode (see Table 5 in \citealt{gehrz1998}).  

The above results have been recently corroborated by \citet{derdzinski2017}, who suggested that dust formation could occur in the cool dense shells behind shocks caused by the interaction of the nova outflow with lower velocity mass ejected earlier in the outburst. By applying classical nucleation theory and a thermodynamic evolution model for the post-shock gas, they find that silicates and carbon grains can form and grow to sizes $\sim 0.1\, \mu$m, and total dust masses of $10^{-10} - 10^{-7} \, M_\odot$, consistent with the observed ones.

Given these figures, CNe are unlikely to be a major contributor to interstellar grains, although their rate is observed to be $35 \pm 11$ per year (possibly 50 per year) in the Galaxy \citep{starrfield2016}. 

\subsection{Type Ia supernovae}

Despite extensive searches, no evidence for dust condensation in type Ia SNe (SNIa) has been found to date.  Spitzer and Herschel observations of type Ia SN remnants, including Kepler and Tycho \citep{blair2007, gomez2012, williams2013}, have provided no evidence for any dust associated with the ejecta, showing that all the IR emission from these remnants come from pre-existing circumstellar dust heated by the explosion blast wave.
Explanations for the lack of newly formed dust in type Ia SNe have been proposed, based on unfavourable physical conditions in the ejecta and/or on the destructive effect caused by the reverse shock \citep{nozawa2011}. Indeed, compared to core-collapse SNe, type Ia are characterized by a larger expansion velocity of the ejecta (up to $10^4$ km/s, one order of magnitude larger than for Type IIp SNe) that translates into lower gas density and less efficient grain condensation and growth. As a result, even if they form, grains have radii that never exceed 0.01 $\mu$m \citep{nozawa2011}. A larger abundance of $^{56}$Ni (0.6 $M_\odot$) compared to core-collapse SNe (about 0.06 $M_\odot$) implies a larger flux of energetic electrons and gamma photons produced during the radioactive decay, that can prevent the formation of grain molecular precursors. 

Despite these unfavourable conditions, theoretical models predict that, depending on the efficiency of molecule formation and on the adopted sticking probability, a mass of dust ranging between $3 \, 10^{-4} M_\odot$ to 0.2 $M_\odot$ is able to form. While present observational limits on type Ia SNe allow for the presence of $\sim 0.03 - 0.075 \, M_\odot$ of silicate grains in the ejecta, carbon grain formation must be significantly smaller than predicted by some of the models, pointing to a larger suppression of C grain condensation by energetic electrons and photons and/or that the outermost C-O layers are almost fully burnt \citep{nozawa2011}.

Finally, the effects of the reverse shock are expected to be more destructive for SNIa than for type IIP SNe (see Sect.~\ref{sec:snrevshock}). Due to the lack of a hydrogen envelope, the reverse shock can sweep up the dust formation region much earlier ($ < 500$ yr) than in envelope-retaining SNe ($ > 1000$ yr). At such early times, the gas density in the shocked ejecta is high enough that dust grains are efficiently decelerated and eroded due to frequent collisions with the gaseous ions. In addition, the radii of newly formed grains are small (0.01 $\mu$m), they are quickly destroyed by thermal sputtering without being injected into the ISM. As a result, if the circumstellar medium density is larger than 0.1 cm$^{-3}$, as suggested by the observed sizes of type Ia SN remnants \citep{borkowski2006,badenes2007} the dust mass decreases to $10^{-5} M_\odot$ in less than 1 Myr.

It is important to mention that the lack of dust formation in type Ia SN ejecta has important implications for the origin of iron grains, as it implies that more than 65\% of iron is injected in the ISM in gaseous form \citep{dwek2016}. Yet, ultraviolet and X-ray observations along many lines of sight in the ISM show that iron is strongly depleted in the gas phase and that about 90\% of the total iron mass must be locked up in interstellar dust. These two evidences suggest that most of the missing iron must have formed outside the traditional stellar condensation sources, likely through accretion of ISM gas onto pre-existing silicate, carbon, or composite grains \citep{dwek2016}. \\

\section{Relative importance of SNe and AGBs as dust polluters}
\label{sec:relativerole}

If we restrict the analysis to the two main classes of stellar dust sources, SNe and AGBs, estimating their relative importance in producing interstellar dust depends on a number of factors: the adopted mass- and metallicity-dependent dust yields, $m_{\rm dust} (m, Z)$, the stellar initial mass function (IMF), $\phi(m)$, and the star formation history, ${\rm SFR}(t)$. 
Following \citet{valiante2009}, we can express the total mass of dust released in the ISM by stars as:
\begin{equation}
M_{\rm d}(t) = \int_{0}^{t} dt' \int_{m_{\tau_m}}^{m_{\rm up}} \, m_{\rm dust} (m, Z) \, \phi(m)\, {\rm SFR} (t' - \tau_m)\, dm, 
\label{eq:dustyield}
\end{equation}
\noindent
where $m_{\tau_m}$ is the mass of a star with lifetime $\tau_m$ which formed at time $t' - \tau_m$, and the metallicity is computed at the stellar formation time, $Z = Z(t' - \tau_m)$. Adopting the SN dust yields from \citet{bianchi2007}, with $\sim 7 \%$ of the newly formed dust surviving the passage of the reverse shock, and the AGB dust yields from \citet{ferrarotti2006}, \citet{valiante2009} compute $M_{\rm d}(t)$ assuming a Larson IMF,
\begin{equation}
\phi(m) = \frac{dN}{dm} = m^{-(\alpha +1)} {\rm e}^{-m_{\rm ch}/m},
\label{eq:larsonIMF}
\end{equation}
\noindent
with $\alpha = 1.35$ and varying the characteristic mass $m_{\rm ch}$ from $0.35 M_\odot$ (equivalent to a Salpeter-like or Kroupa-like IMF), to $10 M_\odot$ (equilavent to a top-heavy IMF). They also explored the effect of changing the star formation history from a single burst to a constant star formation rate, and adopted different (constant) stellar metallicities ($Z = 0$ and $Z_\odot$). Using the mass- and metallicity-dependent stellar lifetimes from \citet{raiteri1996}, they find that -- at early times -- dust production is dominated by SNe as a consequence of the shorter lifetimes of their progenitor stars. AGB
stars start to produce dust after about 30 Myr (i.e. when a $8 M_\odot$ star evolves off the main sequence to the AGB phase). When $m_{\rm ch} = 0.35 M_\odot$, the characteristic timescale at which AGBs dominate dust production ranges between 150 and 500 Myr, depending both on the assumed star formation history and on the initial stellar metallicity. This conclusion is significantly affected by variations of the IMF: for a $m_{\rm ch} = 5 M_\odot$, dust from AGB starts to dominate only on
time-scales larger than 1 Gyr, and SNe are found to dominate dust
evolution when $m_{\rm ch} \geq 10 M_\odot$. Thus, variations of the stellar IMF over cosmic history due to the smaller gas metallicity (see e.g. \citealt{omukai2005, schneider2012a, chon2021}) and/or to the larger cosmic microwave background temperature (see e.g. \citealt{schneider2010, chon2022}) might significantly affect the origin and properties of dust in the high-redshift Universe. Here we revisit their analysis by exploring different combinations of SN dust survival fractions and AGB dust yields.

\subsection{Dependence on the adopted dust yields and stellar metallicity} 

\begin{figure*}
\centerline{\includegraphics[width=6cm]{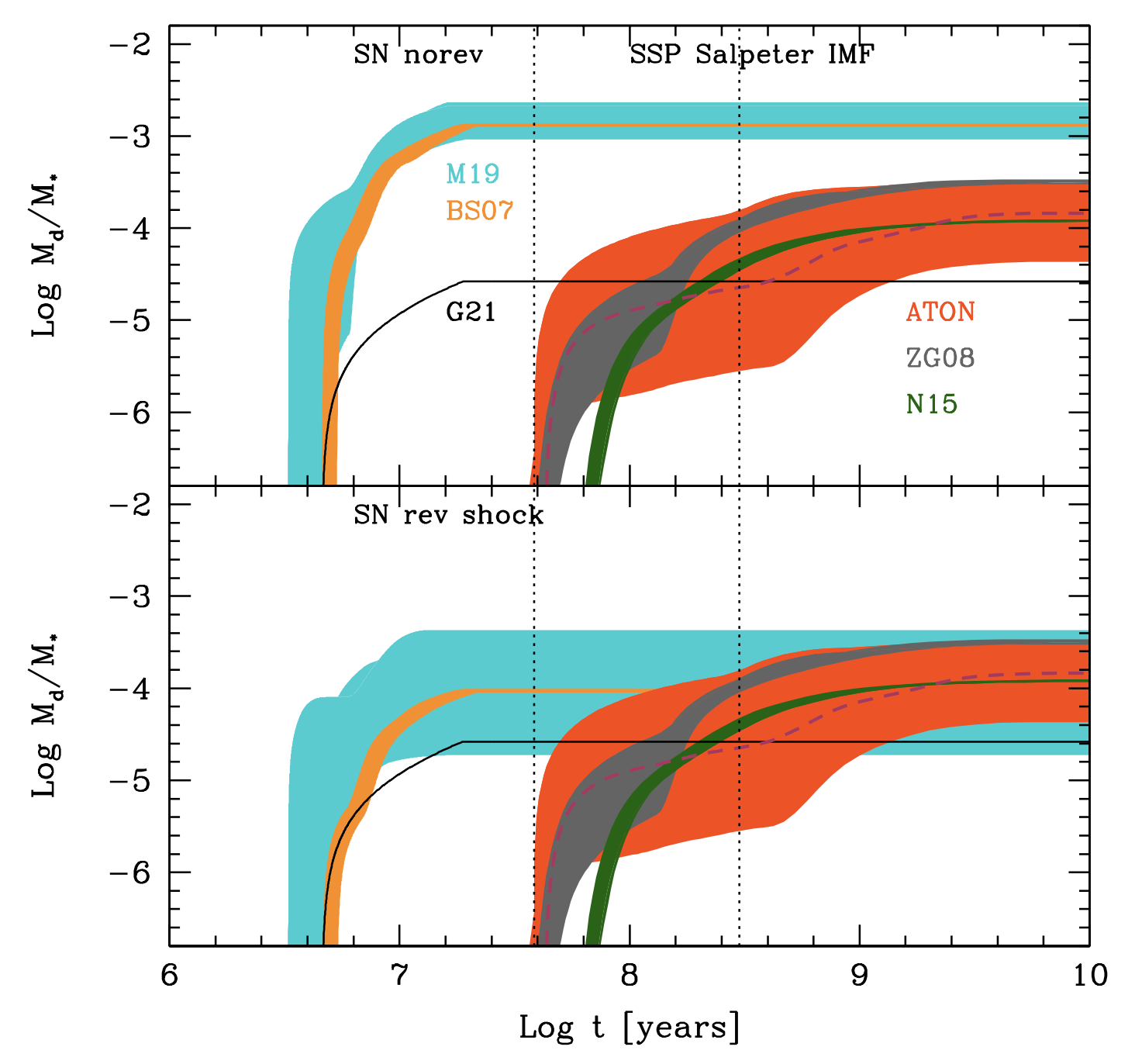}
\includegraphics[width=6.1cm]{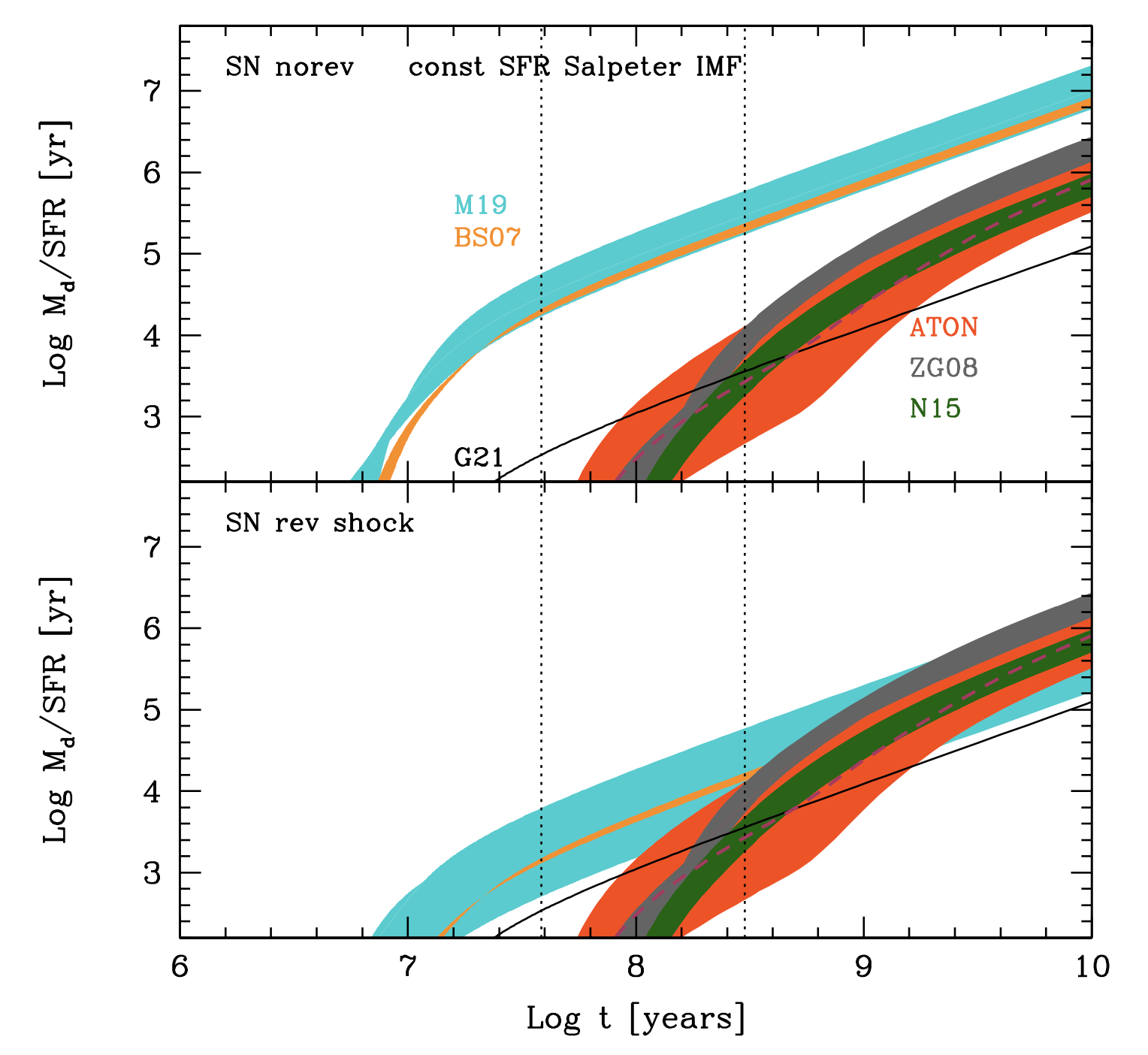}}
\caption{Relative importance of SNe and AGBs as stellar sources of dust, assuming that stars form according to a Salpeter IMF in the mass range $[0.1 - 100] \, M_\odot$ when all stars form in a single burst (SSP at $t = 0$, {\bf left panels}), and when stars form continuously with a constant SFR ({\bf right panels}, see text). Different colours indicate different sets of SN and AGB dust yields: SN dust yields are from \citet{bianchi2007} (orange, BS07), non-rotating models from \citet{marassi2019} (cyan, M19), and the empirical SN dust yield inferred by \citet{galliano2021} from a detailed analysis of a sample of local galaxies (black, G21, see Sect.~\ref{sec:dustpedia}); AGB dust yields are from \citet{zhukovska2008} (gray, ZG08), the COLIBRI model (green, N15), and the ATON model (red, ATON). For theoretical dust yields, the shaded regions encompass variations in the initial stellar metallicity in the range $[10^{-4} - 1]\, Z_\odot$ (as a reference, the dashed dark red line shows the ATON prediction for a stellar metallicity of $Z = 0.1 \, Z_\odot$). {\bf Top and bottom panels} show the results of SN dust yield with no/with reverse shock destruction, adopting a circumstellar medium density of $n_{\rm ISM} = 0.6 \, {\rm cm}^{-3}$ (see Table \ref{table:RSmassfrac}). 
The two vertical dotted lines mark the lifetimes of a $8 \, M_\odot$ (35 Myr) and a $3 \, M_\odot$ (300 Myr) star, which correspond, respectively, to the maximum mass of AGBs, and the transition mass from carbon ($\leq 3 \, M_\odot$) to silicate ($> 3 \, M_\odot$) dust production (see Sect.~\ref{sec:agbmodels}). Independent of the adopted star formation history, supernovae dominate dust production in the first 35--50 Myr of the evolution, and the fractional contribution of AGBs depends on the strength of SN reverse shock destruction, on the adopted set of yields, on the initial stellar metallicity, and SF history (see text).}.
\label{fig:dustevo_salpeter}
\end{figure*}

We first explore in Fig.~\ref{fig:dustevo_salpeter} the relative importance of SNe and AGBs when the stars are formed according to a Salpeter IMF in the mass range $[0.1 - 100] \, M_\odot$. We consider two extreme star formation histories: in the left panels, the stars are formed in a single burst at $t=0$, i.e. a single stellar population (SSP) model, and the time-dependent dust mass released in the ISM, $M_{\rm d} (t)$, is normalised to the total stellar mass formed in the burst, $M_{\star}$. In the right panels, the stars are continuously formed with a constant SFR, and $M_{\rm d}(t)$ is normalised to the constant value of SFR. As a result, in order to compare the two models, we need to assume a specific value of $M_{\star}$ and time, $t_{\star}$, and multiply the SSP model by $M_{\star}$, and the constant SF model by $M_{\star}/t_{\star}$.

The different colours represent different sets of SN and AGB dust yields, with the shaded regions reflecting variations in the adopted initial metallicity in the range $[10^{-4} - 1] \, Z_\odot$. 
In general, the dust mass released by a star is expected to depend on the initial stellar metallicity. However, the extent of this dependence varies among models. For SN dust yields, the \citet{bianchi2007} models predict a moderate dependence, with solar metallicity stars producing, on average, $\sim 30\%$ more dust than stars with initial metallicity of $10^{-4} Z_\odot$ (see the orange shaded region). The metallicity dependence is found to be more significant in the SN dust yields computed by \citet{marassi2019}, partly also as a consequence of the different grid of SN models adopted. Due to the reduced amount of mass lost in stellar winds, $\gtrsim 30 M_\odot$ stars with initial metallicity $\lesssim 10^{-2} Z_\odot$ experience a strong fallback or fail to explode; as a result, their dust (and metal) production is negligble, compared to more metal-rich stars of comparable mass (see \citealt{marassi2019} for more details). Similar considerations apply to AGB dust yields: the metallicity dependence of dust yields depends on the adopted model (see also Sect.~\ref{sec:agbmodels}). For ATON yields, $\gtrsim 3 M_\odot$ stars -- which experience HBB and do not produce carbon dust -- produce a dust mass that decreases by almost 2 dex when their initial metallicity decreases from solar to $\simeq 10^{-2} Z_\odot$ (see Fig.~\ref{fig:agbdust_aton}). The metallicity dependence of stellar dust yields is also reflected in the dust composition, as it will be discussed in Sect.~\ref{subsec:composition}.

The top and bottom panels show, respectively, the results of SN dust production with no/with the effects of the reverse shock destruction, assuming a circumstellar medium density of $n_{\rm ISM} = 0.6 \, {\rm cm}^{-3}$ (see Table \ref{table:RSmassfrac}). In the first 3--5 Myr of evolution, no dust is released in the ISM, until the most massive SN progenitors explode; SNe dominate dust production for the first 35--50 Myr of the evolution, until the most massive AGBs ($\lesssim 8 \, M_\odot$) start to contribute (see the first vertical dotted line at $t = 35$ Myr). On longer timescales, the relative importance of AGBs and SNe depend on a number of factors: the degree of dust destruction by the reverse shock, the adopted set of yields, and the initial metallicity of the stars. 
Assuming no reverse shock destruction, the fractional contribution of AGBs is always less than 10\%, while it grows to 30--40\% ($\simeq 14$ \%) at $t = 300$ Myr (corresponding to the lifetime of a $3 \, M_\odot$ star, see the second vertical dotted line in Fig.~\ref{fig:dustevo_salpeter}), and to 40--70\% ($\simeq 40$\%) at $t = 1.5$ Gyr for a SSP model (continuous SFR model) when SN dust destruction by the reverse shock is taken into account. The range of fractional contribution reflect the variation of the dust yields with initial stellar metallicity, and we have taken the maximum spread among the different dust yields shown in Fig.~\ref{fig:dustevo_salpeter} (ATON models and M19 SN dust yields). The solid black lines in the two panels indicate the empirical SN dust yield inferred by \citet{galliano2021} from a detailed analysis of a sample of local galaxies (see Sect.~\ref{sec:dustpedia} for more details). This empirical value provides an estimate of the effective SN dust yield, and appears to be consistent with the theoretical predictions, when the effect of the reverse shock is accounted for.

\begin{figure*}
\includegraphics[width=6cm]{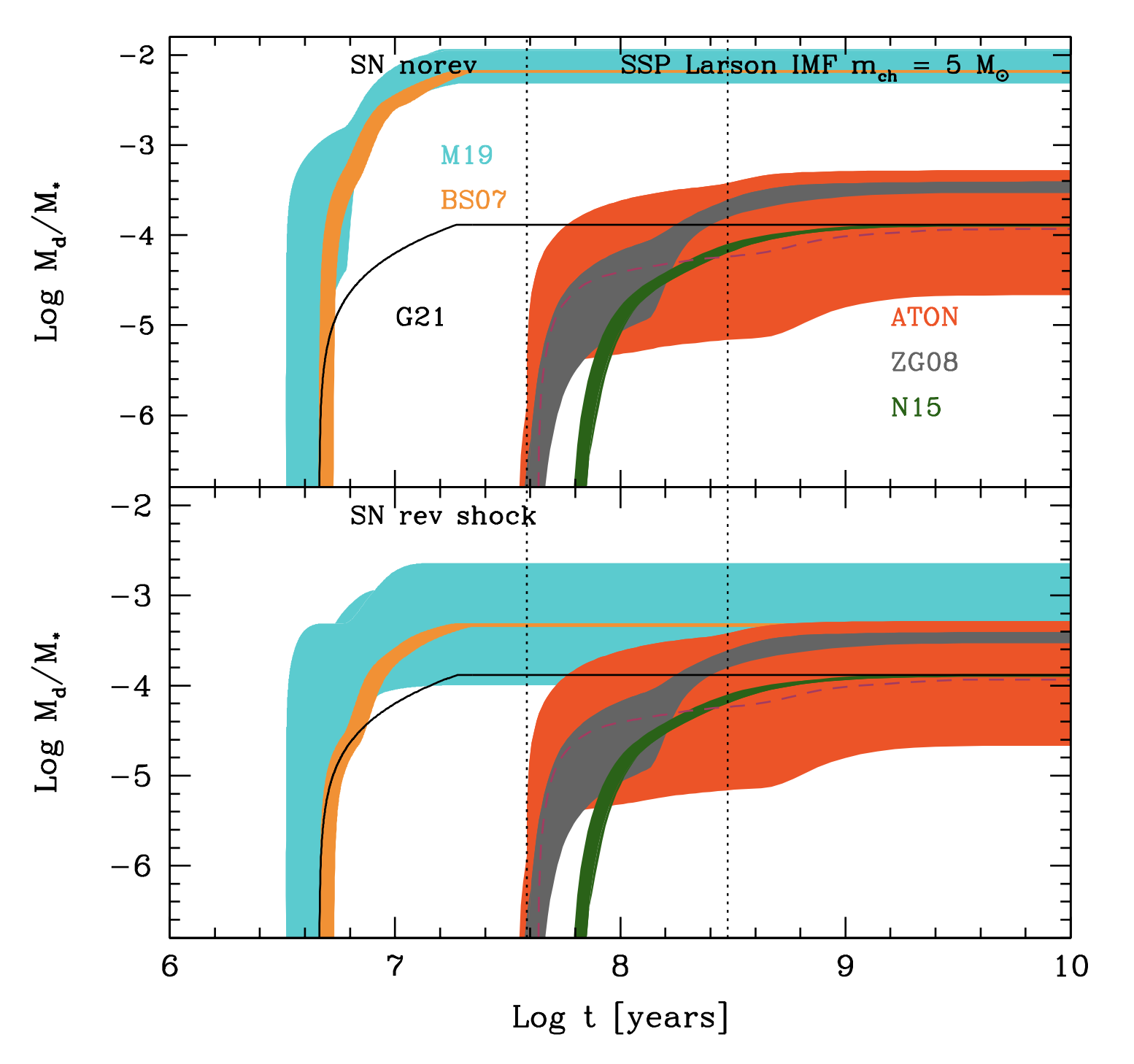}
\includegraphics[width=6cm]{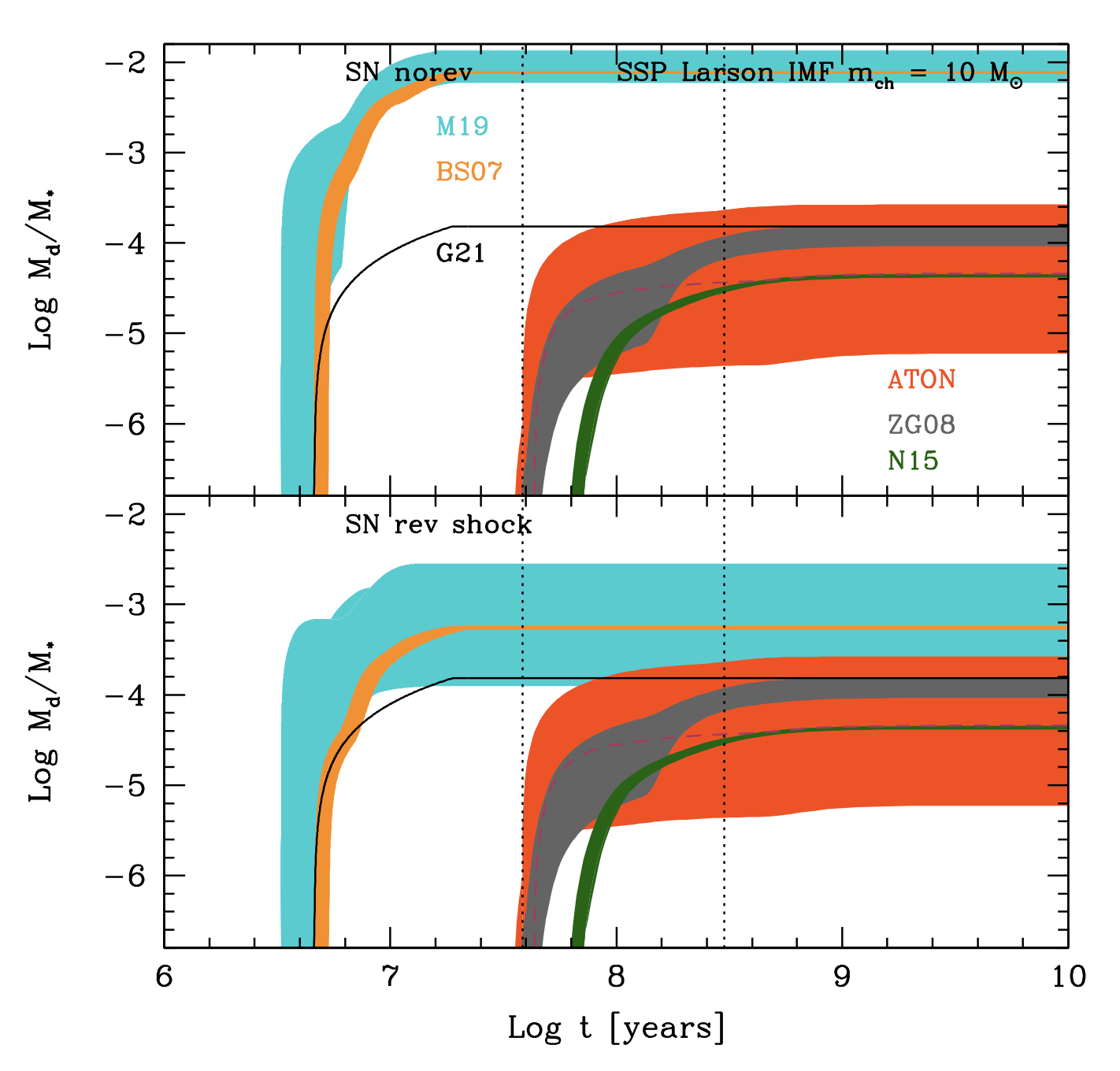}
\caption{Relative importance of SN and AGBs for dust production, assuming that all stars are formed in a single burst at $t=0$ according to a Larson IMF in the mass range $[0.1 - 100]\, M_\odot$ with $m_{\rm ch} = 5 \, M_\odot$ ({\bf left panels}) and $m_{\rm ch} = 10 \, M_\odot$ ({\bf right panels}). {\bf Top and bottom panels} show SN dust production with no/with the effects of the reverse shock, respectively. Coloured shaded regions and lines have the same meaning as in Fig.~\ref{fig:dustevo_salpeter}. Compared to the Salpeter IMF, adopting a top-heavy IMF leads to an increase of the SN dust yield by a factor 4--6, and the fractional contribution of AGB dust yields is always $\leq 16 \%$, adopting the maximum spread among the different dust yields (ATON models and M19 SN dust yields).}
\label{fig:dustevo_larson}
\end{figure*}

\subsection{Dependence on the stellar IMF} 

In Fig.~\ref{fig:dustevo_larson}, we further explore how the relative importance of SNe and AGBs depend on the adopted stellar IMF. Following \citet{valiante2009}, we consider a Larson IMF (see Eq.~\eqref{eq:larsonIMF}) in the mass range $[0.1 - 100] \, M_\odot$ and we vary the characteristic stellar mass, assuming $m_{\rm ch} = 5 \, M_\odot$ (left panels), and $m_{\rm ch} = 10 \, M_\odot$ (right panels), adopting a SSP model; hence, the results can be compared with the left panels of Fig.~\ref{fig:dustevo_salpeter} which are indicative of a Larson IMF with $m_{\rm ch} = 0.35 \, M_\odot$.
Compared to this reference case, increasing the characteristic mass of the IMF leads to an increase of the SN dust yield by a factor $\simeq$4--6; compared to the reference Salpeter IMF model, the AGB dust yields increase by a factor $\simeq 1.6$ when $m_{\rm ch} = 5 \, M_\odot$,  and decrease by a factor $\simeq 2$ when $m_{\rm ch} = 10 \, M_\odot$. The fractional contribution of AGB dust yields is always $\leq 16 \%$, adopting the maximum spread among the different dust yields (ATON models and M19 SN dust yields).

\begin{figure*}
\includegraphics[width=12cm]{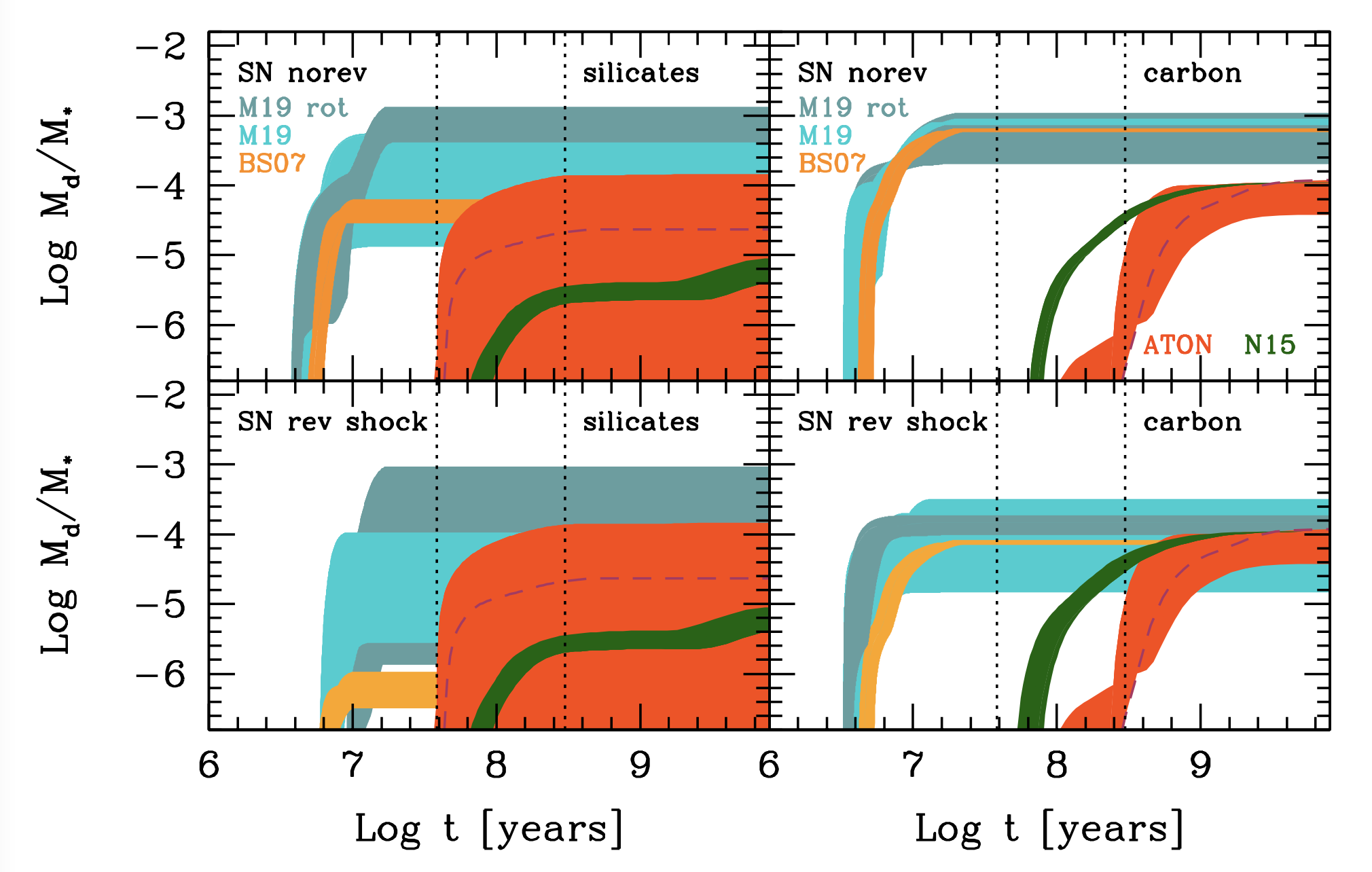}
\caption{Silicate ({\bf left panels}) and carbon ({\bf right panels}) dust production for a SSP formed at $t=0$ with initial metallicity in the range $[10^{-4} - 1] \, Z_\odot$. All stars are assumed to form according to a Salpeter IMF in the mass range $[0.1 - 100]\, M_\odot$, with different coloured shaded regions representing specific sets of SN and AGB dust yields, following the same colour coding as in Figs. \ref{fig:dustevo_salpeter} and \ref{fig:dustevo_larson}, with -- in addition -- the set of dust yields for rotating SN models by \citet{marassi2019} (dark turquoise, M19 rot). As expected, the contribution of AGBs to silicate dust production (that comprises Mg$_2$SiO$_4$, Fe$_2$SiO$_4$, MgSiO$_3$, FeSiO$_3$, and SiO$_2$) strongly depends on the initial metallicity of the stars and on the adopted dust yields, but starts already on timescales $t > 35 - 50$\,Myr (see the dashed dark red line, which shows the contribution of AGBs with initial metallicity $Z = 0.1 \, Z_\odot$ predicted by the ATON model); conversely, the contribution of AGBs to carbon dust production starts only on timescales $\gtrsim 300$\,Myr, as it requires $\lesssim 3 \, M_\odot$ stars to reach the AGB stage.}
\label{fig:dustcomp_salpeter}
\end{figure*}

\subsection{Effects on dust composition} 
\label{subsec:composition}

The relative importance of SNe and AGBs for silicate (left panels) and carbon (right panels) dust production is shown in Fig.~\ref{fig:dustcomp_salpeter}, for a SSP model where all stars are formed according to a Salpeter IMF in the mass range $[0.1 - 100]\, M_\odot$. Similarly to Figs. \ref{fig:dustevo_salpeter} and \ref{fig:dustevo_larson}, top and bottom panels show SN dust production with no/with the effect of the reverse shock, 
with coloured shaded regions representing different sets of SN and AGB dust yields, adopting an initial metallicity of progenitor stars in the range $[10^{-4} - 1] \, Z_\odot$. The figure shows that silicate dust enrichment is very sensitive to the initial metallicity of the stars, as illustrated by the large extent of the shaded regions, both for SNe (particularly, when the effect of the reverse shock is considered), and for AGBs (particularly when considering ATON dust yields). Also, the degree of silicate dust destruction by the reverse shock strongly depends on the adopted SN dust yields: the surviving silicate dust mass fraction is only $\simeq 2\,\%$ for the SN dust yields by \citet{bianchi2007}, $\simeq 16 \%$ for the non rotating SN models by \citet{marassi2019}, and $\simeq 80 \%$ for the least massive rotating SN models with $Z \simeq Z_\odot$ by \citet{marassi2019}. Conversely, carbon dust production is less sensitive to the initial metallicity of the star, but depends on the adopted dust yields: as anticipated in Sect.~\ref{sec:agbmodels}, ATON models predict a sharp transition from silicate dust production (for initial stellar masses $> 3 \, M_\odot$) to carbon dust production (for initial stellar masses $\leq 3 \, M_\odot$), which can start only on timescales longer than $\simeq 300$\, Myr. Hence, we expect SNe to dominate carbon dust production in young galaxies, where most of the stars have ages $t < 300 $ Myr. The recent detection with JWST of the 2175 \AA \, absorption feature, which is attributed to carbonaceous dust grains (specifically polycyclic aromatic hydrocarbons, PAHs, or nano-sized graphitic grains), in the spectrum of a 
galaxy at $z = 6.71$ by \citet{witstok2023}, is a strong indication of rapid carbon dust production, likely by Wolf--Rayet (WR) stars or by SNe.

\subsection{Relevance of stellar sources of dust at $z > 4$}
\label{sec:summary_fig1}

A summary of the typical dust enrichment timescales for some of the dust production mechanisms discussed in the previous sections is shown in Fig.~\ref{fig:dustevo_summary}. 
To provide an indication of their relative importance -- given the time constraints at $z > 4$ -- we also represent with vertical dotted lines some reference values of the Hubble
time at $4 \leq z \leq 18$ (adopting a standard $\Lambda$CDM cosmology with $\Omega_{\rm M} = 0.3$,  $\Omega_{\rm \Lambda} = 0.7$, and $h = 0.67$). These are compared with the
dust enrichment timescales of stars that are formed in a single burst at $z_{\rm form} = 20$ ($t_{\rm age} = 0$) according to a Salpeter IMF with masses in the range 
$[0.1 - 100] ~ M_\odot$. On the vertical axis we show the time-dependent cosmic dust yield, i.e. the dust mass released in the ISM (computed from Eq.~\eqref{eq:dustyield}) 
normalized to the total stellar mass formed (similar to the left panels in Figs. \ref{fig:dustevo_salpeter} and \ref{fig:dustcomp_salpeter}). The dark (light) blue shaded regions 
illustrate the contribution of core-collapse SNe, adopting the yields for non rotating stellar progenitors with initial mass in the range 
$[13 - 120] ~ M_\odot$ from \citet[][see the cyan squares in Fig.~\ref{fig:sn_comp1}]{marassi2019} spanning an initial progenitor metallicity in the range $\rm 0.1<Z/Z_\odot <1$ ($\rm 0.01<Z/Z_\odot <0.1$). For the same yields and the same metallicity ranges, the cyan shaded regions illustrate the effects of the partial dust destruction due to the reverse shock 
(indicated as SN rev. sh. in the legend), adopting a constant circumstellar medium density of $n_{\rm ISM} = 0.6 ~ {\rm cm}^{-3}$ (see Sect.~\ref{sec:snrevshock} and Table 
\ref{table:RSmassfrac}). The figure shows that SNe provide a prompt enrichment channel, which is capable of enriching the ISM of galaxies already at $z \simeq 18$. The degree of dust 
enrichment can be as big as $M_{\rm d}/M_{\star} \simeq 10^{-2.5}$ or as small as $M_{\rm d}/M_{\star} \simeq 10^{-4.8}$, depending mostly on the effects of the reverse 
shock and on the initial metallicity of the stars. SNe provide the dominant contribution to early dust enrichment in the first few tens of Myr, compared to WRs and RSGs (shown as the green dashed and red solid lines in Fig.~\ref{fig:dustevo_summary}). These two contributions have been computed adopting the corresponding yields presented in Sects.~\ref{sec:rsg}--\ref{sec:wr}. In particular,
we assume that stars with initial masses $\geq 40 M_\odot$ become WRs and contribute to dust enrichment as described by Eq.~\eqref{eq:WRyields}, while stars with initial mass in the range 10--25 $M_\odot$ become RSGs and contribute to dust enrichment as described by Eq.~\eqref{eq:RSGyields}. Finally, the dark (light) yellow shaded regions illustrate the contribution of AGBs with initial progenitor metallicity in the range $\rm 0.1<Z/Z_\odot <1$ ($\rm 0.01<Z/Z_\odot <0.1$). Here we have adopted the yields from the ATON model for stars with initial mass in the range 
$[1 - 8]~M_\odot$ from \citet{ventura2012a, ventura2018, dellagli2019}. It is clear that the most massive AGBs start to contribute already after 35 Myr since the onset of star formation, but their dust yield
is very sensitive to the initial metallicity of the stars: at $z \simeq 10$ (300 Myr since the onset of star formation at $z_{\rm form} = 20$) the AGBs contribution can be as large as $10^{-4.6} \lesssim M_{\rm d}/M_{\star} \lesssim 10^{-3.8}$ for stars with initial metallicity in the range $\rm 0.1<Z/Z_\odot <1$, but can drop down to $M_{\rm d}/M_{\star} \simeq 10^{-5.6}$ for stars with initial metallicity $\simeq 0.01 Z_\odot$. 

Hence, at early cosmic times the contribution of AGBs can not be neglected, particularly if the stars have metallicities $Z > 0.01 Z_\odot$. 
However SNe appear to dominate early dust production, in most of the cases, unless the effect of the reverse shock is very significant. This is even more
so if at early cosmic times the stellar IMF deviates from the Salpeter law, becoming progressively more top-heavy (biased towards larger stellar masses),
as suggested by semi-analytical models \citep{omukai2005, schneider2010}, high-resolution hydrodynamical simulations \citep{chon2021, chon2022}, or empirical
models \citep{jermyn2018, steinhardt2022}.

\section{Alternative models: smoking quasar} 
\label{sec:qsodust}
So far we have discussed stellar sources of dust, which are generally considered the primary factories of dust. However, \citet{elvis2002} first pointed out that gas in the circumnuclear region of accreting supermassive black holes, aka Active Galactic Nuclei (AGN), can also have the physical conditions adequate for the formation of dust. Indeed, at least those AGN accreting at substantial rate relative to the Eddington limit ($\dot{M}_{\rm BH} > 0.01~\dot{M}_{\rm Edd}$), are surrounded by very dense ($n \sim 10^{11}~{\rm cm}^{-3}$,  i.e. approaching the densities of stellar atmospheres), ionized clouds, which emit a large number of recombination and collisionally excited nebular lines. These clouds are located within the central parsec and are characterised by large velocities (a few percent of the speed of light), causing their nebular emission lines to be highly Doppler-broadened, which has resulted in this region being dubbed ``Broad Line Region'' (BLR). The temperature of the gas in this region ($\sim 2~10^4$\, K) is too high for the formation of dust. However, there are various indications that at least a fraction of the BLR clouds are in outflow \citep[e.g.][]{Elvis2000,Elitzur2014,Kollatschny2013,Matthews2020}. \citet{elvis2002} pointed out that -- during their outward trajectory, at a few pc from the BH -- the clouds cool down, because of both adiabatic expansion and reduction of the radiation flux from the accretion disk, while maintaining relatively high densities. Therefore, such outflowing clouds are expected to experience a phase in which temperature and density are adequate for the nucleation of dust grains, and far enough from the AGN not to be destroyed by its intense radiation field. The outflow velocities are large enough to escape the central region, so they can enrich with dust the ISM of the host galaxy and, potentially, even its intergalactic medium (IGM). \citet{elvis2002} estimate that the most powerful AGN (quasars), with luminosities of about $10^{46}~{\rm erg~s}^{-1}$, could produce dust at a rate of $\sim 0.01~M_\odot~{\rm yr}^{-1}$.

\begin{figure*}
\centerline{\includegraphics[width=\textwidth]{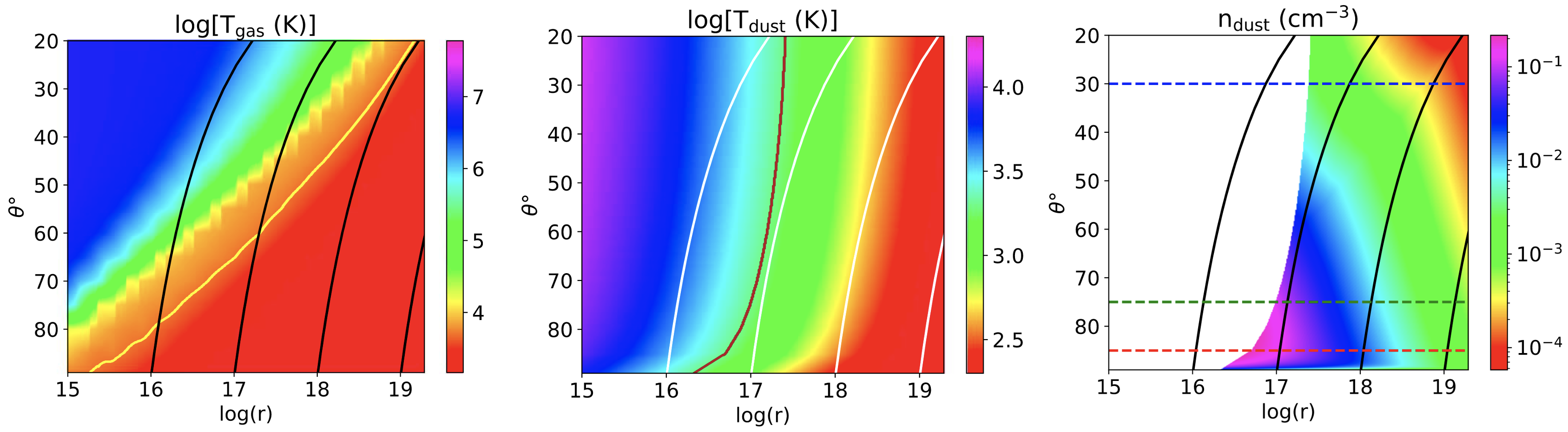}}
\caption{Some of the gas and dust in the quasar-driven wind model presented by \citet{sarangi2019}, on radial-azimutal diagrams relative to the accretion disk. \textbf{Left panel: } gas temperature, where the yellow line indicates the temperature below which dust nucleation can happen ($\sim$ 4000~K). \textbf{Central panel:} dust temperature distribution, where the brown line indicates the temperature ($\sim$ 2000~K) below which dust can survive against sublimation. \textbf{Right panel:} density of dust grains, limited to the region where dust can survive. The black and white lines are the magnetic field lines. Images reproduced with permission from \citet{sarangi2019}, copyright by AAS.}
\label{fig:qso_dust}
\end{figure*}

As we will discuss in the second part of this review, such large production rate can potentially be a rapid channel of dust formation rate in the early universe, and could explain some of the dust in distant galaxies, especially quasar host galaxies, which have been observed to harbour large masses of dust \citep[e.g.][]{Bertoldi2023,Venemans2018,fan2023,Wang2008,Banados2018,Wang2021}. In particular, \citet{maiolino2006} already pointed out that this channel of dust production could potentially explain most of the dust observed in some distant quasars. However, \citet{pipino2011} integrated the quasar dust formation scenario by \citet{elvis2002} into a more comprehensive model of dust evolution in massive galaxies, finding that dust produced in quasar winds contributes little to the global dust budget, and can play a significant role only in the central region of galaxies.

More recently, \citet{sarangi2019} explored this scenario via a more detailed magneto-hydrodynamical modelling of quasar-driven winds. They confirmed that in a significant fraction of such winds the outflowing gas clouds reach temperatures and densities adequate for the nucleation of dust (see Fig.\ref{fig:qso_dust}). Most of the dust formed in these winds is expected not to sublimate and to experience significant growth. As a consequence, they predict a significant fraction of the quasar-driven wind (especially towards the equatorial directions) to be heavily dust-loaded, possibly having most of the metals locked into dust. They suggest that the ``dusty torus'' envisaged by the unified model of AGNs, and observed by interferometric IR observations \citep{Raban2009,Burtscher2013,Tristram2014}, could actually be a dynamic structure resulting from dust formation and growth in such quasar-driven winds, in agreement with previous suggestions on the nature of this nuclear structure \citep{Elitzur2006,Nenkova2008}. From their model, they derive an analytical expression for the dust production rate in  quasar-driven winds, given by:
\begin{equation}
    \dot{M}_{\rm d} \simeq 3.5 \, \dot{m}^{2.25} \, M_8~M_\odot {\rm yr}^{-1}
\label{eq:quasar_mdust_dot}
\end{equation}
where $\dot{m} = \dot{M}_{\rm BH}/\dot{M}_{\rm Edd}$ is the accretion rate in units of the Eddington limit, and $M_8$ is the black hole mass in units of $10^8~M_\odot$. They however clarify that, because of the various assumptions, the expression above is likely an upper limit on the dust production rate. It is also important to note that their model predict that most of the dust formed in these winds is in the form of silicate dust (in agreement with observational results indicating that a large fraction of the nuclear dust is made of silicates), as a consequence of the nuclear region being oxygen rich.

\begin{figure*}
\centerline{\includegraphics[width=0.8\textwidth]{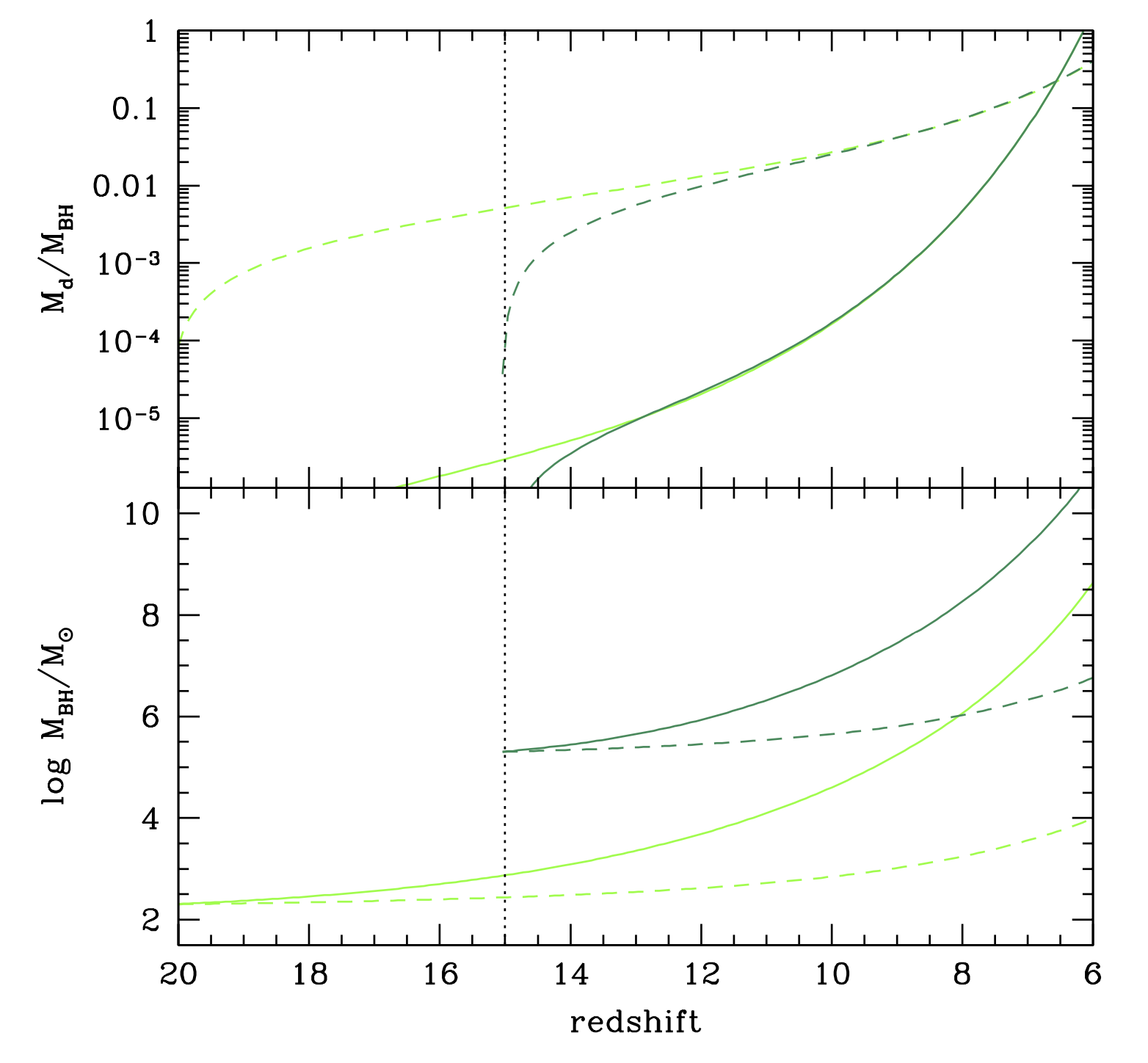}}
\caption{Expected time evolution of the dust mass formed in quasar winds, assuming the dust production rate estimated by \citet{sarangi2019} and reported in Eq.~\eqref{eq:quasar_mdust_dot}.  \textbf{Bottom panel}: adopted BH evolutionary tracks. The light (dark) green lines represent the evolution of BHs formed from light (heavy) BH seeds at $z = 20$ ($15$), and accreting continuously at their Eddington rates (solid lines) or at 0.3 of their Eddington rates (dashed lines). \textbf{Top panel}: for each evolutionary track, the redshift dependent dust mass formed in the winds is computed integrating Eq.~\eqref{eq:quasar_mdust_dot} and is normalized to the final BH mass at $z = 6$. Note that this should be interpreted as an upper limit to the dust mass produced in the wind, as it implicitly relies on the assumption of a pre-enriched outflowing gas. Hence, the actual timescales for dust production have to be convolved with the timescales for production of the refractory elements by stars in the quasar host galaxy.
The vertical dotted line in both panels mark the adopted formation redshift of heavy BH seeds.}  
\label{fig:mdust_quasar}
\end{figure*}

We shall note that, in contrast to stellar sources of dust, this alternative scenario requires the nuclear ISM to be pre-enriched. So the actual timescales for dust production have to be convolved with the timescales for production of the refractory elements. With this caveat in mind, in Fig.~\ref{fig:mdust_quasar} we attempt to quantify the expected dust mass formed by smoking quasars, using Eq.\ref{eq:quasar_mdust_dot} and different BH evolutionary scenarios. The bottom panel shows the redshift evolution of the BH mass, assuming that that BH forms from a light seed with $100 M_\odot$ at $z = 20$ (light green) or from a heavy BH seed with $10^5 M_\odot$ at $z = 15$ (dark green), and then grows continuously with $\dot{m} = 1$ (solid lines) or $0.3$ (dashed lines)\footnote{The adopted BH evolutionary tracks are meant to be indicative of possible evolutionary scenarios which lead to the formation of super-massive black holes powering the observed Active Galactic Nuclei (AGNs) and quasars at $z \simeq 6$. See \citet{Inayoshi2020} and \citet{Volonteri2021} for recent reviews on early BH formation and growth.}. For each of these evolutionary tracks, the top panel shows the corresponding dust mass formed in the winds, normalized to the final BH mass at $z = 6$, showing that quasar winds can lead to dust masses up to $M_{\rm d}/M_{\rm BH} \lesssim 0.2 - 1$. 

While more realistic models for the cosmological evolution of nuclear BHs and their host galaxies are required to better assess the relative contribution of quasar winds and stellar sources to ISM dust enrichment \citep{valiante2012, Valiante2014}, we can provide an approximate estimate assuming a system with $M_{\rm BH}/M_{\star} \simeq 0.01 - 0.1$ at $z \simeq 6$, as suggested by observations of high-redshift quasars \citep{fan2023} and AGNs, including recent JWST observations (see e.g. \citealt{maiolino2023} and references therein). Adopting an indicative value of the dust-to-stellar mass contributed by stellar sources of $M_{\rm d}/M_{\star} \simeq 10^{-3}$ (in the range of what predicted for a Salpeter IMF, see e.g. Fig.~\ref{fig:dustevo_salpeter}), we find that the dust mass formed in quasar wind can be comparable to the dust mass released by stellar sources if $M_{\rm d}/M_{\rm BH} \simeq 0.01 - 0.1$, an efficiency that is well within the upper limits derived above and shown in Fig.~\ref{fig:mdust_quasar}.


\section{Dust reprocessing in the ISM}

Once injected in the ISM, dust grains produced by stellar sources are expected to experience significant reprocessing as they travel in the different phases of the ISM. 

Some grains will be removed from the ISM if they are in star forming regions and will be incorporated in newly formed stars, a process generally known as \emph{astration}. 
Grains that are in the warm and hot ISM phase can undergo thermal and kinetic sputtering when hit by SN shocks, or can be sublimated. This leads to a modification of the size distribution of ISM grains and to a reduction of the ISM dust mass. In regions exposed to intense radiation fields, photo-destruction of small grains can occur. 

Other processes can increase the ISM dust mass. In the cold neutral medium of the ISM, grain growth by accretion of gas-phase elements, a process also referred to as \emph{depletion} of condensible elements from the gas, can lead to an increase in dust mass as well as to a modification of the grain size distribution. 

Finally, there are other processes which do not lead to a change in the total ISM dust mass, but which alter the grain size distribution, such as grain shattering and fragmentation by grain-grain collisions in low-velocity shocks, grain coagulation, or which lead to structural modifications such as impact with high energy photons or cosmic rays. 

Some of these processes depend on the structural  properties of the grains (composition, size, and grain structure), and their relative importance is also tightly dependent on the nature of the environment where they reside, i.e. on the ISM properties of the host galaxy. Hence, a detailed discussion of these processes is beyond the scope of the present review, and we refer the interested readers to \citet{draine2009}, \citet{hirashita2013}, and \citet{galliano2018} for more information on their theoretical descriptions and observational evidences. 

Our purpose here is to provide a synthetic summary of how some of these processes have been incorporated into galaxy evolution models, and how observations of local galaxies have recently provided important clues on their relative importance. This may serve as a valuable support when interpreting the properties of high redshift galaxies.

\subsection{Destruction by interstellar shocks}
The expanding shock waves generated by SN explosions provide the dominant destruction process of pre-existing ISM dust grains, through thermal and kinetic sputtering, aided by shattering due to grain-grain collisions (see e.g. \citealt{jones1996}). Similarly to what has been discussed in Sect.~\ref{sec:snrevshock}, the results of different studies depend on the physical processes implemented, on the treatment of the shocks (steady state shock or time-dependent hydro-dynamical description in 1 or 3D), on the adopted initial grain size distribution and composition. A summary of some of the existing estimates of dust survival fractions can be found in Table 2 of \citet{micelotta2018}. Here we briefly summarize a few recent results obtained by \citet{bocchio2014, slavin2015, hu2019, martinez2019, kirchschlager2023}. 

\begin{figure}
\centerline{\includegraphics[width=12cm]{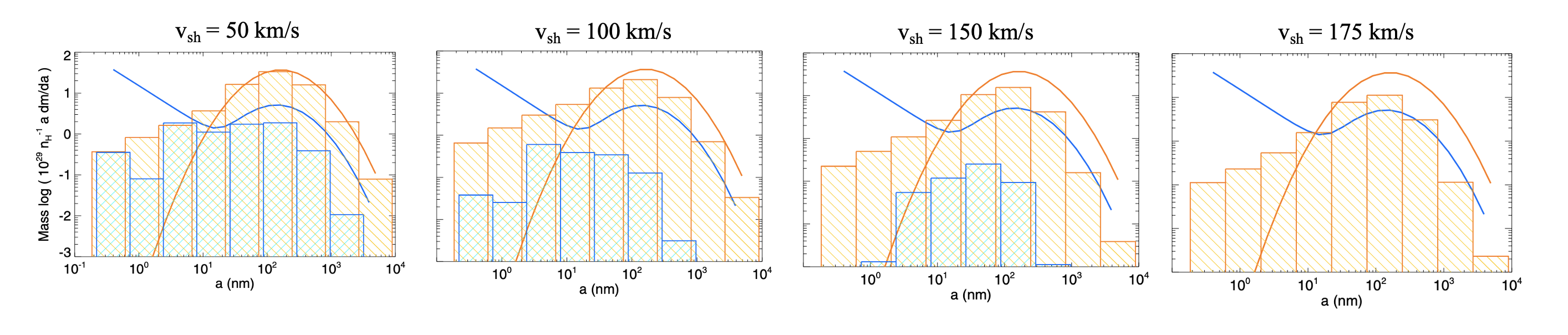}}
\caption{Comparison between the pre-shock (solid lines) and post-shock (histograms) size distributions of carbonaceous (blue) and silicate (orange) grains found by \citet{bocchio2014} assuming different shock velocities ($v_{\rm sh} = 50, 100, 150$, and $175$\, km/s, going from left to right), and an initial grain model from \citet{jones2013}. 
The 200 \, km/s post-shock size distribution is almost identical to the case of a 175 \, km/s shock.
Figure adapted from \citet{bocchio2014}.}
\label{fig:bocchio_forshock}
\end{figure}

\citet{bocchio2014} adopted the grain model by \citet{jones2013} to re-evaluate dust processing by steady state shocks with velocities $v_{\rm sh} = 50 - 200$\, km/s, propagating in a homogeneous ambient medium with density $n_0 = 0.25$\, cm$^{-3}$.
The results of their study is summarized in Fig.~\ref{fig:bocchio_forshock}, where the pre-shock (solid lines) and post-shock (histograms) grain size distributions of carbonaceous (blue) and silicate (orange) grains are compared for increasing shock velocities (from left to right). The smallest carbonaceous grains 
start to be destroyed already by a shock with velocity $50$\, km/s, and are completely destroyed when $v_{\rm sh} \ge 175$\, km/s. Silicate grains are more resistant, and their size distribution is affected by grain destruction and fragmentation when $v_{\rm sh} \ge 100$\, km/s. The following expressions provide a good fit to their estimated  carbonaceous and silicate grains destruction efficiencies as a function of the shock velocity \citep{bocchio2014}:
\[
\epsilon_{\rm carb} (v_{\rm s7}) =
\begin{system}
0.66 + 0.23 \, v_{\rm s7} \quad {\rm for} \quad 0.5 < v_{\rm s7} \le 1.5 \\
1 \quad \quad \quad \quad  \quad \quad \, \, {\rm for} \quad 1.5 < v_{\rm s7} \le 2
\end{system}
\]
\[
\epsilon_{\rm sil} (v_{\rm s7}) =
\begin{system}
0.61 \, v_{\rm s7} - 0.31 \quad {\rm for} \quad 0.5 < v_{\rm s7} \le 1.25 \\
0.11 + 0.28 \, v_{\rm s7}  \quad {\rm for} \quad 1.25 < v_{\rm s7} \le 2
\end{system}
\]
where $v_{\rm s7}$ is the shock velocity in units of 100\,km/s. When $v_{\rm s7} = 1$ this implies that 
$\epsilon_{\rm carb} = 0.89$ and $\epsilon_{\rm sil} = 0.30$. 

This result can be compared with the values of 
$\epsilon_{\rm carb} = 0.10$ and $\epsilon_{\rm sil} = 0.23$ that are found by \citet{slavin2015} by
running 1D hydro-dynamical simulations of SNR evolution in order to follow dust destruction beyond 
the point at which the remnant becomes radiative, where the plane parallel, steady state shock approximation
breaks down. This is because during the late stage remnant phase, relatively slow shocks still destroy dust. In addition, they sweep up increasing volumes of the ISM such that even with a low efficiency of grain destruction, slow shocks can be important for the overall grain destruction in the ISM \citep{slavin2015}.
However, the difference in the dust destruction efficiency, particularly of carbonaceous grains, compared to \citet{bocchio2014}, are due to the initial grain properties adopted by \citet{slavin2015}, who start from an MRN distribution
($dn/da \propto a^{-3.5}$ with 5 nm $\le a \le 250$ nm, \citealt{mathis1977}) for both silicates and carbonaceous grains, with a significantly smaller fraction of dust mass in very small carbon grains, compared to \citet{bocchio2014}.
For silicate grains following the same initial size distribution, \citet{kirchschlager2022} find that 
grain-grain collisions are very important and can significantly increase the dust destruction rates, compared to runs when only sputtering is considered. They evolve the SNR by running 3D hydro-dynamical simulations and find that $\epsilon_{\rm sil} = 0.3 \,  (0.2)$ when the blast wave expands in a homogeneous ambient medium with density $n_0 = 0.1$ (1) cm$^{-3}$.  

The dust destruction efficiencies can be used to evaluate the dust destruction timescale, $\tau_{\rm dest}$, also known as the lifetime of dust against destruction by 
SN shocks \citep{dwek1980}:
\begin{equation}
\frac{1}{\tau_{\rm dest}} = \frac{R_{\rm SN} \, m^{\rm dest}_{\rm gas}}{M_{\rm ISM}}
\label{eq:taudes}
\end{equation}
\noindent
where $R_{\rm SN}$ is the rate of SN explosions, $M_{\rm ISM}$ is the total mass of the ISM, and $m^{\rm dest}_{\rm gas}$ is the ISM mass that is completely cleared out of dust by a single SN explosion. This can be written as:
\begin{equation}
m^{\rm dest}_{\rm gas} = \int \epsilon (v_{\rm s}) \, dM_{\rm s}(v_{\rm s}),
\label{eq:mdust}
\end{equation}
\noindent
where $M_{\rm s} (v_{\rm s})$ is the mass of the ISM shocked to a velocity of at least $v_{\rm s}$.
For a Sedov-Taylor blastwave expanding in a uniform medium, this can be expressed as \citep{McKee1989}:
\begin{equation}
M_{\rm s}(v_{\rm s}) = \frac{E_{\rm SN}}{\sigma v_{\rm s}^2} = 6800 \, M_\odot \frac{E_{\rm 51}}{v_{\rm s7}^2},
\label{eq:STapprox}
\end{equation}
\noindent
where $E_{\rm 51}$ is the SN explosion energy in units of $10^{51}$\,erg, and $\sigma = 0.736$ \citep{ostriker1988}. 

The results of different studies, particularly those based on hydro-dynamical simulations, are sometimes compared in terms of $m^{\rm dest}_{\rm gas}$. These generally depend on the adopted SN explosion energy, ISM density, initial grain size distribution, as well as on the physical processes implemented in the simulations and the total integration time. Table~\ref{table:mdest} reports a compilation of recent results. 

\begin{table}
\caption{Compilation of carbon and silicate dust destruction efficiencies obtained by different studies. The results are reported in terms of
$m^{\rm dest}_{\rm gas}$, in solar masses, defined as the mass of ISM gas that is completely cleared out of dust by a single SN explosion with energy $E_{\rm 51}$, exploding in a homogeneous medium with density $n_{\rm ISM}$ (see Eq.~\eqref{eq:mdust}). For \citet{martinez2019}, we show their result for a mixture of carbonaceous and silicate grains when the SN explosion occurs within a pre-existing stellar wind-blown bubble (WDB), or in a homogeneous medium (see also \citealt{kirchschlager2023} for a comparison between these findings). We note that the quoted values for \citet{bocchio2014} were not available in the original paper and have been estimated from their reported dust destruction timescales.}
\label{table:mdest}
\begin{tabular}{llll}
\hline\noalign{\smallskip}
Reference & carb \, $m^{\rm dest}_{\rm gas} [M_\odot]$  & sil \, $m^{\rm dest}_{\rm gas} [M_\odot]$  & Notes \\
\hline\noalign{\smallskip}
\citet{bocchio2014} & 21100 & 4220 & $n_{\rm ISM} =  0.25$ cm$^{-3}$, $E_{\rm 51} = 1$ \\
\citet{slavin2015} & 1220 & 1990 & $n_{\rm ISM} =  0.25$ cm$^{-3}$, $E_{\rm 51} = 0.5$ \\
\citet{hu2019} & 1330 \, (1050) & 1990 \, (1370) & $n_{\rm ISM} =  0.1 \, (1)$ cm$^{-3}$, $E_{\rm 51} = 1$\\
\citet{kirchschlager2022} & -- & 6470 \,(7090) & $n_{\rm ISM} =  0.1 \, (1)$ cm$^{-3}$, $E_{\rm 51} = 1$ \\
\hline\noalign{\smallskip}
& dust mix $m^{\rm dest}_{\rm gas} [M_\odot]$ &  &\\
\citet{martinez2019} &  $\sim 45$ & WDB &$n_{\rm ISM} =  1$ \, cm$^{-3}$, $E_{\rm 51} = 0.9$ \\
\citet{martinez2019} &  $> 120$ & no WDB & $n_{\rm ISM} =  1$ \, cm$^{-3}$, $E_{\rm 51} = 0.9$ \\
\hline\noalign{\smallskip}
\end{tabular}
\end{table}

By using 3D hydro-dynamical simulations, \citet{hu2019} quantify the amount of ambient dust destroyed by thermal and non-thermal sputtering by a single SN explosion exploring a broad range of ambient densities. In Table~\ref{table:mdest} we only report their result for $n_{\rm ISM} =  0.1$ and 1 cm$^{-3}$, which appear in very good agreement with \citet{slavin2015}, although both of these studies predict smaller $m^{\rm gas}_{\rm des}$ compared to \citet{kirchschlager2022}, which also include the effect of grain-grain collisions. Note also that \citet{kirchschlager2022} adopts a larger SN explosion energy and ambient gas density compared to \citet{slavin2015}, and follow the evolution on a longer timescale. \citet{martinez2019} explore dust destruction when the SN blast wave expands in the tenuous medium excavated by a pre-existing stellar wind, terminating with a wind-driven shell. If the mass of the wind-driven shell (WDS) is $\gtrsim 40$ the mass of the SN ejecta, the SN forward shock is unable to overrun the WDS, and the ambient ISM medium ahead of the WDS remains unaffected. This significantly decreases the dust destruction efficiencies, as shown in Table \ref{table:mdest}. It is important to note, however, that even when the explosion takes place in a homogeneous medium, their estimated $m^{\rm gas}_{\rm des}$ for a mixture of silicate and carbon grains is significantly smaller compared to other studies, due to the shorter timescale at which $m^{\rm gas}_{\rm des}$ is evaluated ($6.1$ \, kyr, compared to $\sim 10^2 - 10^3$ \, kyr, see \citealt{kirchschlager2022}).

The above studies allow to estimate dust destruction when a single SNR interacts with its homogeneous environment. However, in a multi-phase ISM, the effects of the interaction between the SNR and the ISM will be different if dust resides in cold clouds, in the warm medium, or in the hot phase \citep{McKee1989}. In addition, the temporal and spatial correlation of SN explosions needs to be taken into account, as SNe exploding in superbubbles or above the galactic disk will not be effective at destroying dust \citep{McKee1989, dwek1980}. To account for these effects, Eq.~\eqref{eq:taudes} is generally modified as:
\begin{equation}
\frac{1}{\tau_{\rm dest}} = \frac{R_{\rm SN,eff} \, m_{\rm gas,eff}^{\rm dest}}{M_{\rm ISM}}
\label{eq:taudes_eff}
\end{equation}
\noindent
where $R_{\rm SN,eff} = \delta_{\rm SN} R_{\rm SN}$ is the effective SN rate, and $m^{\rm gas,eff}_{\rm dest} = f_{\rm eff} m^{\rm dest}_{\rm gas}$, where the parameter 
$f_{\rm eff}$ accounts for the multi-phase structure of the ISM.
Following \citet{McKee1989}, in a three phase ISM model the filling factor of the hot, warm and cold phases are $f_{\rm h} \sim 0.7$, $f_{\rm w} \sim 0.3$, $f_{\rm c} \sim 0.02$, respectively. Since the density of the hot phase is too low for any dust destruction to occur, and the filling factor of cold clouds is very small, the dominant contribution comes from the warm phase, and $f_{\rm eff} = f_{\rm w}/f_{\rm h} \sim 0.43$\footnote{A derivation of this expression can be found in \citet{McKee1989} and in \citet{slavin2015}. It is based on the idea that the warm gas is confined into clouds (with filling factor $f_{\rm cloud} = f_{\rm w}$) embedded in the hot intercloud medium (with filling factor $f_{\rm h}$), and that the shocked cloud is at the same pressure of the shocked hot gas, so that $\rho_{\rm cloud} v_{\rm s,cloud}^2 \simeq \rho_{\rm h} v_{\rm s,h}^2$, where $\rho_{\rm cloud}$ ($\rho_{\rm h}$) and $v_{\rm s,cloud}$ ($v_{\rm s,h}$) are the mass density and shock velocity in the cloud (hot gas). The cloud shocks can be radiative, but the blast wave in the hot gas is not. Hence, Eq.~\eqref{eq:STapprox} in the hot phase can be written as: $M_{\rm s, h} = f_{\rm h} \, \rho_{\rm h} \, V_{\rm s} = E_{\rm SN}/(\sigma v_{\rm s, h}^2)$, and the shocked cloud mass can be written as: $M_{\rm s, cloud} = f_{\rm cloud} \, \rho_{\rm cloud} \, V_{\rm s} = (f_{\rm cloud}/f_{\rm h}) \, (\rho_{\rm cloud}/\rho_{\rm h}) \, M_{\rm s, h} = (f_{\rm cloud}/f_{\rm h}) \, E_{\rm SN}/(\sigma v_{\rm s, c}^2)$, which then leads to Eq.\ref{eq:taudes_eff}, with $f_{\rm eff} = f_{\rm cloud}/f_{\rm h} = f_{\rm w}/f_{\rm h}$.}.
However, if the ISM is instead dominated by the warm phase ($f_{\rm w} >> f_{\rm h}$), and the part of the shock front that propagates into the hot gas does not destroy any dust, then $f_{\rm eff} = f_{\rm w}$, with $f_{\rm w} \lesssim 0.8$ \citep{slavin2015}. Using observations in our own galaxy, the correction factor for the SN effective rate has been estimated by \citet{McKee1989} to be $\delta_{\rm SN} \sim 0.36$. A very similar value, $\delta_{\rm SN} \sim 0.4$, has been found by \citet{hu2019} by investigating dust sputtering in a multi-phase ISM
that resembles the solar-neighbourhood environment, where SNe are injected stochastically into the ISM. 

\begin{figure}
\centerline{\includegraphics[width=12cm]{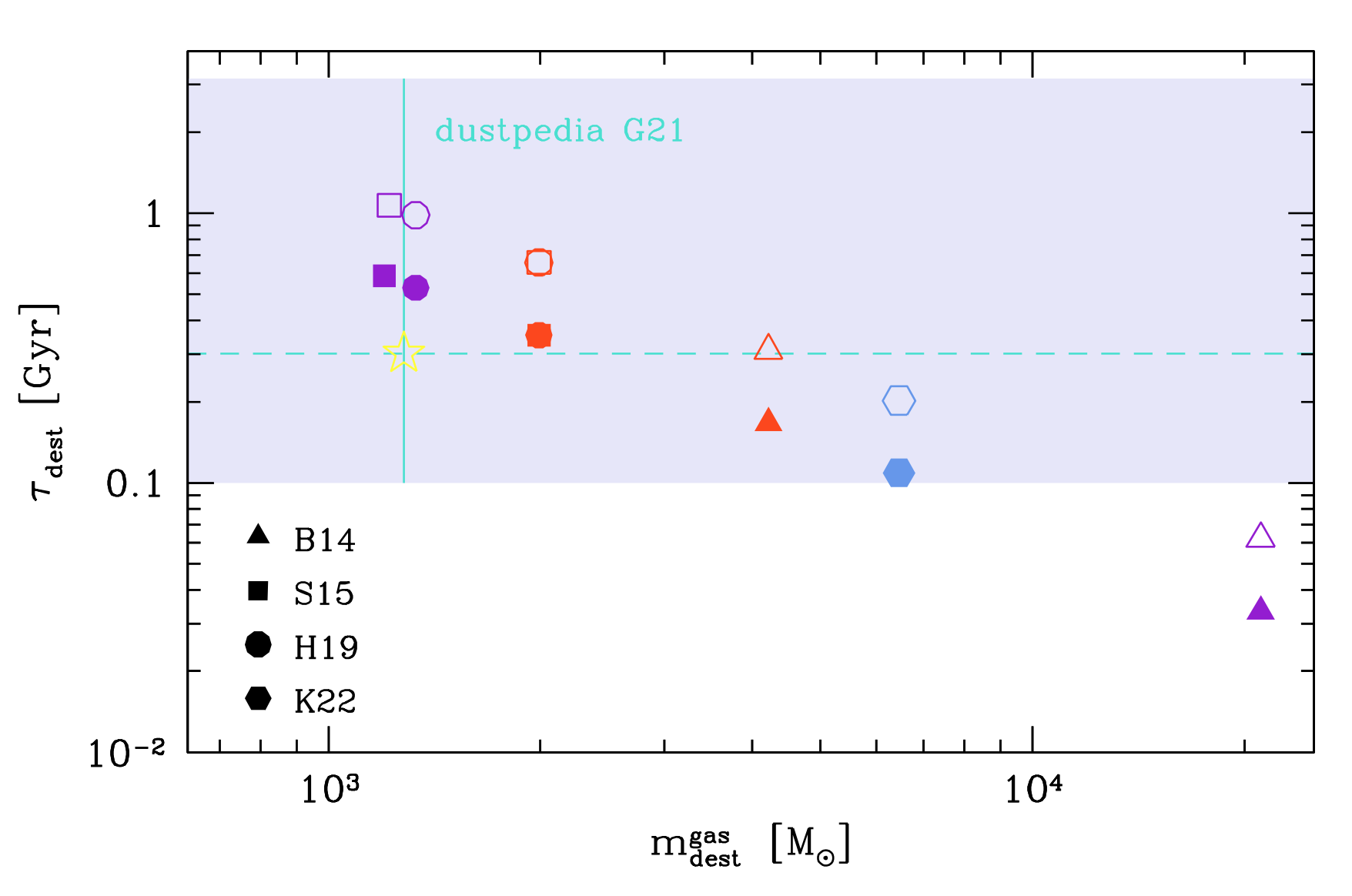}}
\caption{Comparison between the SN dust destruction timescales, $\tau_{\rm dest}$, implied by the different $m_{\rm dest}^{\rm gas}$ reported in Table \ref{table:mdest} for Milky Way conditions ($R_{\rm SN,eff} = 1/125 \, {\rm yr}^{-1}$, $M_{\rm ISM} = 4.5 \times 10^9 \, M_\odot$) and assuming $f_{\rm eff} = 0.43$ (empty symbols), and 0.8 (filled symbols). Where possible, we show separately the results for carbonaceous (violet points) and silicate grains (orange points). B14 = \citet{bocchio2014}, S15 = \citet{slavin2015}, H19 = \citet{hu2019}, K22 = \citet{kirchschlager2022}. Note that the results for silicate grains predicted by S15 and H19 perfectly overlaps. We do not show the results of \citet{martinez2019} as they imply much longer $\tau_{\rm dest}$, larger than the age of the Universe. The vertical solid line is the empirical $m^{\rm dest}_{\rm gas} \simeq 1288 \, M_\odot$ found by \citet{galliano2021} through a Bayesian analysis of a large sample of local galaxies from the DustPedia and Dwarf Galaxy Sample (see Sect.~\ref{sec:dustpedia} and Fig.~\ref{fig:dustpedia1}), and the horizontal dashed line is the $\tau_{\rm dest} = 300$\, Myr for the Milky Way (yellow star), with the shaded region encompassing the range of $\tau_{\rm dest}$ across the galaxies in the sample.}
\label{fig:taudes}
\end{figure}

Assuming $R_{\rm SN,eff} = 1/125 \, {\rm yr}^{-1}$, and $M_{\rm ISM} = 4.5 \times 10^9 \, M_\odot$, one of the two ISM models incorporated in $f_{\rm eff}$, and $m^{\rm dest}_{\rm gas}$ from Table \ref{table:mdest}, it is possible to estimate the dust destruction timescales in the Milky Way predicted by different studies. 
The results are shown in Fig.~\ref{fig:taudes}. 

With the exception of the very small dust destruction timescales predicted for carbonaceous grains by \citet{bocchio2014}, all the other studies\footnote{We do not show the results by \citet{martinez2019} here, as they predict grain destruction timescales that are longer than the age of the Universe, when applied to Milky Way conditions.} predict values of $\tau_{\rm dest}$ in the range 100 Myr--1 Gyr. 

It is important to note that both $f_{\rm eff}$ and $\delta_{\rm SN}$ are likely to be non-universal, i.e. to depend on the star formation history, stellar initial mass function, and ISM properties of the galaxy, which adds additional uncertainty on $\tau_{\rm dest}$. In Fig.~\ref{fig:taudes} we also show the empirical results found by \citet{galliano2021} applying a Bayesian analysis to a large sample of local galaxies of the DustPedia project \citep{davies2017} and of the dwarf galaxy sample (DGS \citealt{madden2013}). The vertical solid line represents the value of $m^{\rm dest}_{\rm gas} \simeq 1288 \, M_\odot$ that they find assuming this to be a universal parameter (the same for all the galaxies in the sample), the horizontal dashed line is the value of $\tau_{\rm dest} = 300$ \, Myr that they infer for the Milky Way (yellow star), and the shaded region provides the range of values of $\tau_{\rm dest}$ found for the galaxies in the sample, which are characterized by a broad range of metallicities, star formation rates, and stellar masses (see \citealt{galliano2021} and Sect.~\ref{sec:dustpedia}).

\subsection{Destruction in the hot gas}

In addition to dust destruction by SN shocks, once the grains enter a hot plasma ($T \gtrsim 10^6$\,K), they are sputtered away by thermal collisions with both protons and helium nuclei. This process has been investigated in the past by many authors \citep{draine1979, seab1987, tielens1994} and it has been implemented in models describing dust evolution in elliptical galaxies \citep{tsai1995, hirashita2015b}, where the hot phase largely dominates the galactic ISM. Recently, thermal sputtering in galactic haloes has also been included in cosmological hydrodynamical simulations which follow dust evolution on-the-fly 
\citep{aoyama2017, mckinnon2017, li2019, vogelsberger2019, Graziani2020}. 

In general, the sputtering yield (target species ejected per incident ion/projectile) for ions of type i depends upon the kinetic energy of the impinging ion, $E_{\rm i}$, and on the angle of incidence between the ion and the surface, $\theta$, $Y_{\rm i}(E_{\rm i}, \theta)$.

The sputtering rate therefore depends upon the angle-averaged yield,
\begin{equation}
\langle Y_{\rm i}(E_{\rm i}) \rangle \, \equiv 2 \, \int_{0}^{\pi/2} Y_{\rm i}(E_{\rm i},\theta) \, {\rm sin} \theta \, {\rm cos} \theta \, d\theta,
\end{equation}
\noindent
which is generally taken to be twice the normal incidence yield, i.e. $\langle Y_{\rm i}(E_{\rm i}) \rangle \, = 2\, Y_{\rm i}(E_{\rm i}, 0)$ \citep{draine1979}. The erosion rate of grain by thermal sputtering is given by:
\begin{equation}
\frac{da}{dt} = - n_{\rm H} \, \frac{m_{\rm sp}}{{2\rho_{\rm gr}}} \, {\Sigma}_{\rm i} \, A_{\rm i} \, \langle Y_{\rm i}(E_{\rm i}) v_{\rm i} \rangle,
\label{eq:erosionrate}
\end{equation}
where $n_{\rm H}$ is the number density of hydrogen, $m_{\rm sp}$ is the average mass of the sputtered species, $\rho_{\rm gr}$ is the density of the grain material, $A_{\rm i}$ is the mass fraction of impinging species i with velocity $v_{\rm i}$, and the average is over a Maxwellian distribution.

\begin{figure}
\centerline{\includegraphics[width=10cm]{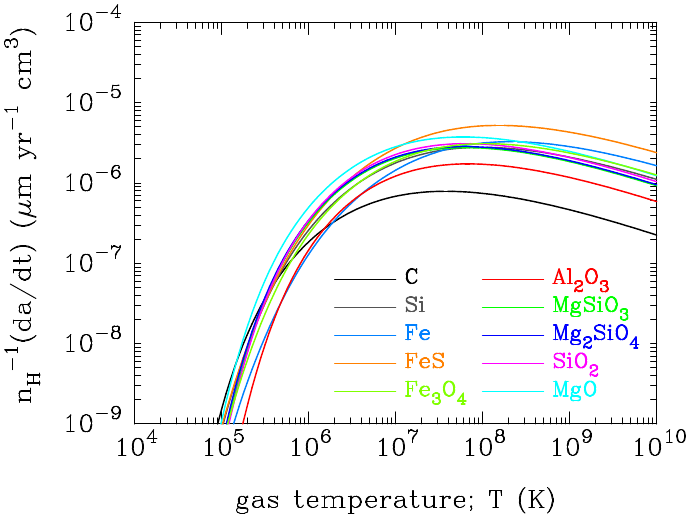}}
\caption{The erosion rate of different grain species by thermal sputtering in a hot gas with a metallicity of $Z = 10^{-4} Z_\odot$ as a function of temperature, computed using Eq.~\eqref{eq:erosionrate} and the sputtering yields calculated by \citet{Nozawa2006}. Image reproduced with permission from \citet{Nozawa2006}, copyright by AAS.}
\label{fig:erosionrate}
\end{figure}

Different methods to evaluate the sputtering yields for projectile -- target combinations of astrophysical interests have been adopted, based either on simulated data (using the TRIM code, see e.g. \citealt{bianchi2005}, or the EDDY code, see e.g. \citealt{Nozawa2006}), on fitting experimental data 
with an analytic formula (the so-called universal relation derived by \citealt{bohdansky1984}, see e.g. \citealt{tsai1995}), or a mix of the two \citep{Nozawa2006}. The tabulated parameters to compute the sputtering yields can be found in \citet{Nozawa2006}, and the corresponding erosion rate of each dust species in a gas with a metallicity of $Z = 10^{-4} Z_\odot$ is shown in Fig.~\ref{fig:erosionrate} as a function of the gas temperature. The temperature dependence reflects the energy dependence of the sputtering yields, and the erosion rates steeply increase from $T \simeq 10^5$K, reach a peak in the range $(4 - 20)\times 10^7$K, and then slowly decline. At $T \geq 2 \times 10^6$K, C grains have the lowest erosion rate, while FeS grains have an erosion rate at $T \gtrsim 10^7$K which is about 1 dex larger, and is the highest among the dust species considered. At $2\times 10^6 \, {\rm K} \leq T \leq 10^7 \, {\rm K}$, the erosion rate varies in the interval $\sim [0.4 - 4] \times 10^{-6} \, n_{\rm H} \, \mu{\rm m \, yr^{-1} cm^{3}}$.

Alternatively, the analytic form proposed by \citet{tsai1995} can be adopted:
\begin{equation}
\frac{da}{dt} = - 3.2 \times 10^{-18} \, {\rm cm^{4} s^{-1}} \, \frac{\rho}{m_{\rm H}} \left [\left(\frac{T_0}{T}\right)^w +1\right]^{-1}
\end{equation}
\noindent
where $\rho$ is the gas mass density, $m_{\rm H}$ is the hydrogen mass, $T_0 = 2 \times 10^6$ K, and $w=2.5$. When recast in the same units, the above expression leads to a sputtering rate of $\sim 0.5 \times 10^{-6} \, n_{\rm H} \, \mu{\rm m \, yr^{-1} cm^{3}}$ at $T=T_0$, in good agreement with the results presented in Fig.~\ref{fig:erosionrate}.

Starting from the above expressions, the sputtering timescale can be defined as:

\begin{equation}
\tau_{\rm sp} \equiv a \left|\frac{da}{dt}\right|^{-1} \simeq {100 \,\rm Myr} \,  \left(\frac{a}{\rm 0.1 \mu m}\right)  \left(\frac{\rm 10^{-3} cm^{-3}}{n_{\rm H}}\right) \left [\left(\frac{T_0}{T}\right)^w +1\right],
\end{equation}
\noindent
where we have assumed a typical grain radius $a = 0.1 \mu$m, and a gas density in the hot phase of $n_{\rm H} = 10^{-3}$\, cm$^{-3}$. Given a grain of density $\rho_{\rm gr}$ and mass $m_{\rm gr} = 4 \pi a^3 \rho_{\rm gr}/3$, the variation of the dust mass due to thermal sputtering in the hot gas can be written as,

\begin{equation}
\frac{dM_{\rm d,sp}}{dt} = N_{\rm gr} \, \frac{dm_{\rm gr}}{dt} = 3 \frac{M_{\rm d}}{a} \frac{da}{dt} = - 3 \frac{M_{\rm d}}{\tau_{\rm sp}}.
\label{eq:dust_sp}
\end{equation}
\noindent
where the number of grains $N_{\rm gr} = M_{\rm d}/m_{\rm gr}$ is assumed to be constant. The above equation is the way thermal sputtering in the hot gas has been generally included in semi-analytical \citep{Popping2017} and hydrodynamical simulations \citep{aoyama2017, mckinnon2017, li2019, vogelsberger2019, Graziani2020}.

\subsection{Grain growth in the interstellar medium}

Given the dust destruction timescales estimated above, many authors, starting from \citet{draine1979}, have reached the conclusion that most of the interstellar dust is not stardust, but was formed in the ISM (see \citealt{draine2009} and references therein). Following \citet{draine2011}, the argument is very simple: the mean residence time of an atom in the ISM of the Milky Way can be computed as $\tau_{\rm SF} = M_{\rm ISM}/{\rm SFR} \sim 10^9$\,yr, where the ISM mass is $M_{\rm ISM} = 4.5 \times 10^9 \, M_\odot$, and the star formation rate is ${\rm SFR} \lesssim 5 \, M_\odot/{\rm yr}$. If we assume that all the Si atoms enter the ISM as stardust grains, only a fraction of $\tau_{\rm dest}/(\tau_{\rm dest} + \tau_{\rm SF}) \sim 0.20$ of these would still be found in dust grains. Yet, observations of gas-phase ISM abundances show that $\gtrsim 90 \%$ of the Si atoms are missing from the gas-phase, i.e. they are depleted onto dust grains \citep{jenkins2009}. Hence, most of the currently observed silicate grains must have formed in the ISM \citep{draine2011}. 

\subsubsection{Empirical evidences}
There are empirical evidences of localized dust processing in the ISM of the Milky Way and Magellanic Clouds, such as variations in the grain emissivity \citep{kohler2015}, that could be explained by coagulation and accretion of grain mantles \citep{ysard2015}. Depletion studies provide evidence that the fraction of heavy elements that is locked up in grains correlates with the density \citep{jenkins2009, tchernyshyov2015, jenkins2017, romanduval2021}. Resolved FIR observations of dust emission in nearby galaxies, combined with atomic and molecular gas maps, show that the dust-to-gas ratio, D/G, increases with density within each galaxy, with grain growth being the most likely explanation \citep{romanduval2017, clark2023}. Finally, the relation between the D/G and metallicity in local galaxies does not follow a linear trend, but shows a knee, at approximately $Z \sim 0.2 \, Z_\odot$, below which the D/G drops sharply \citep[][see also the discussion in Section 8]{remyruyer2014, galliano2018, devis2019, galliano2021}. This is generally interpreted to reflect a threshold above which grain growth starts to dominate over stellar dust production \citep{asano2013, debennassuti2014, feldman2015, zhukovska2016, schneider2016, ginolfi2017, galliano2021, choban2022}\footnote{Even before these studies, the contribution of grain growth to ISM dust enrichment was included in the models by \citet{dwek1998} and \citet{hirashita1999}, and found to be important to explain the observed depletion patterns and dust-to-gas mass ratios in high metallicity galaxies.}, even in high redshift galaxies \citep{mancini2015, Popping2017, Graziani2020, Dicesare2023}, although
other explanations have been proposed \citep{delooze2020, priestley2022b}. In particular, \citet{delooze2020} apply a Bayesian approach similar to \citet{galliano2021} to fit the local sample of JINGLE galaxies \citep{saintonge2018}, finding that their present-day dust masses can be explained primarily by stellar dust, with a contribution of 20--50\% from grain growth in the ISM,   if dust destruction timescales are long ($\tau_{\rm dest} \sim 1 - 2$\,Gyr), and if the survival fraction of newly formed SN dust passing through the reverse shock is high ($\eta = 37 - 89 \%$). This conclusion differs from the results by \citet{galliano2021} based on the DustPedia and Dwarf Galaxy Sample (DGS), which indicate that grain growth is the dominant formation mechanism at metallicity above $Z = 0.2 \, Z_\odot$. The difference between these two analyses is probably due to the broader extent in metallicity (particularly to low-metallicity galaxies, with $Z < 0.2 \, Z_\odot$) of the sample considered by \citet{galliano2021} (see the discussion in \citealt{galliano2021} and Sect.~\ref{sec:dustpedia}). A different conclusion has also been found in the analysis by \citet{nanni2020}, who fit individual galaxies of the DGS with a dust evolution model with the aim of reproducing the observed dust-to-stellar mass evolution as a function of time. They find that dust growth in the ISM is not necessary in order to reproduce the properties of the galaxies in their sample, and, if present, the importance of this process would be counterbalanced by galactic outflows. This shows that the assumptions in the dust evolution model and parameter degeneracies may bias the results, leading to different conclusions even when the same galaxy sample is considered. 

\subsubsection{The physics of grain growth} 
It is to be said that we do not yet understand how grain growth occurs in the ISM (see the seminal work by \citealt{barlow1978}, which already highlighted some of the difficulties in growing grains in the ISM). The collision rate between atoms or ions onto pre-existing grains depends on the physical conditions of the interstellar gas (temperature and density, which affect the thermal velocities of colliding ions and atoms, as well as the degree of turbulence, which may accelerate the grains), and on the geometric cross section of the grains, which will be dominated by the smallest grains. In addition, it will be sensitive to the charge state of the grains, with neutral and negatively charged grains having an increased collision rate with positive ions thanks to Coulomb focusing \citep{draine2009}. \citet{weingartner1999} discussed the collision rates of positively charged ions in different phases of the ISM, accounting for the charge distribution of small grains. They find that in the cold neutral medium (CNM), with $n_{\rm H} \sim 30$\, cm$^{-3}$ and $T \sim 100$K, the accretion timescale, $\tau_{\rm a}$, defined as,
\[
\tau_{\rm a}^{-1} = -\frac{1}{n} \, \frac{dn}{dt},
\]
where $n$ is the number density of the ions, can be as short as $\sim 2 \times 10^5$\, yr, sufficiently rapid to account for the observed depletion factors of elements like Si and Ti \citep{savage1996}. 

It is important to stress that the above quoted value was obtained under the assumption of a sticking coefficient $S = 1$ (a maximum probability of sticking of the ion with the grain surface in each collision), and an idealized phase of the ISM, characterised by constant values of $n_{\rm H}$ and $T$. \citet{zhukovska2014} used 3D hydrodynamical simulations of giant molecular clouds in a Milky-Way like galaxy, and allow for a temperature-dependent sticking coefficient, using the observed Si depletion as a critical constraint of the model. They find that a sticking coefficient that decreases with temperature provides a better match the observed Si depletion, with $\sim 50$ (30) \% of grain growth occurring in gas with $5\,{\rm cm}^{-3} \le n_{\rm H} < 50\,{\rm cm}^{-3}$ ($50\,{\rm cm}^{-3} \le n_{\rm H} < 500\,{\rm cm}^{-3}$, see their Fig. 5, where the show the evolution of the grain growth rate in different bins of density). Also, enhanced collision rates due to Coulomb focusing are important to match the observed depletion -- density relations for both silicate and iron dust \citep{zhukovska2016}, reducing the accretion timescales from $\sim 15$ Myr ($\sim 14$ Myr) to $\sim 1$ Myr ($\sim 0.16$ Myr) for Si (Fe) species in the CNM (assuming typical values of $n_{\rm H} \sim 30$\, cm$^{-3}$ and $T \sim 100$K). 

\begin{figure}
\centerline{\includegraphics[width=12cm]{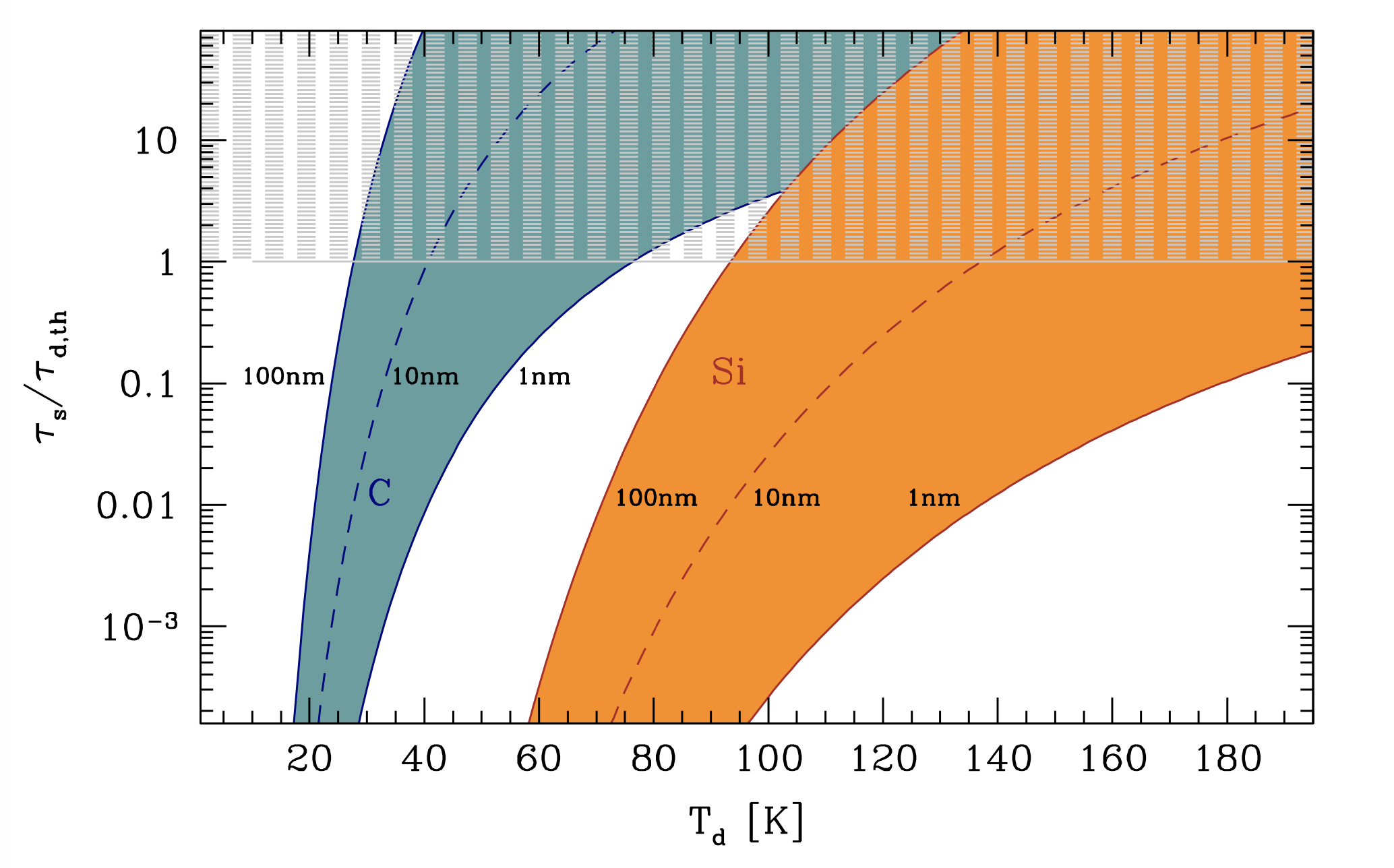}}
\caption{Comparison between the scanning and thermal desorption timescales. We show their ratio as a function
of the grain temperature, for C (blue) and Si (orange), adopting three different grain sizes: $a = 1, 10, 100$ nm (from right to left). We have assumed a binding energy of $E_{\rm b}/k = 800$ K for C and 2700 K for Si, and the corresponding diffusion energy to be $E_{\rm d} = f_{\rm d} E_{\rm b}$, with $f_{\rm d} = 0.5$ (see text). The shaded region is where $\tau_{\rm s} > \tau_{\rm d, th}$, and grain growth is prevented by thermal desorption. The figure shows that for small grains there is a range of grain temperature for which the 
scanning timescale remains smaller than the thermal desorption timescale. The scanning timescale has a very strong dependence on $T_{\rm d}$, and becomes smaller than $1$ Myr when $T_{\rm d}> 25 - 30$ K ($> 10$ K) for for Si (C) atoms scanning grains with sizes $a \leq 100$ nm.}
\label{fig:tauscan}
\end{figure}

Grain growth is not only a matter of kinetics: the colliding atom or ion must also be bound to the surface of the grain in a way that it allows it to be retained against thermal desorption, UV and Cosmic Ray irradiation, which may provide enough energy to the accreted species to eject them from the grain surface and back to the gas phase. A discussion of the relevant processes and of the current uncertainties is provided by \citet{draine2009}, \citet{zhukovska2016}, \citet{ferrara2016}, and \citet{ceccarelli2018}. It is generally assumed that grain growth occurs once the accreted atom or ion (adsorbed species) reaches an active sites of the grain surface, i.e. a site with dangling bonds and high binding energy. The diffusion of adsorbed species on the surface of the grain occurs on the so-called "scanning timescale", 
\[
\tau_{\rm s}^{-1} \sim N_{\rm s}^{-1} \, \nu_0 \, {\rm exp}(-E_{\rm d}/kT_{\rm d}),
\]
\noindent
where $E_{\rm d}$ is the diffusion energy, $T_{\rm d}$ is the dust temperature, $\nu_0 \sim 10^{12} {\rm s}^{-1}$ is the vibrational frequency of the sticking species, and $N_{\rm s} = 4 \pi a^2 n_{\rm s}$ is the number of active sites for a grain with radius $a$, with $n_{\rm s} \sim 1.5 \times 10^{15} {\rm cm}^{-2}$ being the surface density of physisorption sites \citep{hasegawa1992}. Hence, for a grain radius of $a = 100$ nm, $N_{\rm s} \sim 10^6$, while for small grains, with $a = 1 $nm, $N_{\rm s} \sim 180$. The diffusion energy is poorly known, and it is generally taken to be a fraction $f_{\rm d} = 0.3 - 0.8$ of the binding energy $E_{\rm b}$ of the adsorbed species \citep{ferrara2016}. 
The scanning timescale is generally compared to the thermal desorption timescale,
\[
\tau_{\rm d, th}^{-1} \sim  \nu_0 \, {\rm exp}(-E_{\rm b}/kT_{\rm d}).
\]
\noindent
When $\tau_{\rm s} < \tau_{\rm d, th}$, the adsorbed species can scan the entire surface of the grain before being thermally desorbed. Unfortunately, the binding energies of the atoms of interest
(C, Mg, Si, Fe) to surfaces of interest (amorphous silicate or carbonaceous materials) are not known. In \citet{zhukovska2016}, \citet{ferrara2016}, and \citet{ceccarelli2018},  the binding energy
for Si and Fe is taken to be $E_{\rm b}/k = 2700$ K and 4200 K from Table 4 of \citet{hasegawa1993}, respectively. From the same Table, we find the binding energy for C to be 800 K. Adopting these values and 
$f_{\rm d} = 0.5$, in Fig.~\ref{fig:tauscan} we show the ratio of $\tau_{\rm s}/\tau_{\rm d, th}$ as a function
of $T_{\rm d}$ for C and Si assuming three different grain sizes, $a = 1, 10$, and 100 nm. The shaded region is where grain growth is hindered by thermal desorption. The result is very sensitive to the binding energy of the adsorbed species and C atoms can grow only on very small grains, while Si (and Fe) atoms can grow for a range of grain sizes and temperatures (see also the discussion in \citealt{zhukovska2016}). It is important to keep in mind that the scanning timescale has a very steep dependence on $T_{\rm d}$, and becomes smaller than 1 Myr when $T_{\rm d}> 25 - 30$ K ($> 10$ K) for for Si (C) atoms scanning grains with sizes $a \leq 100$ nm. Stochastic heating of very small grains due to UV photon absorption can increase the grain temperature well above the equilibrium value, and drastically reduce the scanning timescales, even when accounting for radiative cooling \citep{zhukovska2016}. 

The above considerations lead us to conclude that very small grains, with sizes $\leq 10$ nm, are likely to play a very important role for grain growth in the cold neutral medium, as they have the largest total surface area, the highest fraction of negatively charged grains, and -- aided by stochastic heating -- their scanning timescales is shorter than the thermal desorption timescale. Once the grains are incorporated in dense molecular clouds, their growth is problematic as the accretion of silicon and carbon-bearing species occurs simultaneously with the
formation of icy mantles \citep{ferrara2016, ceccarelli2018}. Icy mantles are weakly bound to the grain surface and rapidly evaporate when the grains return to the diffuse phase.

It is important to stress that despite the uncertainties that persist on the surface reactions leading to the growth of silicate and amorphous carbon grains in the ISM, the condensation of complex silicates with pyroxene composition at temperatures between 10 and 20 K by accretion of molecules and atoms on cold surfaces
and subsequent reactions between them has been proven experimentally \citep{rouille2015}. The experiments clearly demonstrate an efficient silicate formation at low temperatures, with final fluffy aggregates consisting of small nanometer-sized primary grains. More recently, both silicate and carbonaceous materials have been experimentally proved to condense from cold precursors in the absence of radiations such as interstellar UV photons and cosmic rays, and that species from one of two groups that consist respectively of silicate precursors and carbonaceous matter precursors do not react at cryogenic temperatures with those belonging to the other group \citep{rouille2020}. Such a finding constitutes a clue as to the separation between silicate and carbonaceous materials in the ISM, and supports 
the hypothesis that dust grains can be grown in the ISM.

\subsubsection{The grain growth timescale}
Chemical evolution models generally adopt a very simplified description of the process, capturing the kinetics of the process. Different expressions have been proposed for the grain growth timescales. Following \citet{asano2013}, the dust mass growth rate in the ISM can be expressed as:
\begin{equation}
\left(\frac{dM_{\rm d}}{dt}\right)_{\rm growth} = N_{\rm d} \, \pi <a^2> \, S \, \rho^{\rm gas}_{\rm Z} \, \langle v \rangle,
\end{equation}
\noindent
where $N_{\rm d}$ is the number of grains, $<a^2>$ is the second moment of the grain size, $S$ is the sticking coefficient, $\rho^{\rm gas}_{\rm Z}$ is the gas-phase metal density, and $\langle v \rangle$ is the mean velocity of gas-phase metals. The number of grains can be estimated as $N_{\rm d} = M_{\rm d}/m_{\rm d}$, where 
$m_{\rm d} = 4 \pi <a^3> \sigma/3$ is the average mass of spherically symmetric grains with density $\sigma$, and $<a^3>$ is the 3rd moment of the grain size. We can write, $\rho^{\rm gas}_{\rm Z} = \, Z \, \rho^{\rm gas} = Z \, \mu \, m_{\rm H} \, n_{\rm H}$, where $Z$ is the gas metallicity, $n_{\rm H}$ is the number density of the ISM phase where grain growth occurs, and $\mu$ is the mean molecular weight. Assuming $\mu = 1.4$, $\sigma = 3\, \rm g cm^{-3}$ (for silicate grains), $S = 1$, a mean mass of colliding species of $A m_{\rm H}$ with $A = 20$, and that $A m_{\rm H} <v^2> = k T$, the grain growth timescale can be written as:
\begin{equation}
\tau_{\rm growth} = M_{\rm d} \, \left(\frac{dM_{\rm d}}{dt}\right)_{\rm growth}^{-1} = 6.7 \times 10^6 \, {\rm yr} \, \left(\frac{\overline{a}}{10 \, \rm nm}\right) \, \left(\frac{n_{\rm H}}{30 \, \rm cm^{-3}}\right)^{-1} \, \left(\frac{T}{100 \, K}\right)^{-1/2} \, \left(\frac{Z}{Z_\odot}\right)^{-1},
\label{eq:taugg1}
\end{equation}
\noindent
where $\overline{a}$ is the typical size of the grains, defined as $\overline{a} = <a^3>/<a^2>$ \citep{hirashita2011}, and $Z_\odot = 0.0134$. For the CNM, the above expression leads to a grain growth
timescale of $\tau_{\rm growth} \sim 6.7 $Myr assuming an average grain radius of 10 nm and solar metallicity. This is approximately consistent with the value reported by \citet{zhukovska2016} for silicate grains in the CNM (15 Myr), although they show that the enhanced collision rate due to Coulomb focusing reduces the grain growth timescale to 1 Myr (see above). We shall note that while this simple parametrization allows to reproduce galaxy-integrated dust-to-metals (D/Z) ratios, more sophisticated chemical models
that track the evolution of specific dust species are needed to reproduce the observed scaling relation between individual element depletions and D/Z with column density and local gas density (see e.g. \citealt{zhukovska2014, zhukovska2016, choban2022}).

\begin{figure}
\centerline{\includegraphics[width=6cm]{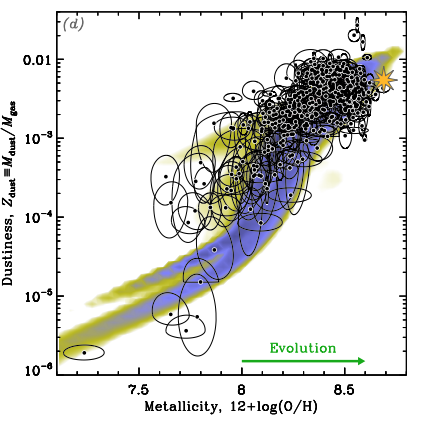}
\includegraphics[width=6cm]{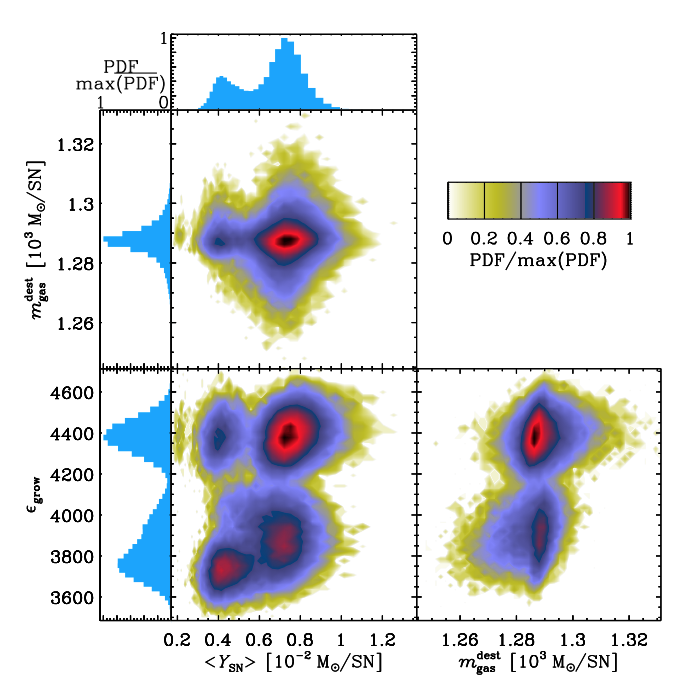}}
\caption{{\bf Left panel}: example of fitted evolutionary tracks to the dustiness (dust-to-gas mass ratio) vs metallicity relation. Each of the 556 galaxies of the sample investigated by \citet{galliano2021} represented as skewed uncertainty ellipses, that are the 1$\sigma$ contour of a bivariate distribution having the same means, variances, skewnesses, and correlation coefficient as the posterior distribution. The central dot shows the mode of this bivariate distribution (see \citealt{galliano2021} for more details on this representation). The yellow star is the Milky Way galaxy. The coloured contours represent the posterior PDF of dust evolutionary tracks, marginalizing over the star formation history of each galaxy, and assuming a Salpeter IMF.
{\bf Right panel}: posterior distribution of the tuning parameters adopted in the hierarchical bayesian analysis of the local sample of galaxies analysed by \citet{galliano2021}: $m^{\rm dest}_{\rm gas}$ is the ISM mass that is completely cleared out of dust by a single SN explosion, $\epsilon_{\rm growth}$ is the grain growth parameter, and $\langle Y_{\rm SN} \rangle$ is the IMF-averaged dust yield per single SN (see Sect.~\ref{sec:dustpedia}) assuming a Salpeter IMF. 
The coloured contours display the bidimensional posterior PDF of pairs of parameters, marginalising over the other ones. The histograms show the posterior PDF of each tuning parameter. Images reproduced with permission from \citet{galliano2021}, copyright by the author(s).}
\label{fig:dustpedia1}
\end{figure}

A different parametrization of the grain growth timescale has been proposed by \citet{mattson2012}. 
They start from the assumption that grain growth occurs predominantly in molecular clouds, and define the 
grain growth timescale as:
\begin{equation}
\tau_{\rm growth} = \Sigma_{\rm d} \, \left(\frac{d\Sigma_{\rm d}}{dt}\right)_{\rm growth}^{-1} = \frac{m_{\rm d} d_{\rm c}}{S \pi a^2 \Sigma_{\rm Z} \langle v \rangle},
\end{equation}
\noindent
where $\Sigma_{\rm d}$ is the dust surface density, $\Sigma_{\rm Z} = Z \Sigma_{\rm mol}$ is the molecular cloud surface density of gas-phase metals, and $d_{\rm c}$ is the typical size of molecular clouds. Assuming that $\Sigma_{\rm mol} \approx \Sigma_{\rm H_2}$, and that $\Sigma_{\rm SFR} = \alpha \Sigma_{\rm H_2}$, the grain growth timescale is written as:
\begin{equation}
\tau_{\rm growth} = \frac{m_{\rm d} d_{\rm c} \alpha}{S \pi a^2 Z \Sigma_{\rm SFR} \langle v \rangle} = \frac{\Sigma_{\rm gas}}{\epsilon_{\rm growth} \, Z \, \Sigma_{\rm SFR}},
\label{eq:tauG_mattsson}
\end{equation}
\noindent
where
\begin{equation}
\epsilon_{\rm growth} = \frac{S \pi a^2 \Sigma_{\rm gas} \langle v \rangle}{m_{\rm d} d_{\rm c} \alpha}
\end{equation}
\noindent
is the grain growth parameter, a dimensionless factor that depends on the physical properties of the ISM environment where 
grain growth occurs, and it is assumed to be constant \citep{mattson2012}. Often, Eq.~\eqref{eq:tauG_mattsson} is written 
in terms of the total gas mass and SFR \citep{delooze2020, galliano2021},
\begin{equation}
\tau_{\rm growth} = \frac{M_{\rm gas}}{\epsilon_{\rm growth} \, Z \, {\rm SFR}},
\label{eq:taugg2}
\end{equation}
\noindent
and $\epsilon_{\rm growth}$ is left as a tuning parameter, to be determined by observations. By applying a Bayesian analysis of a large sample of local galaxies from the DustPedia and Dwarf Galaxy Sample (see Sect.~\ref{sec:dustpedia}), \citet{galliano2021} find an empirical value of $\epsilon_{\rm growth} \simeq 4045$, and a distribution of values for $\tau_{\rm growth}$ that is quite scattered, ranging
from $\sim 1$ Gyr, for the lowest metallicity galaxies in the sample, to $\sim 45$ Myr around solar metallicity. Yet, there are significant variations in $\tau_{\rm growth}$ for similar values of $Z$, suggesting that even in the simple parametrization described
by Eq.~\eqref{eq:taugg2}, the grain growth timescale depends on the specific conditions prevailing in each galaxy. When compared to
Eq.~\eqref{eq:taugg1}, we find that -- for a solar metallicity galaxy -- the two parametrizations lead to comparable results if 
we assume that the average radius may vary in the range $10 {\rm nm} \leq \overline{a} \leq 67$ nm, and that the CNM may be characterised by a broader range of gas densities, with $5 {\rm cm}^{-3} \leq n_{\rm H} \leq 30 {\rm cm}^{-3}$.
The significant scatter in the grain growth timescales inferred for galaxies with comparable metallicities by \citet{galliano2021} reflect their different star formation histories and ages.

\section{A nearby galaxy perspective on dust evolution processes}
\label{sec:dustpedia}

Recently, \citet{galliano2021} have conducted an empirical statistical study based on the sample of $\sim 800$ nearby galaxies of the DustPedia project \citep{davies2017} and of the dwarf galaxy sample (DGS \citealt{madden2013}), which spans a broad range of metallicities, gas fractions, specific star formation rates, and galaxy types. By adopting a hierarchical bayesian approach \citep{galliano2018}, they infer the dust properties of each object from its spectral energy distribution, and then fit theoretical tracks to their sample to derive empirical constraints on key dust evolution processes. Following \citet{rowlands2014} and \citet{devis2017}, they adopt a one-zone dust evolution model where dust is produced by SN and AGBs\footnote{Following \citet{devis2017}, AGBs are assumed to condense 15\% of their heavy elements into dust.}, can grow in the ISM, can be destroyed by SN shocks, and can be removed from the ISM by astration and galaxy outflows. The efficiencies of individual processes are described by simple parametrizations, that depend on three main tuning parameters: the IMF-averaged SN dust yield per SN, defined as,
\[
\langle Y_{\rm SN} \rangle = \frac{\int_{8 M_\odot}^{40 M_\odot} \phi(m) \, m_{\rm dust}^{\rm SN}(m) \,dm}{\int_{8 M_\odot}^{40 M_\odot} \phi(m) \, dm},
\]
\noindent
that enters in the definition of the dust condensation timescale,
\begin{equation}
\frac{1}{\tau_{\rm cond}} = \langle Y_{\rm SN} \rangle \frac{R_{\rm SN}}{M_{\rm d}},
\label{eq:taucond}
\end{equation}
\noindent
where $R_{\rm SN}$ is the SN rate and $M_{\rm d}$ the total dust mass in the ISM. The other tuning parameters are the grain growth efficiency $\epsilon_{\rm growth}$ that enters in the definition of the grain growth timescale given by Eq.\ref{eq:taugg2}, and mass of gas that is cleared out of dust by a single SN explosion, $m^{\rm dest}_{\rm gas}$, that enters in the definition of the dust destruction timescale given by Eq.\ref{eq:taudes}. These tuning parameters are assumed to be universal, i.e. they are fitted to the whole galaxy sample assuming their values to be the same for each galaxy\footnote{There is another set of model parameters that are fitted to each individual galaxy, and that control its particular star formation history (see Table 6 in \citealt{galliano2021}).}. Hence, the main assumption in this approach is that differences between galaxies are due to their particular star formation history.

Figure~\ref{fig:dustpedia1} shows the posterior distribution of the three tuning parameters assuming a Salpeter IMF. From this analysis, they infer the following values: $\langle Y_{\rm SN} \rangle \simeq 7.3^{+0.2}_{-0.3} \times 10^{-3} M_\odot/$SN, $\epsilon_{\rm growth} \simeq 4045^{+404}_{-354}$, $m^{\rm dest}_{\rm gas} \simeq 1288^{+7}_{-8} M_\odot$. 

\begin{figure}
\centerline{\includegraphics[width=8cm]{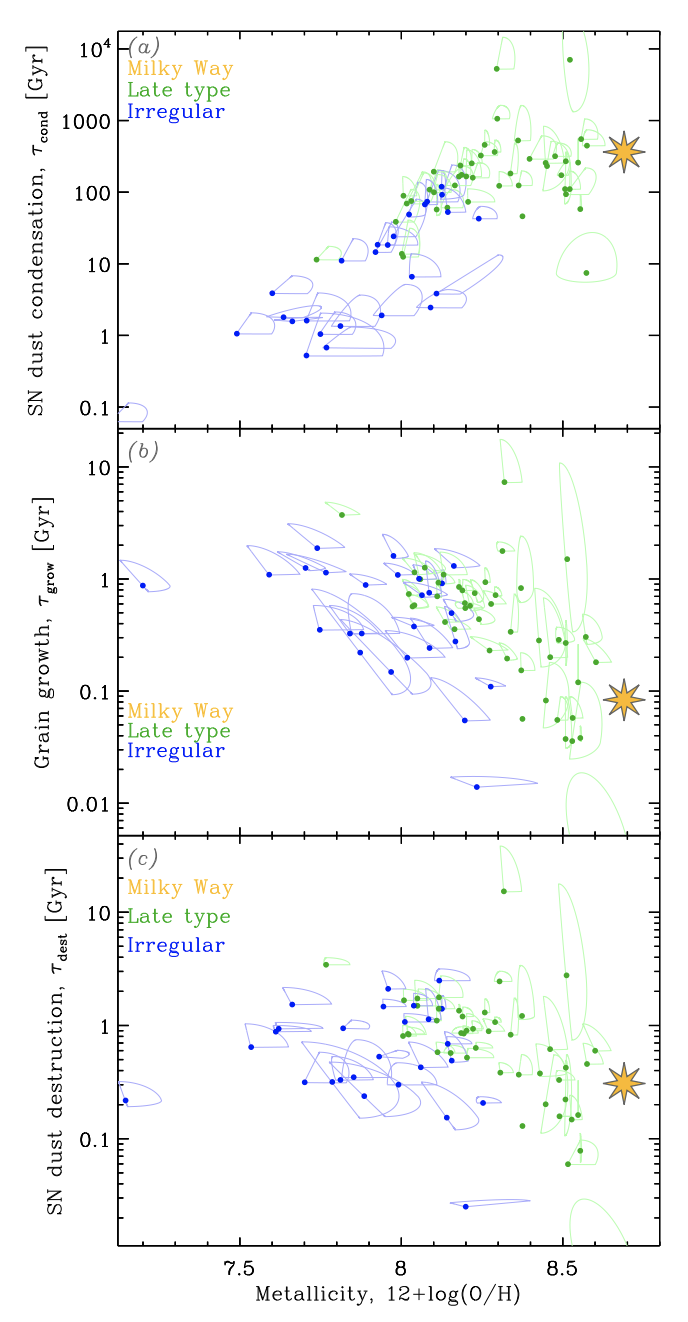}}
\caption{Dust evolution timescales for each galaxy of the sample analysed by \citet{galliano2021} as a function of metallicity.
The three panels display the posterior PDF for each galaxy, represented as skewed uncertainty ellipses, that are the 1$\sigma$ contour of a bivariate distribution having the same means, variances, skewnesses, and correlation coefficient as the posterior distribution. The central dot shows the mode of this bivariate distribution (see \citealt{galliano2021} for more details on this representation). 
The galaxies are color coded according to their Hubble stage, with early-type ($T \leq 0$) in red, late-types ($0< T < 9$) in green, and irregulars ($T \geq 9$) in blue.  
The yellow star represents the Milky Way, at the maximum a posteriori of the three tuning parameters. Image reproduced with permission from \citet{galliano2021}, copyright by the author(s).}
\label{fig:dustpedia2}
\end{figure}

From these values, it is possible to infer the posterior PDF of the dust evolution timescales for each galaxy, using Eqs. \ref{eq:taudes}, \ref{eq:taugg2}, and \ref{eq:taucond}. The results are shown in Fig.\ref{fig:dustpedia2}, where the three panels show $\tau_{\rm cond}$, $\tau_{\rm grow}$, and $\tau_{\rm dest}$ (from top to bottom) for each galaxy as a function of the metallicity, expressed as the oxygen over hydrogen abundance ratio\footnote{Following \citet{galliano2021}, we adopt here a solar metallicity value of 12 + log(O/H) = $8.69 \pm 0.05$, so that $Z/Z_\odot = 2.04\times 10^{-9} \times 10^{\rm (12+ log O/H)}$.}.  Although there is a scatter between galaxies with similar metallicity, which results from variations in their star formation histories, some clear trends appear: the dust condensation timescale increases with metallicity, and ranges from $\tau_{\rm cond} \sim 0.1 - 1$ Gyr for the most metal-poor systems to $\tau_{\rm cond} \sim 1000$ Gyr for solar metallicity galaxies, indicating that this process is too inefficient to account for the dust mass in chemically evolved galaxies. The grain growth timescale decreases with metallicity, ranging from $\tau_{\rm growth} \sim 1$ Gyr for the most metal-poor systems to $\tau_{\rm growth} \sim 0.05$ Gyr for solar metallicity galaxies, with an average value at 12 + log(O/H) $\geq$ 8.5 ($Z \geq 0.65 \, Z_\odot$) of 
$\sim 45$\, Myr, indicating that in this metallicity range dust formation is dominated by grain growth.
Finally, the dust destruction timescale is also very scattered but appears to stay approximately constant and with a value of $\tau_{\rm dest} \sim 0.3$ Gyr accross the metallicity range encompassed by the sample. These results imply that dust production is dominated by grain growth in
the ISM above a metallicity of 12 + log(O/H) $\simeq$ 8.0 ($Z \simeq 0.20 \, Z_\odot$).

It is important to stress that a key aspect of this analysis is that the galaxy sample extends to very low metallicity systems, in the range $0.03 \leq Z/Z_\odot \leq 0.2$. These systems appear to be very critical in constraining the main dust evolution tuning parameters, as they sample a regime where dust production is dominated by stellar sources. The steep non-linear dustiness-metallicity relation reported in the left panel of Fig.~\ref{fig:dustpedia1} supports the evidence that stardust can not be the dominant dust formation process in solar metallicity systems. This also explains the different conclusions drawn by \citet{delooze2020} by means of a bayesian analysis of the JINGLE galaxy sample, which have a more limited metallicity coverage (only 1 source between $0.13 \leq Z/Z_\odot \leq 0.2$, and none below this range). 
According to their analysis, the average galaxy scaling relations, including the dustiness vs metallicity, can be reproduced using closed-box galaxy evolution models (no gas infall or outflows) with a high fraction of SN dust that survives the reverse shock (37--89\%), low grain growth parameters $\epsilon_{\rm growth} \simeq 30 - 40$, and long dust destruction timescales, $\tau_{\rm dest} \simeq 1 - 2$ Gyr (these parameters are not assumed to be universal in their analysis). Hence, low-metallicity dwarfs appear key to disentangle the two main mechanism of dust production, effectively constraining the stellar dust yields \citep{galliano2021}.

While the above results appear promising, the analyses conducted so far rely on relatively simplified modeling of the relevant physical processes. More realistic star formation histories, metal-dependent stellar dust yields, as described in Sects.~\ref{sec:snmodels} and \ref{sec:agbmodels}, metal-dependent dust destruction efficiencies (as recently suggested by \citealt{priestley2022b}) need to be explored and incorporated in dust evolution models, particularly when attempting to constrain the origin of dust by comparing galaxy samples with a broad range of physical properties.

\section{Implications for dust formation scenarios in the early Universe}

In the second part of this review we will discuss in detail how these various channels are relevant for the production of dust in the early Universe, and especially at $z > 4$, by illustrating how they have been incorporated in cosmological models and by comparing them with available observational constraints.
However, here we provide an overview of the relevance and implications of these sources of dust for the enrichment in the early Universe.

Core collapse SNe are the most probable candidates for producing the bulk of, at least, dust seeds in the early Universe. Indeed, the dust yields of their ejecta is observed (and predicted) to be fairly high (more that $\rm 10^{-3}$ dust masses per every stellar mass formed); in combination with the high supernova rate in the earliest phases of stellar evolution (within the first few tens Myr), this potentially implies that grains produced by SNe might dominate the dust content in galaxies in the earliest cosmic epochs. However, the main source of uncertainty is the effect of the reverse shock, which, may destroy most of the dust produced in the ejecta. Models span a wide range of predictions about the dust survival in the ejecta, from nearly total disruption to most of the grains surviving. Additionally, even for what concerns the dust formed in the ejecta, models can span a wide range of yields and dust composition, depending on progenitor mass, metallicity, and model assumptions. These predictions can be directly tested with observations of supernovae only locally, hence spanning a limited parameter space. In the early Universe the metallicities and progenitors masses were much different than probed locally, hence implying that the local obsravational constraints may not fully apply. However, observational constraints on dust properties  and composition at high-$z$ ($M_{\rm dust}/M_{\rm star}$, temperature, emissivity and extinction curve) can shed light on the fraction of dust produced by SNe in the early Universe, and potentially even discriminate between different models.

Contrary to some early claims, AGBs can also be an early and effective source of dust at $z > 4$. Indeed, the first AGBs start producing dust as soon as $\sim$35~Myr after the onset of star formation. In the next review we will show that models can ascribe to AGBs even more than half of the dust observed in massive systems already in place at $z \sim 6$.
AGBs from progenitors more massive than $\sim 3~M_\odot$ (leaving the Main Sequence at 300~Myr, hence more likely to contribute to dust at high redshift), have grain production yields that plummets strongly with metallicity, which may be an issue at $z > 4$. In this regime the bulk of the dust produced is dominated by silicates, which could potentially be tested via observations. The production of dust in AGBs from lower mass progenitors is less dramatically sensitive on metallicity, and in this range the composition of most of grains is carbonaceous in nature. However, already for progenitors with masses of $2~M_\odot$ the lifetime on the main sequence is half a Gigayear, implying that the timescale for dust production would start to be in tension with the age of galaxies observed at $z > 6$ (generally no more than a few hundreds Myr) or even with the age of the Universe.

Red Super Giant stars (RSGs), which originate from  massive stars, can potentially be a fast dust enrichment channel. Their dust yields can approach that of SNe after reverse shock, hence they could potentially contribute to the dust enrichment in the early Universe. However, it is very likely that the supernova explosion at the end of the RSG phase destroys most of the dust produced. Therefore, RSGs are a less plausible channel for the dust production at high-$z$, and overall across most of the cosmic epochs.

Wolf--Rayet stars (WRs) are also a very fast factory of dust, which can therefore be potentially relevant in the earliest epochs, with yields even higher than RSGs. However, they are subject to the same issue as the latter, in the sense that the SNIb/c following the WR phase is likely to destroy most of the dust produced. Additionally, dust is observed to form only in the WC subclass of WR, and WC are not observed to form in low metallicity systems. The latter is an additional major problem, which makes WC an unlikely source of dust at $z > 4$, where the bulk of the galaxy population have subsolar metallicities.

Dense AGN driven winds are predicted to provide the optimal environment for dust production.
Given the large population of AGN revealed by JWST at $z > 4$, this is a potential channel of early enrichment that should be considered seriously. It does require the ISM to be pre-enriched, but observations show that most galaxies (including those hosting AGN) have already reached metallicities $\sim 0.1~Z_\odot$, even at $z \sim 11$. According to models, quasar winds can potentially produce as much dust as the black hole mass by $z \sim 6$, hence potentially accounting for the large dust masses observed in quasar hosts at this early epochs. However, the predicted dust mass is likely an upper limit, as it would require the black hole accreting at a substantial rate (Eddington or super-Eddington) for a few hundred million years. However, the primary problem remains that this scenario would require a direct observational evidence of dust formation in quasar winds.

Finally, the dust reprocessing in the ISM is likely much more relevant in the early Universe.
For instance the high SN rate, in the compact star forming regions characterizing galaxies in the first billion years, may result into an enhanced destruction rate by shocks in the ISM. The strong radiation pressure from young stars is also likely to preferentially expel dust grains from galaxies.

On the other hand, the high gas density of the ISM in  early galaxies may result into very short timescales ($\sim 10^6-10^7$~yr) for the growth of dust grains in the ISM at $z > 4$. Therefore, once dust seeds are produced through various possible channels (SNe, AGBs or AGN winds), growth in the ISM might likely provide the bulk of the dust mass. Of course, this requires the ISM to be enriched; however, as already mentioned, the ISM of galaxies is already fairly enriched by $z \sim 10$.

We will discuss more in detail these various scenarios in the second part of our review.

\section{Conclusions and outlook}

In this review we have attempted to summarise the findings on dust production mechanisms, which are relevant for understanding the formation of dust in the early Universe, and which will provide the backbone of the second part of the review, which will discuss the observational findings and models of dust at high redshift.
We have focused on theoretical models of dust sources, but also discussing comparison with observations.

Summarising the various models of dust formation is not simple, given that, as we have seen throughout the review, there is a large variety of models and assumptions, which can result into completely  different dust yield and dust properties for the same class of objects. With this caveat in mind, in the following we can provide some general summarizing consideration about the possible sources of dust.

\begin{itemize}

\item[$\bullet$] {\bf Core-Collapse Supernovae}

\begin{itemize}

\item Different models have been developed to describe the formation of dust in SN ejecta, from Classical Nucleation Theory (CNT) to Kinetic and Molecular Nucleation Theories (KNT, MNT). In these frameworks theories have spanned a broad range of assumptions in terms of  properties of the SN explosion,
dynamical evolution of the SN remnant, physical processes implemented in
the models, and late-time evolution of the grains when they cross the forward shock and are
slowed down in the ISM.

\item Depending on the assumptions and model prescriptions, the predicted dust masses produced by SN ejecta (before reverse shock) range between $\sim 0.03 \, M_\odot$ and $\sim 1~ \, M_\odot$.

\item These predictions embrace the mass of dust typically observed in SN remnants (about $0.4 \, M_\odot$, $\sim$50~years after the explosion); although we warn that there is a large dispersion in the measurements, and that the dust masses inferred from the observations can change significantly based on the different assumption to convert observational quantities to dust masses.

\item The reverse shock ($\sim 10^4$ yr after explosion) is recognised to be a crucial phenomenon that can drastically reduce the yield of dust production of SNe and reshape the distribution of grain sizes and species.
Yet, different models predict quite different survival rates, spanning from scenarios in which the disruption is essentially total to models that predict a survival rate of up to 80\%.

\item Despite these uncertainties it is generally acknowledged that supernova ejecta are likely the main channel for dust production associated with stars more massive than 8~$M_\odot$.

\end{itemize}

\item[$\bullet$] {\bf Asymptotic Giant Branch stars (AGBs)}

\begin{itemize}

\item This is the dominant dust production for low and intermediate mass stars ($0.8 \, M_\odot < m_{\rm star} < 8 \, M_\odot$). 

\item Most models adopt a simplified approach of stationary, spherically symmetric winds, due to the difficulties of integrating self-consistently the dust production in the complex AGB atmosphere models.  Yet depending on the detailed assumption about the structure of the stellar envelope and energetics of nucleosynthesis in the various evolutionary steps, the dust mass production can vary by more than one order of magnitude in certain stellar mass ranges, and in certain metallicity ranges.

\item Carbon grains are mostly produced in AGBs with masses in the range $2 \, M_\odot < m_{\rm star} < 3 - 3.5 \, M_\odot$ (which become C-rich during the TP-AGB phase).  This is the range for which the AGB dust production is maximum, reaching about $10^{-2} \, M_\odot$ per star, which, convolved with the IMF, results into the main dust production mechanisms in the local Universe and for evolved systems.

\item The limited mass range in which carbon dust is produced has also implications for observations in the early Universe, as AGBs are unlikely to produce carbon-rich dust in systems younger than about $\rm 300~Myr$

\item Silicate dust production requires initial metallicities $Z > 0.07 \, Z_\odot$, and it is formed primarily by low-mass AGBs ($m_{\rm star} < 2 \, M_\odot$), and by stars more massive than 4--5 $M_\odot$ (suffering Hot Bottom Burning). This implies that in young systems dust produced by AGBs is mostly in the form of silicates.

\end{itemize}

\item[$\bullet$] {\bf  Red Super Giant Stars (RSGs)}\\
RSGs with masses $10 < m_{\rm star}/M_\odot < 25$ are estimated to produce and eject a dust mass in the range $ -3.6 \leq {\rm log}\, m_{\rm dust}/M_\odot  \leq - 2.6$, which is comparable to the expected dust masses from core-collapse SNe, after reverse shock. However, it is likely that most of this dust is destroyed in the subsequent supernova explosion.

\item[$\bullet$] {\bf Wolf--Rayet (WR) stars}\\
 WR stars, formed out of stars more massive than $40 \, M_{\odot}$, are expected and observed to produce and expel gas, but only in the more evolved WC phase. The estimated dust production during this phase is about $10^{-2}\, M_\odot$. However, most of this dust is expected to be destroyed by the subsequent SN explosion. Moreover, WR stars are very rare (at least for a standard IMF) and are therefore expected to be a minor dust contributor. Finally, WCs are known to be absent in low-metallicity galaxies, hence unlikely to contribute to the dust production in the early Universe.

\item[$\bullet$] {\bf Classical Novae}\\
Classical Novae are observed to be another interesting source of dust, which form in the dense shells behind the shocks produced in the outburst. However, although the rate of these events is about 35 per year in the Milky-Way, each event is estimated to produce dust masses of only $\rm \sim 10^{-10}-10^{-7}~M_\odot$. Therefore, they are unlikely to be major contributors of dust. 

\item[$\bullet$] {\bf Type Ia SNe}\\
Although theoretical models predict that type Ia supernovae can produce dust masses in the range between $3 \times 10^{-4} - 0.2 \, M_\odot$, no observational evidence has been found for dust formed in their ejecta, with upper limits of $\sim 0.03 - 0.075 \, M_\odot$. The much higher energetics (and lower densities) involved in type Ia SN explosions are likely responsible for the lack of dust formation (and rapid destruction). The lack of iron dust formed in type Ia SNe, implies that most of the depletion of iron in dust grains must happen via accretion from the ISM.

\item[$\bullet$] {\bf Quasar winds}\\
The dense clouds ejected from quasars are expected to pass through a phase in which density and temperature are optimal for the nucleation of dust grains. In powerful quasars models expect that 
the dust production rate could be up to a few 
$M_\odot$/yr, 
hence they could potentially contribute to the dust enrichment of the host galaxy, even in the early Universe. However, it should be taken into account that this mechanism requires the pre-enrichment of the ISM in the nuclear region.

\end{itemize}

Clearly, the emerging picture is that the primary source of dust in galaxies are SN ejecta and AGB winds. Assessing the relative contribution of these two sources of dust at different times, hence at different cosmic epochs, is less obvious. While the production by SNe is essentially prompt, the production of dust by the first AGBs starts as early as 35 Myr after the beginning of star formation, and catches up fairly quickly (Fig.~\ref{fig:dustevo_summary}). Whether and when AGB dust dominates over the SN production (including reverse shock) depends on various factors, primarily the stellar IMF, metallicity, the degree of dust destruction by the SN reverse shock, and other parameters affecting the dust yield in the two phenomena. For instance, with a standard IMF AGB-dust may start dominating the dust content within a few 100 Myr, while in the case of a top heavy IMF the AGB dust may need more than 1~Gyr to dominate over the SN-produced dust. Such susceptibility on the IMF (but also on metallicity) has obviously important implications at high redshift if, as expected by various models, the IMF evolves with time (along with the chemical enrichment).

We have finally discussed that, regardless of the source of dust, its fate is unavoidably affected by its {\bf evolution and reprocessing in the ISM}. Interstellar shocks generated by SNe can easily destroy dust grains in the ISM and change their size distribution, depending on the shock velocity and on the initial size distribution of the grains. Another fundamental process is the growth of grains through the accretion of gas-phase metals. This process is still poorly understood, but in the cold neutral medium of the ISM can be very efficient and very rapid (with timescales as short as a few Myr), 
implying that possibly most of the dust mass in the ISM of evolved (metal rich) galaxies is actually primarily resulting from growth in the ISM. In the early Universe, the dense environments in high-redshift galaxies may enhance dust growth. However, the lower gas metallicity can result into grain growth timescales comparable to the age of the Universe at those epochs.  

The results reported in this review have highlighted the tremendous progress in modelling the dust formation and processing mechanisms, as well the impressive observational results obtained on dust sources through various cutting edge facilities. New models and numerical simulations, that consistently treat the dust formation with the complex physical processes and with less idealised assumptions, together with extensive observing programs, especially probing dust thermal emission in the far-IR/submm (e.g. with ALMA and NOEMA), and in the mid-IR (e.g. with JWST and, in the near future, with the ELT), will certainly provide tighter constraints on the relative role of the various dust production sources and processing, as well as on the resulting population of dust grains in different environments. However, the results obtained so far already provide an adequate backbone for the extensive models that are being developed to interpret the puzzling observations on dusty (and non-dusty) galaxies at high redshift.  These exciting observational results, that are being obtained with facilities such as ALMA and JWST, as well as the  models that have been proposed to explain them, will be the topic of the second part of this review. 

\begin{acknowledgements}
We dedicate this review to the memory of Stefania Marassi, a brilliant and enthusiastic scientist, who has
given important contributions to the understanding of dust formation and survival in supernovae. She will be greatly missed. \\
We are grateful to the Referees for their careful reading and useful suggestions.
We warmly thank Simone Bianchi, Ilse De Looze, Florian Kirchschlager, Fred Galliano, Mikako Matsuura, Hiroyuki Hirashita, Arka Sarangi, Isabelle Cherchneff, John Slavin for their constructive and insightful comments, and for allowing us to reproduce some of their figures.
RS thanks the Kavli Institute for Cosmology in Cambridge for their support and kind hospitality.
RM acknowledges support by the Science and Technology Facilities Council (STFC), by the ERC through Advanced Grant 695671 ``QUENCH'', and by the UKRI Frontier Research grant RISEandFALL. RM also acknowledges funding from a research professorship from the Royal Society.
\end{acknowledgements}

{\footnotesize\noindent
\textbf{Competing interests} The authors declare no competing interests.
}

\phantomsection
\addcontentsline{toc}{section}{References}
\bibliographystyle{spbasic-FS}
\bibliography{dustreview_cosmic_sources_rev}

\end{document}